\documentclass[pdflatex,sn-mathphys-num,iicol]{sn-jnl}

\usepackage{amsmath,amssymb,bm}
\usepackage{graphicx}
\usepackage{booktabs}
\usepackage{array}
\usepackage{dcolumn}
\usepackage{xcolor}
\usepackage{cuted}
\usepackage{hyperref}
\hypersetup{hidelinks}
\allowdisplaybreaks[2]

\newcommand{\wE}{\omega_{\oplus}}

\newcommand{\SCF}{\mathrm{Sun}}
\newcommand{\lab}{\mathrm{lab}}
\newcommand{\dd}{\mathrm{d}}
\newcommand{\LV}{\mathrm{LV}}
\newcommand{\SM}{\mathrm{S}}
\newcommand{\TM}{\mathrm{T}}

\begin{document}

\title[Multistripe torsion balances for Lorentz violation]{Centerline Registration and Dimension-Dependent Optimization of Multistripe Torsion Balances for Nonminimal Lorentz Violation}

\author[1]{\fnm{Pan-Pan} \sur{Wang}}
\author*[2]{\fnm{Rui} \sur{Luo}}\email{ruiluo@hust.edu.cn}
\author[3]{\fnm{Weisheng} \sur{Huang}}
\author[2]{\fnm{Tao} \sur{Jin}}
\author*[4]{\fnm{Cheng-Gang} \sur{Shao}}\email{cgshao@hust.edu.cn}

\affil[1]{\orgdiv{College of Physics}, \orgname{Chongqing University}, \orgaddress{\city{Chongqing}, \postcode{401331}, \country{China}}}
\affil[2]{\orgdiv{National Gravitation Laboratory, MOE Key Laboratory of Fundamental Physical Quantities Measurement, and School of Physics}, \orgname{Huazhong University of Science and Technology}, \orgaddress{\city{Wuhan}, \postcode{430074}, \country{China}}}
\affil[3]{\orgdiv{School of Mathematics and Statistics}, \orgname{South-Central Minzu University}, \orgaddress{\city{Wuhan}, \postcode{430074}, \country{China}}}
\affil[4]{\orgdiv{School of Physics and Optoelectronic Engineering}, \orgname{Yangtze University}, \orgaddress{\city{Jingzhou}, \postcode{434023}, \country{China}}}

\abstract{\normalfont\unboldmath
Higher-dimension operators in the gravity sector of the Standard-Model
Extension generate anisotropic interactions that grow rapidly at short range.
Building on earlier striped-balance studies, we develop a finite-volume
response and covariance-based optimization for the \(d=6\) and \(d=8\)
pure-gravity sectors.  We compare three, five, and seven stripes at \(d=8\)
and left, centered, and right pattern registrations in both sectors, using
horizontal and vertical stripe axes.  Volume integration gives
azimuth-dependent transfer matrices, while a Gaussian likelihood accounts for
coefficient correlations and the unequal uncertainties of constant and
sidereal amplitudes.  Three stripes give the strongest aggregate \(d=8\)
response.  Reoptimizing two azimuths for each registration increases the
noise-weighted \(D\)-optimal score by up to \(7.3\%\) at \(d=6\) and
\(35.6\%\) at \(d=8\).  The response matrices permit simultaneous fits of
all 14 or 22 real coefficient components.  The best marginalized
\(1\sigma\) sensitivities reach \(10^{-9}\,\mathrm{m}^{2}\) and
\(10^{-15}\,\mathrm{m}^{4}\), about two orders of magnitude beyond earlier
inverse-square-law limits for the most favorable components.  These results
identify centerline registration as an additional route to improved
short-range tests of Lorentz symmetry.
}

\keywords{Lorentz symmetry, short-range gravity, torsion balance, Standard-Model Extension}

\begingroup
\let\newpage\relax
\maketitle
\endgroup
\raggedbottom

\section{Introduction}
\label{sec:introduction}

Local Lorentz invariance (LLI), which requires the outcome of a local nongravitational experiment to be independent of the orientation and uniform velocity of a freely falling reference frame, is a central element of the Einstein equivalence principle and hence a foundational assumption of general relativity (GR) \cite{Will2014}. Lorentz symmetry is likewise built into the Standard Model (SM) of particle physics. Despite their remarkable empirical success, GR and the SM are widely regarded as low-energy limits of a more fundamental framework incorporating quantum gravity. Several candidate scenarios permit Lorentz symmetry to be broken spontaneously, for example through tensor fields acquiring nonzero vacuum expectation values \cite{KosteleckySamuel1989,Kostelecky2004}. The corresponding low-energy effects can be highly suppressed while remaining accessible to precision measurements. Searches for LLI violation therefore provide a sensitive and comparatively model-independent route for testing the foundations of GR and probing physics beyond the SM \cite{KosteleckyRussell2011,Tasson2016}. Updated coefficient compilations and recent gravity-sector studies continue to extend the scope and precision of experimental tests of Lorentz symmetry \cite{KosteleckyRussell2011,DongWangShao2024}.

A comprehensive effective-field-theory description of possible Lorentz and CPT violation is provided by the Standard-Model Extension (SME) \cite{ColladayKostelecky1997,ColladayKostelecky1998,Kostelecky2004}. In the gravity sector, possible effects beyond GR are organized according to the mass dimension $d$ of the corresponding operators. The minimal sector is dominated by $d=4$ operators, while higher-dimensional operators belong to the nonminimal sector and can produce short-range, direction-dependent corrections to the Newtonian interaction between nearby masses \cite{BaileyKostelecky2006,KosteleckyMewes2017}. Since the laboratory rotates with Earth, a fixed Lorentz-violating background would generally appear as a small torque signal varying with sidereal time. This makes short-range torsion-balance experiments a natural setting for searching for such effects.

Experimental coefficients are conventionally reported in the approximately inertial Sun-centered celestial-equatorial frame. The rotation of the Earth transforms the constant coefficients in this frame into time-dependent laboratory coefficients, producing characteristic variations at the sidereal frequency $\omega_{\oplus}$ and its harmonics. For operators of fixed mass dimension $d$, harmonics can occur through $(d-2)\omega_{\oplus}$ \cite{KosteleckyMewes2017}. In particular, the $d=6$ potential scales as $r^{-3}$ and produces an inverse-quartic force, with nine sidereal amplitudes depending on 14 independent coefficients with $j=2,4$. At $d=8$, the potential scales as $r^{-5}$ and produces an inverse-sixth-power force, with 13 sidereal amplitudes depending on 22 coefficients with $j=4,6$ \cite{Shao2019InverseSixth}. The rapid increase of these interactions at small separation gives short-range gravity experiments a pronounced advantage for testing nonminimal operators. Short-range experiments also probe coefficient combinations that are inaccessible to observations of gravitational-wave propagation, making the two approaches complementary \cite{KosteleckyMewes2017}.

Short-range searches have already demonstrated substantial sensitivity to these effects. Analyses of torsion-balance and resonant-oscillator data from the Huazhong University of Science and Technology (HUST) and Indiana University (IU), followed by a combined analysis, constrained the 14 observable $d=6$ coefficients at approximately the $10^{-9}\,\mathrm{m}^{2}$ level \cite{LongKostelecky2015,Shao2015ISL,Shao2016Combined}. A subsequent combined analysis constrained all 22 $d=8$ coefficients at approximately the $10^{-12}\,\mathrm{m}^{4}$ level \cite{Shao2019InverseSixth}. No statistically significant Lorentz-violating signal was found. These studies also exposed a limitation of conventional inverse-square-law experiments. Their planar test and source masses are designed primarily to maximize sensitivity to isotropic Yukawa-type interactions and to suppress the Newtonian background. For anisotropic Lorentz-violating interactions, however, volume averaging over parallel plates produces substantial cancellation. In the ideal limit of an infinite uniform plate, the Lorentz-violating force vanishes, so the measurable signal from finite plates is dominated by edge effects \cite{Shao2015ISL,Shao2016Enhanced}.

To enhance this finite-edge contribution, striped source and test masses were
proposed for short-range Lorentz-symmetry tests
\cite{Shao2016Enhanced,ChenTanShao2017}.  Repeated density discontinuities
increase the effective edge length, while a suitable offset between the facing
patterns can add the Lorentz-violating torques constructively and suppress the
leading Newtonian torque.  Subsequent work compared geometries with three,
five, and seven stripes for the \(d=6\) sector and identified a five-stripe
baseline \cite{Jin2026Optimization}.  A separate \(d=8\) study evaluated the
22 spherical-coefficient responses and two-azimuth sensitivity of a fixed
five-stripe design in its horizontal and vertical orientations
\cite{Jin2024d8}.  Li et al.\ studied two-azimuth coefficient estimation
for a \(d=6\) striped balance \cite{Li2025CQG}.  Together, these studies
established the striped geometry, azimuth-selection strategies, and
sector-specific baselines.

In this work, we formulate a finite-volume and covariance-based comparison
of stripe count, orientation, and centerline registration.  The \(d=8\)
stripe-count study selects three stripes.  For \(d=6\), we use the
five-stripe baseline of Ref.~\cite{Jin2026Optimization} and focus on the
previously unexplored left--center--right registration.  Each sector is
evaluated independently, and the operating azimuths are reoptimized for each
registration using a noise-weighted Fisher criterion.  For each configuration, the interaction is
integrated over the finite mass volumes to obtain transfer matrices connecting
the Sun-centered Newton spherical coefficients to the observable sidereal
torque amplitudes.  Unweighted aggregate transfer functions first expose the
contributions of the independent harmonic blocks.  A Gaussian likelihood and
the associated Fisher matrix then quantify the simultaneous coefficient
sensitivity, including estimator correlations and the different uncertainties
of the constant and sidereal channels.  This separation shows when a larger
geometric response does, and does not, produce a smaller joint covariance.
The resulting optimization over stripe number, centerline registration,
stripe orientation, and apparatus azimuth provides a design strategy for
next-generation searches for nonminimal Lorentz violation and other
anisotropic short-range interactions.

The remainder of this paper is organized as follows. Section~\ref{sec:theory}
derives the torsion-balance observables for the \(d=6\) and \(d=8\) sectors.
Section~\ref{sec:geometry} describes the striped mass configurations and
Newtonian background. Section~\ref{sec:calculation} presents the parameter
estimator, covariance, and response criteria. Section~\ref{sec:results}
compares centerline registrations in both sectors and stripe counts at
\(d=8\), then evaluates coefficient sensitivity.
Section~\ref{sec:conclusion} summarizes the results.

\section{Theoretical framework}
\label{sec:theory}

\subsection{Lorentz-violating short-range gravitational potential}
\label{subsec:lvpotential}

The theoretical basis of this work is the pure-gravity sector of the SME, in which possible violations of LLI are organized as effective operators of mass dimension \(d\) \cite{BaileyKostelecky2006,KosteleckyMewes2017}.  In the quasistatic weak-field limit appropriate for a laboratory torsion balance, the potential between two point masses may be written as
\begin{equation}
  V(\bm r,T)
  =
  -\frac{Gm_1m_2}{r}
  +
  \sum_d V_{\LV}^{(d)}(\bm r,T),
  \label{eq:total-potential}
\end{equation}
where \(\bm r=\bm r_1-\bm r_2\), \(r=|\bm r|\), and \(T\) denotes local sidereal time.  For the nonminimal even-\(d\) operators considered here, the Lorentz-violating part can be expressed in the Newton spherical-coefficient basis as
\begin{equation}
  V_{\LV}^{(d)}(\bm r,T)
  =
  -Gm_1m_2
  \sum_{jm}
  \frac{Y_{jm}(\hat{\bm r})}{r^{d-3}}\,
  k_{jm}^{N(d),\lab}(T).
  \label{eq:point-potential}
\end{equation}
The spherical harmonics make the anisotropic angular structure explicit, while the radial scaling shows why short source-test separations are especially valuable.  The two cases studied in this proposal are \(d=6\), for which \(V_{\LV}^{(6)}\propto r^{-3}\), and \(d=8\), for which \(V_{\LV}^{(8)}\propto r^{-5}\).  With the usual reality and trace relations imposed, these sectors contain 14 and 22 real observable coefficient combinations, respectively.  The conventional Newtonian term in Eq.~\eqref{eq:total-potential} is treated as a background torque and is evaluated separately from the Lorentz-violating signal.

In the approximately inertial Sun-centered frame introduced above, the SME coefficients are time independent, whereas the laboratory-frame combinations appearing in Eq.~\eqref{eq:point-potential} acquire a sidereal modulation from Earth's rotation.  The transformation can be written with Wigner rotation matrices,
\begin{equation}
  k_{jm}^{N(d),\lab}(T)
  =
  \sum_{m'=-j}^{j}
  D^{(j)}_{m m'}(\alpha,\beta,\gamma)
  k_{jm'}^{N(d),\SCF},
  \label{eq:wigner}
\end{equation}
where the Euler angles are fixed by the laboratory colatitude \(\chi\), the azimuthal orientation of the apparatus, and the local sidereal phase \(\wE T\).  It is convenient to collect the same rotation into a geometry-dependent laboratory response matrix,
\begin{equation}
  k_{jm}^{N(d),\lab}(T)
  =
  \sum_{m'=-j}^{j}
  {\cal R}^{(j)}_{m m'}(\chi,\theta,\wE T)
  k_{jm'}^{N(d),\SCF},
  \label{eq:rotation-symbolic}
\end{equation}
with \(\theta\) denoting the horizontal orientation of the pendulum axis relative to the local reference direction. Here \(m\) labels the laboratory-frame spherical component and \(m'\) labels the corresponding Sun-frame projection; the prime distinguishes the two frame labels and does not denote differentiation. The phase factor associated with Earth's rotation implies that the Sun-frame sector \(|m'|\) contributes at the sidereal harmonic \(|m'|\wE\). Consequently the \(d=6\) signal contains harmonics through \(4\wE\), while the \(d=8\) signal contains harmonics through \(6\wE\).

A torsion balance measures the torque about the suspension fiber rather than the potential itself.  If \(\varphi\) is the controlled horizontal relative angle between the pendulum and source mass, the Lorentz-violating torque from a point-mass pair is
\begin{equation}
  \tau_{\LV,\mathrm{pt}}^{(d)}(\bm r,T,\varphi)
  =
  -\frac{\partial V_{\LV}^{(d)}(\bm r,T,\varphi)}{\partial \varphi}.
  \label{eq:torque-definition}
\end{equation}
After inserting Eq.~\eqref{eq:rotation-symbolic}, the measurable sidereal signal has the finite Fourier form
\begin{equation}
\begin{split}
  \tau_{\LV}^{(d)}(T,\varphi)
  ={}&
  C_0^{(d)}(\varphi)
  \\
  &+
  \sum_{n=1}^{n_{\max}}\Bigl[
  C_n^{(d)}(\varphi)\cos(n\wE T)
  \\
  &\hspace{3.0em}
  +
  S_n^{(d)}(\varphi)\sin(n\wE T)
  \Bigr].
\end{split}
  \label{eq:sidereal-fourier}
\end{equation}
where \(n_{\max}=4\) for \(d=6\) and \(n_{\max}=6\) for \(d=8\).  The amplitudes \(C_n^{(d)}\) and \(S_n^{(d)}\) are not fundamental constants.  They are response amplitudes determined by the finite mass geometry, the source-test gap, the density distribution, the laboratory colatitude, and the apparatus orientation. Different sidereal harmonics therefore probe different Sun-frame projection sectors. The constant channel is associated with \(m'=0\), the first harmonic with \(|m'|=1\), and so on.

\subsection{Spherical-harmonic response and transfer amplitudes}
\label{subsec:finite-response}

The torque and sidereal expansion in Eqs.~\eqref{eq:torque-definition} and
\eqref{eq:sidereal-fourier} show how a torsion balance converts the
Lorentz-violating potential into measurable Fourier amplitudes. We now make
the angular response explicit. In this subsection,
\(\theta_1\equiv\varphi\) denotes the horizontal relative angle between the
pendulum and source mass, whereas \(\theta\) remains the azimuth of the
apparatus relative to local south.

The spherical basis is convenient for rotations between the Sun-centered and
laboratory frames. The finite test and source masses, however, are assembled
from rectangular prisms, so a Cartesian representation is useful for the
volume integrations. Let
\begin{equation}
  \bm r=(x,y,z),\qquad r^2=x^2+y^2+z^2,
  \label{eq:local-separation-vector}
\end{equation}
where \(x,y,z\) are the Cartesian components of the separation vector between
a source-mass element and a test-mass element. For a rank-\(j\) multi-index
\(J=a_1\cdots a_j\), the spherical harmonic is first written in the
symmetric-trace-free (STF) form \cite{LedesmaMewes2020,Shao2019InverseSixth}
\begin{equation}
  Y_{jm}(\hat{\bm r})
  =c_{jm}^{*\langle J\rangle}n_{\langle J\rangle}(x,y,z).
  \label{eq:stf-kernel}
\end{equation}
Here the direction tensor in Eq.~\eqref{eq:stf-kernel} is generated from the
Newtonian kernel according to
\begin{equation}
  n_{\langle J\rangle}(x,y,z)
  =\frac{r^{j+1}}{(-1)^j(2j-1)!!}
  \partial_{\langle J\rangle}\!\left(\frac{1}{r}\right),
  \label{eq:stf-direction-tensor}
\end{equation}
while the constant STF tensor that selects a definite \((j,m)\) component is
fixed by the inverse projection
\begin{equation}
  c_{jm}^{\langle J\rangle}
  =\frac{(2j+1)!!}{4\pi j!}
  \int n^{\langle J\rangle}Y_{jm}^{*}(\hat{\bm r})\,\dd\Omega.
  \label{eq:stf-coefficient-tensor}
\end{equation}
Equations~\eqref{eq:stf-direction-tensor} and
\eqref{eq:stf-coefficient-tensor} therefore define, respectively, the
geometry-dependent direction tensor and the numerical projection tensor that
appear in the first relation.

Using the Condon--Shortley phase convention, the \(j=2\) harmonics required
for the \(d=6\) sector are
\begin{align}
Y_{2,-2}&=\sqrt{\frac{15}{2\pi}}\frac{(x-iy)^2}{4r^2},
\nonumber\\
Y_{2,-1}&=\sqrt{\frac{15}{2\pi}}\frac{(x-iy)z}{2r^2},
\nonumber\\
Y_{2,0}&=-\sqrt{\frac{5}{\pi}}
\frac{x^2+y^2-2z^2}{4r^2},
\nonumber\\
Y_{2,1}&=-\sqrt{\frac{15}{2\pi}}\frac{(x+iy)z}{2r^2},
\nonumber\\
Y_{2,2}&=\sqrt{\frac{15}{2\pi}}\frac{(x+iy)^2}{4r^2}.
\label{eq:d6-y2-harmonics}
\end{align}
The corresponding \(j=4\) harmonics are
\begin{align}
Y_{4,-4}&=3\sqrt{\frac{35}{2\pi}}\frac{(x-iy)^4}{16r^4},
\nonumber\\
Y_{4,-3}&=3\sqrt{\frac{35}{\pi}}\frac{(x-iy)^3z}{8r^4},
\nonumber\\
Y_{4,-2}&=-3\sqrt{\frac{5}{2\pi}}
\frac{(x-iy)^2(x^2+y^2-6z^2)}{8r^4},
\nonumber\\
Y_{4,-1}&=-3\sqrt{\frac{5}{\pi}}
\nonumber\\[-0.2ex]
&\quad\times
\frac{(x-iy)z(3x^2+3y^2-4z^2)}{8r^4},
\nonumber\\
Y_{4,0}&=\frac{3}{16\sqrt\pi}
\left(3-30\frac{z^2}{r^2}+35\frac{z^4}{r^4}\right),
\nonumber\\
Y_{4,1}&=3\sqrt{\frac{5}{\pi}}
\nonumber\\[-0.2ex]
&\quad\times
\frac{(x+iy)z(3x^2+3y^2-4z^2)}{8r^4},
\nonumber\\
Y_{4,2}&=-3\sqrt{\frac{5}{2\pi}}
\frac{(x+iy)^2(x^2+y^2-6z^2)}{8r^4},
\nonumber\\
Y_{4,3}&=-3\sqrt{\frac{35}{\pi}}\frac{(x+iy)^3z}{8r^4},
\nonumber\\
Y_{4,4}&=3\sqrt{\frac{35}{2\pi}}\frac{(x+iy)^4}{16r^4}.
\label{eq:d6-y4-harmonics}
\end{align}
The remaining components follow from
\(Y_{j,-m}=(-1)^mY_{jm}^{*}\). The same reality relation for the Newton
spherical coefficients, \(k_{j,-m}^{N(d)}=(-1)^m
k_{jm}^{N(d)*}\), permits the response to be expressed in terms of real and
imaginary parts with \(m\geq0\).

To make this pairing explicit, introduce the rotated angular component
\begin{equation}
  {\cal A}_{jm'}(\hat{\bm r};\chi)
  \equiv
  \sum_{m=-j}^{j}Y_{jm}(\hat{\bm r})
  d_{m m'}^{j}(-\chi),
  \label{eq:A-factor-definition}
\end{equation}
where \(d_{m m'}^{j}\) is a Wigner small-\(d\) matrix and the sign of
\(\chi\) follows the Euler-angle convention in Eq.~\eqref{eq:wigner}. For
\(m'>0\), define the two real quadrature functions
\begin{equation}
\begin{aligned}
  A_{jm'}^{(c)}&\equiv
  {\cal A}_{jm'}+(-1)^{m'}{\cal A}_{j,-m'},\\
  A_{jm'}^{(s)}&\equiv
  i\left[{\cal A}_{jm'}-(-1)^{m'}{\cal A}_{j,-m'}\right],\\
  A_{j0}&\equiv {\cal A}_{j0}.
\end{aligned}
  \label{eq:A-quadrature-definition}
\end{equation}
For \(d=6\), the response functions used in each sidereal channel can be
collected as
\begin{align}
 \bm A_0&=(A_{20},A_{40}),
 \nonumber\\
 \bm A_1&=(A_{21}^{(c)},A_{21}^{(s)},
 A_{41}^{(c)},A_{41}^{(s)}),
 \nonumber\\
 \bm A_2&=(A_{22}^{(c)},A_{22}^{(s)},
 A_{42}^{(c)},A_{42}^{(s)}),
 \nonumber\\
 \bm A_3&=(A_{43}^{(c)},A_{43}^{(s)}),
 \nonumber\\
 \bm A_4&=(A_{44}^{(c)},A_{44}^{(s)}).
 \label{eq:d6-A-groups}
\end{align}
Here the subscript on \(\bm A_{m'}\) labels the harmonic at
\(m'\wE\), while its entries enumerate the allowed \(j\) values and the two
quadratures. The same construction extends to \(m'=0,\ldots,6\) for
\(d=8\), with \(j=4,6\).

The phase factors \(e^{\pm im'\wE T}\), together with
\(k_{j,-m'}^{N(d)}=(-1)^{m'}k_{jm'}^{N(d)*}\), give the explicit mapping
\begin{align}
 \operatorname{Re}k_{jm'}^{N(d)}&\longmapsto
 \nonumber\\
 &A_{jm'}^{(c)}\cos(m'\wE T)
  +A_{jm'}^{(s)}\sin(m'\wE T),
 \nonumber\\
  \operatorname{Im}k_{jm'}^{N(d)}&\longmapsto
  \nonumber\\
  &A_{jm'}^{(s)}\cos(m'\wE T)
  -A_{jm'}^{(c)}\sin(m'\wE T).
 \label{eq:real-imaginary-A-rule}
\end{align}
This relation also explains the complex notation adopted below. At
\(\chi=0\), \(d_{m m'}^{j}(0)=\delta_{m m'}\), and therefore
\begin{equation}
\begin{aligned}
 A_{jm'}^{(c)}\big|_{\chi=0}
 &=2\operatorname{Re}Y_{jm'},\\
 A_{jm'}^{(s)}\big|_{\chi=0}
 &=-2\operatorname{Im}Y_{jm'}.
\end{aligned}
 \label{eq:A-to-spherical-harmonic}
\end{equation}
Combining these two quadratures gives
\(A_{jm'}^{(c)}-iA_{jm'}^{(s)}=2Y_{jm'}\) at \(\chi=0\). Thus each pair of
real point-pair functions is packaged as
\(\gamma_{2q-1}^{(d)}-i\gamma_{2q}^{(d)}=2Y_{jm'}\). The negative-\(m\)
choice used later for the odd \(d=8\) block is the equivalent basis obtained
from the same reality relation.

For \(d=6\), it is convenient to combine the 14 real point-pair functions
into two real functions and six complex pairs. In the notation used below,
the superscript on the lowercase \(\gamma\) identifies the operator
dimension. The functions are
\begingroup
\setlength{\mathindent}{1em}
\begin{align}
\gamma_1^{(6)}=Y_{2,0}
&=\sqrt{\frac{5}{16\pi}}
\left(-1+3\frac{z^2}{r^2}\right),
\nonumber\\
\gamma_2^{(6)}=Y_{4,0}
&=\frac{3}{16\sqrt\pi}
\nonumber\\
&\quad\times
\left(3-30\frac{z^2}{r^2}+35\frac{z^4}{r^4}\right),
\nonumber\\
 \gamma_3^{(6)}-i\gamma_4^{(6)}
 &=2Y_{2,2}
 \nonumber\\
 &=\sqrt{\frac{15}{2\pi}}\frac{(x+iy)^2}{2r^2},
\nonumber\\
\gamma_5^{(6)}-i\gamma_6^{(6)}=2Y_{4,2}
&=\frac{3}{4}\sqrt{\frac{5}{2\pi}}
\frac{(x+iy)^2}{r^2}
\nonumber\\[-0.2ex]
&\quad\times
\left(-1+7\frac{z^2}{r^2}\right),
\nonumber\\
\gamma_7^{(6)}-i\gamma_8^{(6)}=2Y_{4,4}
&=\frac{3}{8}\sqrt{\frac{35}{2\pi}}
\frac{(x+iy)^4}{r^4},
\nonumber\\
 \gamma_9^{(6)}-i\gamma_{10}^{(6)}
 &=2Y_{2,1}
 \nonumber\\
 &=-\sqrt{\frac{15}{2\pi}}\frac{(x+iy)z}{r^2},
\nonumber\\
\gamma_{11}^{(6)}-i\gamma_{12}^{(6)}=2Y_{4,1}
&=\frac{3}{4}\sqrt{\frac{5}{\pi}}
\frac{(x+iy)z}{r^2}
\nonumber\\[-0.2ex]
&\quad\times
\left(3-7\frac{z^2}{r^2}\right),
\nonumber\\
 \gamma_{13}^{(6)}-i\gamma_{14}^{(6)}
 &=2Y_{4,3}
 \nonumber\\
 &=-\frac{3}{4}\sqrt{\frac{35}{\pi}}
\frac{(x+iy)^3z}{r^4}.
\label{eq:d6-gamma-functions}
\end{align}
\endgroup

The \(d=8\) sector contains 22 real point-pair functions associated with
\(j=4\) and \(j=6\). We adopt the ordering
\begin{equation}
\begin{aligned}
\gamma_1^{(8)}&=Y_{4,0},\\
\gamma_2^{(8)}&=Y_{6,0},\\
\gamma_3^{(8)}-i\gamma_4^{(8)}&=2Y_{4,2},\\
\gamma_5^{(8)}-i\gamma_6^{(8)}&=2Y_{6,2},\\
\gamma_7^{(8)}-i\gamma_8^{(8)}&=2Y_{4,4},\\
\gamma_9^{(8)}-i\gamma_{10}^{(8)}&=2Y_{6,4},\\
\gamma_{11}^{(8)}-i\gamma_{12}^{(8)}&=2Y_{6,6},\\
\gamma_{13}^{(8)}-i\gamma_{14}^{(8)}&=2Y_{4,-1},\\
\gamma_{15}^{(8)}-i\gamma_{16}^{(8)}&=2Y_{6,-1},\\
\gamma_{17}^{(8)}-i\gamma_{18}^{(8)}&=2Y_{4,-3},\\
\gamma_{19}^{(8)}-i\gamma_{20}^{(8)}&=2Y_{6,-3},\\
\gamma_{21}^{(8)}-i\gamma_{22}^{(8)}&=2Y_{6,-5}.
\end{aligned}
\label{eq:d8-gamma-basis}
\end{equation}
The negative-\(m\) entries in the odd block are a basis convention for the
point-pair functions. They are equivalent to a positive-\(m\) coefficient
basis through the reality relations stated above. The Cartesian expressions
follow directly by substituting the \(j=4\) harmonics in
Eq.~\eqref{eq:d6-y4-harmonics} and the \(j=6\) harmonics listed in
Appendix~\ref{app:j6-harmonics}. In Cartesian form, the
functions in Eq.~\eqref{eq:d8-gamma-basis} are
\begin{align}
\gamma_1^{(8)}
&=\frac{1}{16\sqrt\pi}
\left(9-90\frac{z^2}{r^2}+105\frac{z^4}{r^4}\right),
\nonumber\\
\gamma_2^{(8)}
&=-\frac{1}{32}\sqrt{\frac{13}{\pi}}
\biggl(5-105\frac{z^2}{r^2}+315\frac{z^4}{r^4}
\nonumber\\[-0.2ex]
&\hspace{7.4em}-231\frac{z^6}{r^6}\biggr),
\nonumber\\
\gamma_3^{(8)}-i\gamma_4^{(8)}
&=-\frac{3}{4}\sqrt{\frac{5}{2\pi}}
\frac{(x+iy)^2}{r^2}
\nonumber\\[-0.2ex]
&\quad\times
\left(1-7\frac{z^2}{r^2}\right),
\nonumber\\
\gamma_5^{(8)}-i\gamma_6^{(8)}
&=\frac{1}{32}\sqrt{\frac{1365}{\pi}}
\frac{(x+iy)^2}{r^2}
\nonumber\\[-0.2ex]
&\quad\times
\left(1-18\frac{z^2}{r^2}+33\frac{z^4}{r^4}\right),
\nonumber\\
\gamma_7^{(8)}-i\gamma_8^{(8)}
&=\frac{3}{8}\sqrt{\frac{35}{2\pi}}
\frac{(x+iy)^4}{r^4},
\nonumber\\
\gamma_9^{(8)}-i\gamma_{10}^{(8)}
&=-\frac{3}{16}\sqrt{\frac{91}{2\pi}}
\frac{(x+iy)^4}{r^4}
\nonumber\\[-0.2ex]
&\quad\times
\left(1-11\frac{z^2}{r^2}\right),
\nonumber\\
\gamma_{11}^{(8)}-i\gamma_{12}^{(8)}
&=\frac{1}{32}\sqrt{\frac{3003}{\pi}}
\frac{(x+iy)^6}{r^6},
\nonumber\\
\gamma_{13}^{(8)}-i\gamma_{14}^{(8)}
&=-\frac{3}{4}\sqrt{\frac{5}{\pi}}
\frac{(x-iy)z}{r^2}
\nonumber\\[-0.2ex]
&\quad\times
\left(3-7\frac{z^2}{r^2}\right),
\nonumber\\
\gamma_{15}^{(8)}-i\gamma_{16}^{(8)}
&=\frac{1}{8}\sqrt{\frac{273}{2\pi}}
\frac{(x-iy)z}{r^2}
\nonumber\\[-0.2ex]
&\quad\times
\left(5-30\frac{z^2}{r^2}+33\frac{z^4}{r^4}\right),
\nonumber\\
\gamma_{17}^{(8)}-i\gamma_{18}^{(8)}
&=\frac{3}{4}\sqrt{\frac{35}{\pi}}
\frac{(x-iy)^3z}{r^4},
\nonumber\\
\gamma_{19}^{(8)}-i\gamma_{20}^{(8)}
&=-\frac{1}{16}\sqrt{\frac{1365}{\pi}}
\frac{(x-iy)^3z}{r^4}
\nonumber\\[-0.2ex]
&\quad\times
\left(3-11\frac{z^2}{r^2}\right),
\nonumber\\
\gamma_{21}^{(8)}-i\gamma_{22}^{(8)}
&=\frac{3}{16}\sqrt{\frac{1001}{\pi}}
\frac{(x-iy)^5z}{r^6}.
\label{eq:d8-gamma-functions}
\end{align}

Equations~\eqref{eq:d6-gamma-functions}, \eqref{eq:d8-gamma-basis}, and
\eqref{eq:d8-gamma-functions} use the ordinary
components \(x,y,z\) in Eq.~\eqref{eq:local-separation-vector}. If the
laboratory rotation is absorbed into the angular functions, the same
expressions are evaluated at the rotated arguments
\((x_\theta,y_\theta,z_\theta)\), where
\begin{align}
x_\theta\equiv\tilde x
&=x\cos\theta\cos\chi-y\sin\theta\cos\chi+z\sin\chi,
\nonumber\\
y_\theta\equiv\tilde y
&=x\sin\theta+y\cos\theta,
\label{eq:rotated-cartesian-coordinates}\\
z_\theta\equiv\tilde z
&=-x\cos\theta\sin\chi+y\sin\theta\sin\chi+z\cos\chi.
\nonumber
\end{align}
A tilde therefore denotes a rotated component of the same separation vector,
not a new coordinate.

The lowercase \(\gamma_i^{(d)}\) describe one point-mass pair. The uppercase
transfer coefficients \(\Gamma_i^{(d)}\) are obtained after integrating over
the finite source and test bodies. For \(d\in\{6,8\}\), define
\begin{equation}
\begin{split}
\Gamma_i^{(d)}(\theta,\theta_1)={}&
G\sum_{a\in\TM}\sum_{b\in\SM}\rho_a\rho_b
\int_{V_a}\!\dd^3x\int_{V_b}\!\dd^3x'\\
&\times\frac{\partial}{\partial\theta_1}
\left[
\frac{\gamma_i^{(d)}(\theta,\hat{\bm r}(\theta_1),\chi)}
{r^{d-3}(\theta_1)}
\right],
\end{split}
\label{eq:gamma-volume-definition}
\end{equation}
where \(\bm r=\bm x-\bm x'\), \(i=1,\ldots,14\) for \(d=6\), and
\(i=1,\ldots,22\) for \(d=8\). This definition separates the
geometry-dependent response from the fundamental Sun-frame coefficients.

For either operator dimension, the relation between the coefficient vector
and the measured sidereal amplitudes is linear.
\begin{equation}
\bm a^{(d)}(\theta,\theta_1)
=\mathbf G^{(d)}(\theta,\theta_1)\bm k^{(d)},
\qquad d\in\{6,8\}.
\label{eq:transfer-matrix}
\end{equation}
For \(d=6\), we order the coefficient and amplitude vectors as
\begin{equation}
\begin{split}
\bm k^{(6)}=(&k_{2,0}^{(6)},k_{4,0}^{(6)},
\operatorname{Re}k_{2,2}^{(6)},\operatorname{Im}k_{2,2}^{(6)},
\operatorname{Re}k_{4,2}^{(6)},\operatorname{Im}k_{4,2}^{(6)},\\
&\operatorname{Re}k_{4,4}^{(6)},\operatorname{Im}k_{4,4}^{(6)},
\operatorname{Re}k_{2,1}^{(6)},\operatorname{Im}k_{2,1}^{(6)},\\
&\operatorname{Re}k_{4,1}^{(6)},\operatorname{Im}k_{4,1}^{(6)},
\operatorname{Re}k_{4,3}^{(6)},\operatorname{Im}k_{4,3}^{(6)})^T,
\end{split}
\label{eq:d6-coeff-basis}
\end{equation}
\begin{equation}
\begin{split}
\bm a^{(6)}=(&C_0^{(6)},C_2^{(6)},S_2^{(6)},C_4^{(6)},S_4^{(6)},\\
&C_1^{(6)},S_1^{(6)},C_3^{(6)},S_3^{(6)})^T,
\end{split}
\label{eq:amplitude-vector-general}
\end{equation}
where \(k_{j,m}^{(6)}\equiv k_{jm}^{N(6),\SCF}\). The entries of
Eq.~\eqref{eq:transfer-matrix} are then
\begin{align}
C_0^{(6)}={}&\Gamma_1^{(6)} k_{2,0}^{(6)}
+\Gamma_2^{(6)} k_{4,0}^{(6)},
\nonumber\\
C_2^{(6)}={}&\Gamma_3^{(6)}\operatorname{Re}k_{2,2}^{(6)}
+\Gamma_4^{(6)}\operatorname{Im}k_{2,2}^{(6)}
\nonumber\\
&+\Gamma_5^{(6)}\operatorname{Re}k_{4,2}^{(6)}
+\Gamma_6^{(6)}\operatorname{Im}k_{4,2}^{(6)},
\nonumber\\
S_2^{(6)}={}&\Gamma_4^{(6)}\operatorname{Re}k_{2,2}^{(6)}
-\Gamma_3^{(6)}\operatorname{Im}k_{2,2}^{(6)}
\nonumber\\
&+\Gamma_6^{(6)}\operatorname{Re}k_{4,2}^{(6)}
-\Gamma_5^{(6)}\operatorname{Im}k_{4,2}^{(6)},
\nonumber\\
C_4^{(6)}={}&\Gamma_7^{(6)}\operatorname{Re}k_{4,4}^{(6)}
+\Gamma_8^{(6)}\operatorname{Im}k_{4,4}^{(6)},
\nonumber\\
S_4^{(6)}={}&\Gamma_8^{(6)}\operatorname{Re}k_{4,4}^{(6)}
-\Gamma_7^{(6)}\operatorname{Im}k_{4,4}^{(6)},
\label{eq:d6-response-even}
\end{align}
and
\begin{align}
C_1^{(6)}={}&\Gamma_9^{(6)}\operatorname{Re}k_{2,1}^{(6)}
+\Gamma_{10}^{(6)}\operatorname{Im}k_{2,1}^{(6)}
\nonumber\\
&+\Gamma_{11}^{(6)}\operatorname{Re}k_{4,1}^{(6)}
+\Gamma_{12}^{(6)}\operatorname{Im}k_{4,1}^{(6)},
\nonumber\\
S_1^{(6)}={}&\Gamma_{10}^{(6)}\operatorname{Re}k_{2,1}^{(6)}
-\Gamma_9^{(6)}\operatorname{Im}k_{2,1}^{(6)}
\nonumber\\
&+\Gamma_{12}^{(6)}\operatorname{Re}k_{4,1}^{(6)}
-\Gamma_{11}^{(6)}\operatorname{Im}k_{4,1}^{(6)},
\nonumber\\
C_3^{(6)}={}&\Gamma_{13}^{(6)}\operatorname{Re}k_{4,3}^{(6)}
+\Gamma_{14}^{(6)}\operatorname{Im}k_{4,3}^{(6)},
\nonumber\\
S_3^{(6)}={}&\Gamma_{14}^{(6)}\operatorname{Re}k_{4,3}^{(6)}
-\Gamma_{13}^{(6)}\operatorname{Im}k_{4,3}^{(6)}.
\label{eq:d6-explicit-response}
\end{align}

For \(d=8\), the 22 independent coefficient combinations and 13 sidereal
amplitudes are ordered as
\begin{equation}
\begin{split}
\bm k^{(8)}=(&k_{4,0}^{(8)},k_{6,0}^{(8)},\\
&\operatorname{Re}k_{4,2}^{(8)},\operatorname{Im}k_{4,2}^{(8)},\\
&\operatorname{Re}k_{6,2}^{(8)},\operatorname{Im}k_{6,2}^{(8)},\\
&\operatorname{Re}k_{4,4}^{(8)},\operatorname{Im}k_{4,4}^{(8)},\\
&\operatorname{Re}k_{6,4}^{(8)},\operatorname{Im}k_{6,4}^{(8)},\\
&\operatorname{Re}k_{6,6}^{(8)},\operatorname{Im}k_{6,6}^{(8)},\\
&\operatorname{Re}k_{4,1}^{(8)},\operatorname{Im}k_{4,1}^{(8)},\\
&\operatorname{Re}k_{6,1}^{(8)},\operatorname{Im}k_{6,1}^{(8)},\\
&\operatorname{Re}k_{4,3}^{(8)},\operatorname{Im}k_{4,3}^{(8)},\\
&\operatorname{Re}k_{6,3}^{(8)},\operatorname{Im}k_{6,3}^{(8)},\\
&\operatorname{Re}k_{6,5}^{(8)},\operatorname{Im}k_{6,5}^{(8)})^T,
\end{split}
\label{eq:d8-coeff-basis}
\end{equation}
\begin{equation}
\begin{split}
\bm a^{(8)}=(&C_0^{(8)},C_2^{(8)},S_2^{(8)},C_4^{(8)},S_4^{(8)},
C_6^{(8)},S_6^{(8)},\\
&C_1^{(8)},S_1^{(8)},C_3^{(8)},S_3^{(8)},C_5^{(8)},S_5^{(8)})^T,
\end{split}
\label{eq:d8-amplitude-vector}
\end{equation}
where \(k_{j,m}^{(8)}\equiv k_{jm}^{N(8),\SCF}\). The corresponding response
relations are \cite{Shao2019InverseSixth}
\begin{align}
C_0^{(8)}={}&\Gamma_1^{(8)} k_{4,0}^{(8)}
+\Gamma_2^{(8)} k_{6,0}^{(8)},
\nonumber\\
C_2^{(8)}={}&\Gamma_3^{(8)}\operatorname{Re}k_{4,2}^{(8)}
+\Gamma_4^{(8)}\operatorname{Im}k_{4,2}^{(8)}
\nonumber\\
&+\Gamma_5^{(8)}\operatorname{Re}k_{6,2}^{(8)}
+\Gamma_6^{(8)}\operatorname{Im}k_{6,2}^{(8)},
\nonumber\\
S_2^{(8)}={}&\Gamma_4^{(8)}\operatorname{Re}k_{4,2}^{(8)}
-\Gamma_3^{(8)}\operatorname{Im}k_{4,2}^{(8)}
\nonumber\\
&+\Gamma_6^{(8)}\operatorname{Re}k_{6,2}^{(8)}
-\Gamma_5^{(8)}\operatorname{Im}k_{6,2}^{(8)},
\nonumber\\
C_4^{(8)}={}&\Gamma_7^{(8)}\operatorname{Re}k_{4,4}^{(8)}
+\Gamma_8^{(8)}\operatorname{Im}k_{4,4}^{(8)}
\nonumber\\
&+\Gamma_9^{(8)}\operatorname{Re}k_{6,4}^{(8)}
+\Gamma_{10}^{(8)}\operatorname{Im}k_{6,4}^{(8)},
\nonumber\\
S_4^{(8)}={}&\Gamma_8^{(8)}\operatorname{Re}k_{4,4}^{(8)}
-\Gamma_7^{(8)}\operatorname{Im}k_{4,4}^{(8)}
\nonumber\\
&+\Gamma_{10}^{(8)}\operatorname{Re}k_{6,4}^{(8)}
-\Gamma_9^{(8)}\operatorname{Im}k_{6,4}^{(8)},
\nonumber\\
C_6^{(8)}={}&\Gamma_{11}^{(8)}\operatorname{Re}k_{6,6}^{(8)}
+\Gamma_{12}^{(8)}\operatorname{Im}k_{6,6}^{(8)},
\nonumber\\
S_6^{(8)}={}&\Gamma_{12}^{(8)}\operatorname{Re}k_{6,6}^{(8)}
-\Gamma_{11}^{(8)}\operatorname{Im}k_{6,6}^{(8)},
\label{eq:d8-response-even}
\end{align}
and
\begin{align}
C_1^{(8)}={}&\Gamma_{13}^{(8)}\operatorname{Re}k_{4,1}^{(8)}
+\Gamma_{14}^{(8)}\operatorname{Im}k_{4,1}^{(8)}
\nonumber\\
&+\Gamma_{15}^{(8)}\operatorname{Re}k_{6,1}^{(8)}
+\Gamma_{16}^{(8)}\operatorname{Im}k_{6,1}^{(8)},
\nonumber\\
S_1^{(8)}={}&\Gamma_{14}^{(8)}\operatorname{Re}k_{4,1}^{(8)}
-\Gamma_{13}^{(8)}\operatorname{Im}k_{4,1}^{(8)}
\nonumber\\
&+\Gamma_{16}^{(8)}\operatorname{Re}k_{6,1}^{(8)}
-\Gamma_{15}^{(8)}\operatorname{Im}k_{6,1}^{(8)},
\nonumber\\
C_3^{(8)}={}&\Gamma_{17}^{(8)}\operatorname{Re}k_{4,3}^{(8)}
+\Gamma_{18}^{(8)}\operatorname{Im}k_{4,3}^{(8)}
\nonumber\\
&+\Gamma_{19}^{(8)}\operatorname{Re}k_{6,3}^{(8)}
+\Gamma_{20}^{(8)}\operatorname{Im}k_{6,3}^{(8)},
\nonumber\\
S_3^{(8)}={}&\Gamma_{18}^{(8)}\operatorname{Re}k_{4,3}^{(8)}
-\Gamma_{17}^{(8)}\operatorname{Im}k_{4,3}^{(8)}
\nonumber\\
&+\Gamma_{20}^{(8)}\operatorname{Re}k_{6,3}^{(8)}
-\Gamma_{19}^{(8)}\operatorname{Im}k_{6,3}^{(8)},
\nonumber\\
C_5^{(8)}={}&\Gamma_{21}^{(8)}\operatorname{Re}k_{6,5}^{(8)}
+\Gamma_{22}^{(8)}\operatorname{Im}k_{6,5}^{(8)},
\nonumber\\
S_5^{(8)}={}&\Gamma_{22}^{(8)}\operatorname{Re}k_{6,5}^{(8)}
-\Gamma_{21}^{(8)}\operatorname{Im}k_{6,5}^{(8)}.
\label{eq:d8-response-odd}
\end{align}

The constant channel corresponds to the Sun-frame sector \(m'=0\), whereas
each sector with \(|m'|=n>0\) generates the sine-cosine pair at \(n\wE\).
Thus one apparatus orientation
provides nine amplitudes for the 14 \(d=6\) coefficients and 13 amplitudes for
the 22 \(d=8\) coefficients. At least two linearly independent response
matrices are therefore needed for simultaneous coefficient extraction.

\section{Experimental concept and geometry}
\label{sec:geometry}

\subsection{Planar cancellation, apparatus, and coordinate conventions}
\label{subsec:edge}
\label{subsec:coordinate}

The multistripe geometry is motivated directly by the planar limit.  For two
uniform infinite parallel plates, in-plane translational and rotational
symmetries pair every separation direction with an equivalent contribution of
opposite angular weight.  The anisotropic Lorentz-violating force and torque
therefore vanish after integration.  Finite boundaries break this cancellation,
so the residual signal from realistic plates is dominated by their edges
\cite{LongKostelecky2015,Shao2016Enhanced}.  Alternating tungsten and glass
regions introduce repeated density boundaries while retaining an approximately
planar configuration.  More stripes provide more edges, but also reduce the
stripe width and can strengthen cancellation between adjacent boundaries.
Consequently the response must be determined from the full finite-volume
integral rather than from the total edge length alone.

\begin{figure*}[t]
 \centering
 \includegraphics[width=0.64\textwidth]{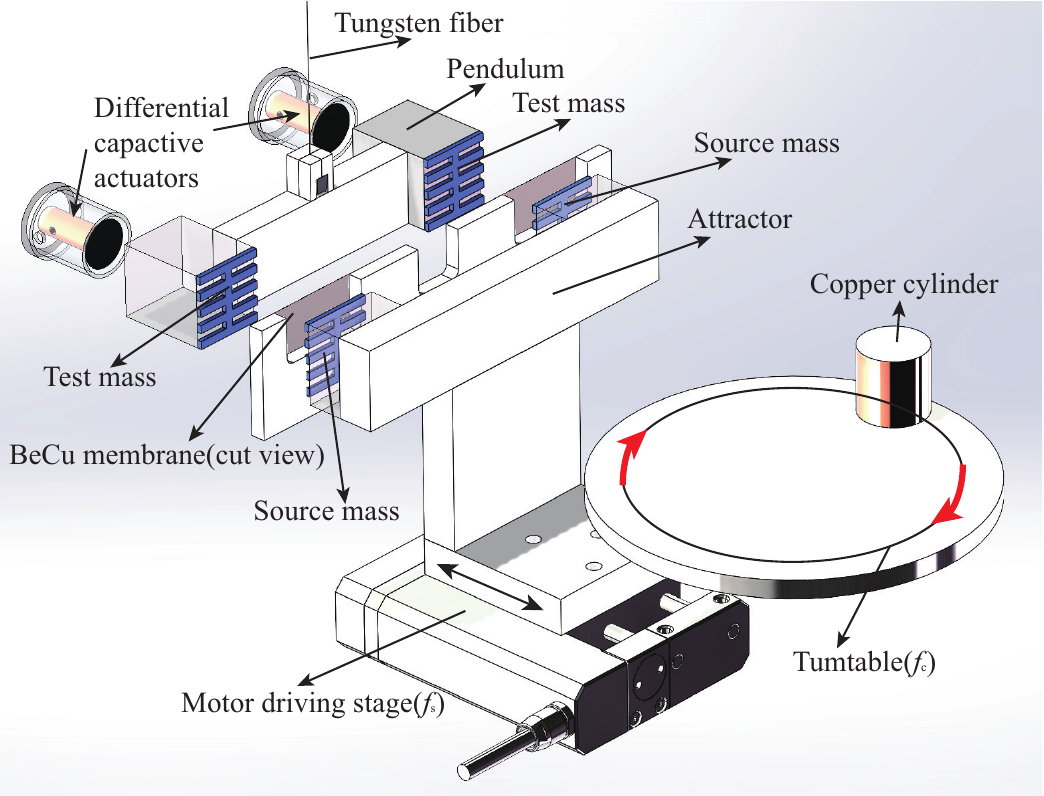}
 \caption{Schematic of the proposed short-range torsion-balance apparatus.
 The I-shaped pendulum and patterned source mass form two facing interaction
 regions; capacitive actuators control alignment, an autocollimator reads the
 pendulum twist, and a motor-driven translation stage modulates the separation.
 A rotating copper cylinder provides an independent torque calibration.}
 \label{fig:apparatus-schematic}
\end{figure*}

The proposed apparatus consists of an I-shaped torsion pendulum facing a
geometrically matched attractor.  The origin of the right-handed laboratory
frame is at the center of the pendulum bar.  The \(x\) axis follows the pendulum
arm, the \(y\) axis is normal to the facing surfaces and points toward the
attractor, and the \(z\) axis points vertically upward.  The azimuth \(\theta\)
is measured from local south to the \(x\) axis and is the apparatus orientation
entering Eq.~\eqref{eq:rotation-symbolic}; the controlled relative rotation of
the pendulum and attractor remains \(\theta_1\), as defined in
Sect.~\ref{subsec:finite-response}.
The principal mechanical elements and readout are summarized in
Fig.~\ref{fig:apparatus-schematic}.

The pendulum has two glass end blocks of nominal dimensions
\(19.8\times19.8\times16~\mathrm{mm}^3\), joined by a
\(70.4\times10\times8~\mathrm{mm}^3\) glass bar.  A patterned tungsten layer
with an overall area of \(19.8\times19.8~\mathrm{mm}^2\) and thickness
\(1.3~\mathrm{mm}\) is mounted on the facing surface of each end block.  The
assembled pendulum has a mass of approximately \(72~\mathrm{g}\) and is
suspended by a \(25~\mu\mathrm{m}\)-diameter tungsten fiber.  The attractor
carries matched tungsten patterns.  Its glass end blocks have in-plane
dimensions enlarged by one elementary pattern width and a nominal thickness of
\(24~\mathrm{mm}\), while its connector bar has dimensions
\((70.4-p_x)\times12.001\times8~\mathrm{mm}^3\).

Unless stated otherwise, the numerical calculations use tungsten and glass
densities
\begin{equation}
 \rho_w=19170.0~\mathrm{kg\,m^{-3}},
 \qquad
 \rho_g=2520.0~\mathrm{kg\,m^{-3}},
 \label{eq:material-densities}
\end{equation}
and a laboratory colatitude
\begin{equation}
 \chi=\frac{59.5}{180}\pi=59.5^{\circ}.
 \label{eq:laboratory-colatitude}
\end{equation}

In this paper, stripe orientation is defined by the long axis on the
facing \(x\)-\(z\) surface.  A horizontal stripe extends along \(x\), whereas a
vertical stripe extends along \(z\).  Let \(N_h=2n_h+1\) and
\(N_v=2n_v+1\) denote the corresponding odd numbers of horizontal and vertical
tungsten stripes.  For active dimensions \(W_t\) along \(z\) and \(L_t\) along
\(x\), the elementary cell widths are
\begin{equation}
  p_z=\frac{W_t}{4n_h+1},
  \qquad
  p_x=\frac{L_t}{4n_v+1}.
  \label{eq:stripe-width}
\end{equation}
For the square patterns used here, \(L_t=W_t=L=19.8~\mathrm{mm}\) and
\(n_h=n_v=n\), so
\(p_x=p_z=p=L/(4n+1)=L/(2N_s-1)\), with
\(N_s=3,5,7\) corresponding to \(n=1,2,3\).  The horizontal and vertical
designs are related by interchanging the two in-plane pattern axes.  This
interchange changes the directional moments in
Eq.~\eqref{eq:gamma-volume-definition} and therefore provides complementary
response matrices.

Each pendulum end forms a facing test-source pair.  The left attractor pattern
is shifted upward and toward the left by half a cell, while the right pattern is
shifted downward and toward the right by half a cell.  With positive \(x\)
toward the left pendulum end, the relative displacements are
\begin{equation}
\begin{aligned}
  (\Delta x_L,\Delta z_L)
  &=\left(+\frac{p_x}{2},+\frac{p_z}{2}\right),\\
  (\Delta x_R,\Delta z_R)
  &=\left(-\frac{p_x}{2},-\frac{p_z}{2}\right).
\end{aligned}
  \label{eq:half-stripe-shift}
\end{equation}
The two registrations thus differ by one complete cell in both in-plane
directions.  Under reflection about the suspension axis, the leading Newtonian
torques from the two ends have opposite signs, whereas the selected anisotropic
torques have the same sign.  The conventional torque is therefore suppressed
to first order while the Lorentz-violating contribution adds constructively
\cite{Shao2016Enhanced}.

The attractor is supported symmetrically on a motor-driven translation stage,
which modulates the surface-to-surface gap between
\begin{equation}
 g_{\min}=0.4\times10^{-3}~\mathrm{m},
 \qquad
 g_{\max}=1.0\times10^{-3}~\mathrm{m}.
 \label{eq:gap-values}
\end{equation}
We define the measured torque amplitude by
\begin{equation}
  \Delta\tau
  =
  \frac{\tau(g_{\min})-\tau(g_{\max})}{2}.
  \label{eq:gap-modulation}
\end{equation}
The half-difference rejects static offsets; reversing its order changes only
the overall phase convention.  Differential capacitive actuators control the
relative alignment, an autocollimator records the pendulum twist, and a copper
calibration mass on a turntable provides an independent torque scale.

\subsection{Block-indexing and centerline rules}
\label{subsec:antisymmetric}

The finite bodies are decomposed into uniform rectangular prisms.  To make the
geometrical construction unambiguous, a block \(q\) is represented by
\begin{equation}
  {\cal B}_q
  =
  \bigl(a_q,b_q,c_q;x_q,y_q,z_q;\rho_q\bigr),
  \label{eq:block-parameter}
\end{equation}
where \((a_q,b_q,c_q)\) are the half-dimensions along \((x,y,z)\),
\((x_q,y_q,z_q)\) is the center, and \(\rho_q\) is the density.  We use
\(\rho_w\) and \(\rho_g\) for tungsten and glass, respectively.

\begin{figure*}[t]
 \centering
 \includegraphics[width=0.94\textwidth]{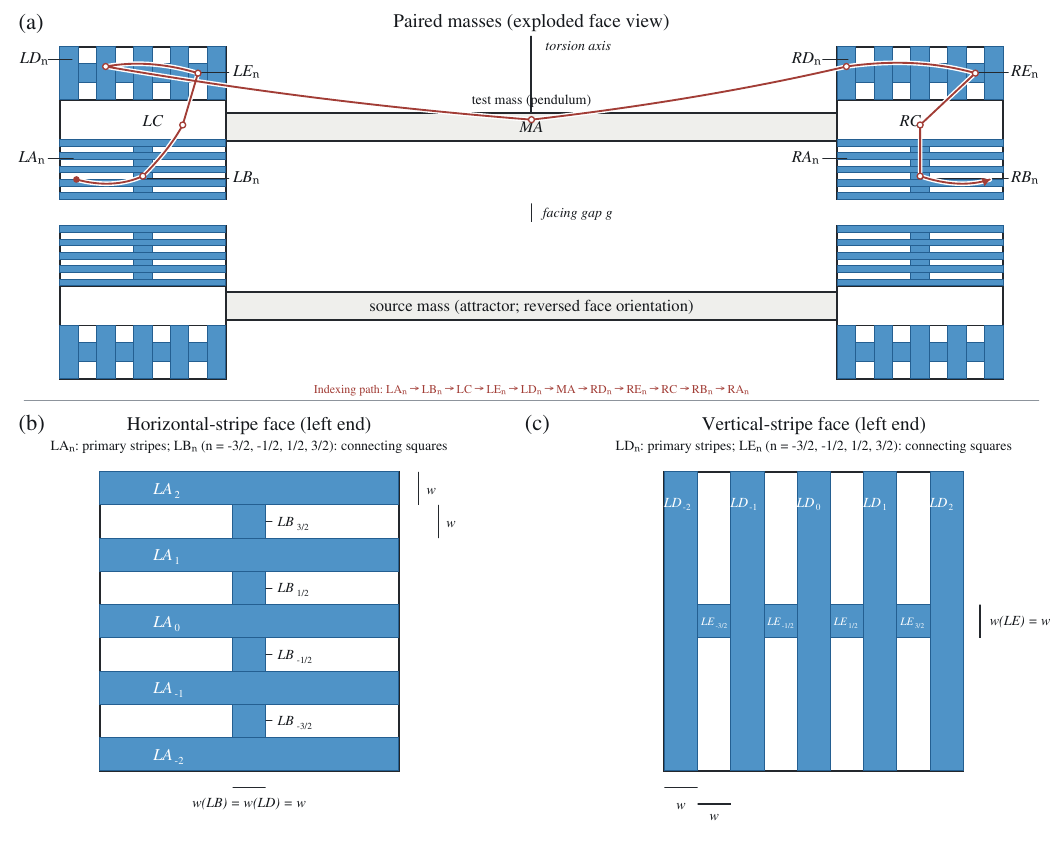}
 \caption{Block decomposition and indexing of the multistripe geometry.
 Panel (a) shows the ordered test- and source-mass blocks and the path used to
 assign the labels in Eq.~\eqref{eq:block-ordering}.  Panels (b) and (c) show
 the left-end horizontal- and vertical-stripe families, respectively.  Their
 stripe long axes extend along \(x\) and \(z\), and the panels include the
 integer stripe indices and the narrow connector blocks.}
 \label{fig:stripe-indexing}
\end{figure*}

The labels follow the path around the two pendulum ends used to assemble the
ordered block list, as illustrated in Fig.~\ref{fig:stripe-indexing}.  On the
left end the sequence is
\((LA,LB,LC,LE,LD)\), followed by the central bar \(MA\); the reflected right
end is \((RD,RE,RC,RB,RA)\).  Here \(A\) and \(D\) denote the two complete
stripe families, \(B\) and \(E\) their narrow connecting prisms, and \(C\) the
glass substrate.  Attractor blocks use the same order with the suffix \(S\),
giving
\begin{equation}
\begin{aligned}
  \mathcal{T}={}&(LA,LB,LC,LE,LD,MA,\\[-0.2ex]
  &\quad RD,RE,RC,RB,RA),\\
  \mathcal{S}={}&(LAS,LBS,LCS,LES,LDS,MAS,\\[-0.2ex]
  &\quad RDS,RES,RCS,RBS,RAS).
\end{aligned}
  \raisetag{1.2\baselineskip}
  \label{eq:block-ordering}
\end{equation}
The suffix labels a geometrical family rather than one prism; each stripe
family expands into the index sets
\begin{equation}
\begin{aligned}
  I_n&=\{-n,-n+1,\ldots,n\},\\
  H_n&=\left\{-n+\frac12,\ldots,n-\frac12\right\}.
\end{aligned}
  \label{eq:block-index-sets}
\end{equation}
Thus the five-stripe case has integer labels
\(-2,-1,0,1,2\) for the complete stripes and half-integer labels
\(-3/2,-1/2,1/2,3/2\) for the four connectors.

The rule is most compactly stated for the left test mass.  Let \(d_{xt}\) be
the distance from the suspension axis to an end-block center, \(d_{yt}\) the
distance from the glass center to either tungsten face, \(T_t\) the tungsten
thickness, and \(T_{t,g}\) the glass thickness.  With
\({\cal B}(a,b,c;\bm r;\rho)\) denoting the tuple in
Eq.~\eqref{eq:block-parameter}, the five families are
\begin{align}
 {\cal B}_{LA}(q)={}&
 {\cal B}\!\biggl(\frac{L_t}{2},\frac{T_t}{2},\frac{p_z}{2};
 \nonumber\\
 & (d_{xt},d_{yt},2qp_z);\rho_w\biggr),
 \quad q\in I_{n_h},
 \nonumber\\
 {\cal B}_{LB}(q)={}&
 {\cal B}\!\biggl(\frac{p_x}{2},\frac{T_t}{2},\frac{p_z}{2};
 \nonumber\\
 & (d_{xt}+2n_v\ell_t p_x,d_{yt},2qp_z);
 \nonumber\\
 &\rho_w\biggr),\quad q\in H_{n_h},
 \nonumber\\
 {\cal B}_{LC}={}&
 {\cal B}\!\biggl(\frac{L_t}{2},\frac{T_{t,g}}{2},\frac{W_t}{2};
 \nonumber\\
 &(d_{xt},0,0);\rho_g\biggr),
 \nonumber\\
 {\cal B}_{LE}(q)={}&
 {\cal B}\!\biggl(\frac{p_x}{2},\frac{T_t}{2},\frac{p_z}{2};
 \nonumber\\
 &(d_{xt}+2qp_x,-d_{yt},2n_h\ell_t p_z);
 \nonumber\\
 &\rho_w\biggr),\quad q\in H_{n_v},
 \nonumber\\
 {\cal B}_{LD}(q)={}&
 {\cal B}\!\biggl(\frac{p_x}{2},\frac{T_t}{2},\frac{W_t}{2};
 \nonumber\\
 &(d_{xt}+2qp_x,-d_{yt},0);\rho_w\biggr),
 \quad q\in I_{n_v}.
 \label{eq:left-test-blocks}
\end{align}
The central bar is centered at the origin and can be written as
\begin{equation*}
 {\cal B}_{MA}
 =
 {\cal B}\!\left(
 \frac{L_{t,\mathrm{bar}}}{2},
 \frac{T_{t,\mathrm{bar}}}{2},
 \frac{W_{t,\mathrm{bar}}}{2};
 \bm 0;\rho_g
 \right).
\end{equation*}
The right-end families are obtained
from Eq.~\eqref{eq:left-test-blocks} by the reflection
\((x,q)\mapsto(-x,-q)\), with
\(LA\leftrightarrow RA\), \(LB\leftrightarrow RB\),
\(LC\leftrightarrow RC\), \(LE\leftrightarrow RE\), and
\(LD\leftrightarrow RD\).  This convention makes the sign change in the
\(x\) coordinate explicit while preserving the same physical ordering of the
stripes on both ends.

The discrete parameters \(\ell_t,\ell_s\in\{-1,0,+1\}\) specify the left,
middle, and right locations of the narrow centerline connectors in the test
and source patterns.  In Eq.~\eqref{eq:left-test-blocks}, for example,
\(\ell_t\) moves the \(LB\) connector center to
\(x=d_{xt}+2n_v\ell_t p_x\) and the \(LE\) connector center to
\(z=2n_h\ell_t p_z\).  This changes only the connector registration; the
outer dimensions, stripe number, and material volumes remain fixed.

The attractor is indexed by the same integer and half-integer sets after the
replacement \(t\to s\) and \(\ell_t\to\ell_s\).  Its four dimensionless
offset parameters obey
\begin{equation}
  o_{Lx}=o_{Lz}=o_{Rx}=o_{Rz}=\frac12,
  \label{eq:source-offset-parameters}
\end{equation}
so the centers of all left and right attractor pattern blocks receive,
respectively,
\begin{equation}
\begin{aligned}
  \delta\bm r_L&=(o_{Lx}p_x,0,o_{Lz}p_z),\\
  \delta\bm r_R&=(-o_{Rx}p_x,0,-o_{Rz}p_z),
\end{aligned}
  \label{eq:source-block-offsets}
\end{equation}
which reproduces Eq.~\eqref{eq:half-stripe-shift}.  For one assembly, the
number of blocks generated by these rules is
\begin{equation}
\begin{split}
 N_{\rm block}={}&2\bigl[(2n_h+1)+2n_h+1\\
 &+2n_v+(2n_v+1)\bigr]+1.
\end{split}
  \label{eq:block-count}
\end{equation}
For five stripes in each direction this gives
\(2(5+4+1+4+5)+1=39\) prisms, including the central bar.

Once the two ordered block lists are constructed, the finite-volume transfer
coefficients are obtained by summing the pair responses defined in
Eq.~\eqref{eq:gamma-volume-definition}.
\begin{equation}
  \Gamma_i^{(d)}(\theta,\theta_1,g)
  =
  \sum_{a\in{\cal T}}\sum_{b\in{\cal S}}
  \Gamma_{i,ab}^{(d)}(\theta,\theta_1,g).
  \label{eq:gamma-block-sum}
\end{equation}
This bookkeeping rule is common to the horizontal and vertical designs; only
the assignments of \(p_x\), \(p_z\), and the corresponding block dimensions
are interchanged.

\subsection{Newtonian background and systematic torque}
\label{subsec:blockmodel}

The same prism decomposition is used to evaluate the conventional gravitational
background.  A point-mass approximation or a low-order multipole expansion is
not adequate when the source-test gap is comparable to a prism dimension.  We
therefore begin with the complete volume integral and reduce it analytically,
following the rectangular-solid construction of Adelberger, Collins, and Hoyle
\cite{AdelbergerCollinsHoyle2006}.

Consider two uniform rectangular prisms with densities \(\rho_1\) and
\(\rho_2\), centers \(\bm R_1\) and \(\bm R_2\), and half-dimensions
\(a_{\alpha\mu}\), where \(\alpha=1,2\) labels the prism and
\(\mu=x,y,z\).  Their center displacement is
\begin{equation}
  \bm D=(X,Y,Z)=\bm R_1-\bm R_2,
  \label{eq:cuboid-displacement}
\end{equation}
and a volume point is \(\bm r_\alpha=\bm R_\alpha+\bm u_\alpha\), with
\(|u_{\alpha\mu}|\leq a_{\alpha\mu}\).  The differential force between two
volume elements is
\begin{equation}
  \dd\bm F_{12}
  =-G\rho_1\rho_2
  \frac{\bm r_1-\bm r_2}{|\bm r_1-\bm r_2|^3}
  \dd V_1\dd V_2,
  \label{eq:newtonian-force-element}
\end{equation}
and hence
\begin{equation}
  \bm F_{12}
  =-G\rho_1\rho_2
  \int_{V_1}\!\dd V_1\int_{V_2}\!\dd V_2
  \frac{\bm r_1-\bm r_2}{|\bm r_1-\bm r_2|^3}.
  \label{eq:newtonian-force-volume}
\end{equation}

For a torsion balance the relevant observable is the torque about the
suspension axis.  The exact six-dimensional expression is
\begin{equation}
\begin{aligned}
 \tau_{N,z}^{(12)}
 ={}&\int_{V_1}\!\int_{V_2}
 \bigl(\bm r_1\times\dd\bm F_{12}\bigr)_z
 \\
 ={}&-G\rho_1\rho_2
 \int_{V_1}\!\dd V_1\int_{V_2}\!\dd V_2
 \\
 &\quad\times
 \frac{x_1(y_1-y_2)-y_1(x_1-x_2)}
 {\lvert\bm r_1-\bm r_2\rvert^3}
 \\
 ={}&-G\rho_1\rho_2
 \int_{V_1}\!\dd V_1\int_{V_2}\!\dd V_2
 \\
 &\quad\times
 \frac{y_1x_2-x_1y_2}
 {\lvert\bm r_1-\bm r_2\rvert^3}.
\end{aligned}
\label{eq:newtonian-torque-integral}
\end{equation}
The second equality uses
\(x_1(y_1-y_2)-y_1(x_1-x_2)=y_1x_2-x_1y_2\), the simplification specified by
the geometry convention above.

Introduce the relative and sum coordinates
\begin{equation}
\begin{aligned}
 (x,y,z)&=(x_1-x_2,y_1-y_2,z_1-z_2),\\
 (x_\Sigma,y_\Sigma,z_\Sigma)&=(x_1+x_2,y_1+y_2,z_1+z_2).
\end{aligned}
  \label{eq:relative-sum-coordinates}
\end{equation}
Because the Newton kernel depends only on the relative coordinates, the sum
coordinates can be integrated out exactly.  The result is a signed boundary
sum, equivalent to convolving the Newton Green function with the two box
functions.  Define
\begin{equation}
  \begin{aligned}
  u_\mu(\sigma_\mu,\eta_\mu)
  &=D_\mu+\sigma_\mu a_{1\mu}+\eta_\mu a_{2\mu},\\
  &\hspace{-1.5em}\sigma_\mu,\eta_\mu=\pm1.
  \end{aligned}
 \label{eq:cuboid-boundaries}
\end{equation}
Thus, for example,
\(x_i=X\pm a_{1x}\pm a_{2x}\), with analogous expressions for \(y_j\) and
\(z_k\).  We collect the three boundary arguments in
\(\bm u_{\bm\sigma\bm\eta}
=[u_x(\sigma_x,\eta_x),u_y(\sigma_y,\eta_y),
u_z(\sigma_z,\eta_z)]\). The prism-pair torque can then be written as
\begin{equation}
\begin{aligned}
 \tau_{N,z}^{(12)}
 ={}&G\rho_1\rho_2
 \sum_{\bm\sigma,\bm\eta\in\{-1,+1\}^3}
 \\
 &\quad\times
 \left(\prod_{\mu=x,y,z}\sigma_\mu\eta_\mu\right)
 T_N\!\left[\bm u_{\bm\sigma\bm\eta}\right].
\end{aligned}
\label{eq:newtonian-boundary-sum}
\end{equation}
where \(T_N\) is a primitive of the torque kernel; its dependence on the
transverse prism centers is left implicit.  Equation~\eqref{eq:newtonian-boundary-sum}
is the explicit form of the alternating \(i,j,k\) boundary sum.

The analytic primitive is generated by the following elementary integrations,
where \(r=\sqrt{x^2+y^2+z^2}\).
\begin{align}
 \int\frac{\dd z}{r^3}
 &=\frac{z}{(x^2+y^2)r},
 \nonumber\\
 \int\frac{\dd y}{r}
 &=\ln(y+r),
 \nonumber\\
 \int\frac{\dd x}{x^2+a^2}
 &=\frac{1}{a}\arctan\frac{x}{a}.
 \label{eq:newtonian-elementary-primitives}
\end{align}
Accordingly, the complete kernel has the general structure
\begin{equation}
\begin{split}
 T_N(x,y,z)={}&Q_1^{(N)}\!(x,y,z)r\\
  &+Q_2^{(N)}\!(x,y,z)\ln(z+r)\\
  &+Q_3^{(N)}\!(x,y,z)\arctan\!\left(\frac{xy}{zr}\right)\\
  &+\text{cyclic terms},
\end{split}
 \label{eq:newtonian-kernel-structure}
\end{equation}
where \(Q_a^{(N)}\), with \(a=1,2,3\), are low-order polynomials fixed by the
Newtonian torque numerator.
The square-root, logarithmic, and arctangent form is the torque analogue of the
closed rectangular-solid force expression in
Ref.~\cite{AdelbergerCollinsHoyle2006}.

The background from the complete apparatus is the pairwise sum
\begin{equation}
  \tau_N(g,\bm q)
  =\sum_{a\in{\cal T}}\sum_{b\in{\cal S}}
  \tau_{N,z}^{(ab)}(g,\bm q),
  \label{eq:newtonian-block-sum}
\end{equation}
and its modulated amplitude is obtained from
Eq.~\eqref{eq:gap-modulation}.  Here \(\bm q\) collects the dimensions and
relative coordinates of all prisms.  The ideal left-right reflection suppresses
\(\Delta\tau_N\), but fabrication and alignment errors break the cancellation.
For small deviations,
\begin{equation}
  \delta(\Delta\tau_N)
  =
  \sum_\mu
  \frac{\partial\Delta\tau_N}{\partial q_\mu}\,
  \delta q_\mu
  +O(\delta q_\mu^2),
  \label{eq:systematic-linear}
\end{equation}
with the principal coordinates
\begin{equation}
 \delta\bm q=
 (\delta T_w,\delta H_w,\delta L_w,
  \delta x_{st},\delta z_{st},\delta g).
 \label{eq:principal-geometry-errors}
\end{equation}
These terms describe, respectively, tungsten-thickness, stripe-height,
stripe-length, horizontal-registration, vertical-registration, and gap errors.
The derivatives in Eq.~\eqref{eq:systematic-linear} determine both the
fabrication tolerance budget and the Newtonian nuisance templates used in the
likelihood analysis.

\section{Likelihood and response-optimization method}
\label{sec:calculation}

\subsection{Linear measurement model and maximum-likelihood estimator}
\label{subsec:estimator}

We consider \(N_{\rm seg}\) statistically independent repetitions performed
with one fixed experimental configuration.  The mass geometry, separation,
and apparatus orientation are unchanged, so the response matrix \(\mathbf A\)
is identical for every repetition and only the noise realization varies.  Let
\(\bm v^{(q)}\in\mathbb R^{p}\) denote the demodulated sidereal-amplitude vector
obtained from the \(q\)th repetition, let \(\bm k\in\mathbb R^{n}\) contain the
SME coefficients to be estimated, and let
\(\mathbf A\in\mathbb R^{p\times n}\) be the response matrix constructed from
Eq.~\eqref{eq:transfer-matrix}.  The repeated-measurement model is
\begin{equation}
  \bm v^{(q)}=\mathbf A\bm k+\bm n^{(q)},
  \qquad q=1,\ldots,N_{\rm seg},
  \label{eq:repeated-linear-model}
\end{equation}
with independent Gaussian noise
\begin{equation}
  \bm n^{(q)}\sim{\cal N}(\bm 0,\mathbf C),
  \qquad
  \operatorname{Cov}(\bm n^{(q)},\bm n^{(q')})
  =\delta_{qq'}\mathbf C.
  \label{eq:repeated-noise-model}
\end{equation}
The unit-covariance model used to derive the analytic design formulas below is
recovered by setting \(\mathbf C=\mathbf I_p\).  For measured vectors
\(\{\bm v^{(q)}\}_{q=1}^{N_{\rm seg}}\), write the residual as
\(\bm r^{(q)}=\bm v^{(q)}-\mathbf A\bm k\).  The likelihood and log likelihood are
\begin{align}
  L(\bm k)={}&(2\pi)^{-pN_{\rm seg}/2}
  (\det\mathbf C)^{-N_{\rm seg}/2}
  \nonumber\\
  &\times\exp\!\left[-\frac12\sum_{q=1}^{N_{\rm seg}}
  (\bm r^{(q)})^T\mathbf C^{-1}\bm r^{(q)}\right],
  \label{eq:gaussian-likelihood}\\
  \ln L(\bm k)={}&\text{const.}
  -\frac12\sum_{q=1}^{N_{\rm seg}}
  (\bm r^{(q)})^T\mathbf C^{-1}\bm r^{(q)}.
  \label{eq:gaussian-log-likelihood}
\end{align}
Defining the sample mean
\begin{equation}
  \overline{\bm v}=\frac{1}{N_{\rm seg}}
  \sum_{q=1}^{N_{\rm seg}}\bm v^{(q)},
  \label{eq:sample-mean}
\end{equation}
the quadratic form separates as
\begin{align}
 &\sum_{q=1}^{N_{\rm seg}}
 (\bm v^{(q)}-\mathbf A\bm k)^T\mathbf C^{-1}
 (\bm v^{(q)}-\mathbf A\bm k)
 \nonumber\\
 &\quad=\sum_{q=1}^{N_{\rm seg}}
 (\bm v^{(q)}-\overline{\bm v})^T\mathbf C^{-1}
 (\bm v^{(q)}-\overline{\bm v})\nonumber\\
 &+N_{\rm seg}(\overline{\bm v}-\mathbf A\bm k)^T\mathbf C^{-1}
 (\overline{\bm v}-\mathbf A\bm k).
 \label{eq:quadratic-decomposition}
\end{align}
The cross term vanishes because
\(\sum_q(\bm v^{(q)}-\overline{\bm v})=\bm0\).  If \(\mathbf A\) has full
column rank, maximizing Eq.~\eqref{eq:gaussian-log-likelihood} gives the normal
equations and estimator
\begin{align}
 \mathbf A^T\mathbf C^{-1}\mathbf A\,\widehat{\bm k}
 &=\mathbf A^T\mathbf C^{-1}\overline{\bm v},
 \label{eq:normal-equations}\\
 \widehat{\bm k}
 &=\left(\mathbf A^T\mathbf C^{-1}\mathbf A\right)^{-1}
 \mathbf A^T\mathbf C^{-1}\overline{\bm v}.
 \label{eq:generalized-mle}
\end{align}
For a square nonsingular response matrix with \(\mathbf C=\mathbf I_n\), this
reduces to \(\widehat{\bm k}=\mathbf A^{-1}\overline{\bm v}\), the form used
in the analytic special cases below.

\subsection{Covariance and confidence ellipsoid}
\label{subsec:covariance}

The sample mean obeys
\begin{equation}
  \overline{\bm v}=\mathbf A\bm k+\overline{\bm n},
  \qquad
  \overline{\bm n}\sim{\cal N}\!\left(
  \bm0,\frac{\mathbf C}{N_{\rm seg}}\right).
  \label{eq:mean-distribution}
\end{equation}
Consequently, the estimator is unbiased and normally distributed,
\begin{equation}
 \widehat{\bm k}\sim{\cal N}(\bm k,\bm\Sigma_k),
 \qquad
 \bm\Sigma_k=\frac{1}{N_{\rm seg}}
 \left(\mathbf A^T\mathbf C^{-1}\mathbf A\right)^{-1}.
 \label{eq:mle-covariance}
\end{equation}
The corresponding information matrix is
\begin{equation}
 \mathbf F_k=\bm\Sigma_k^{-1}
 =N_{\rm seg}\mathbf A^T\mathbf C^{-1}\mathbf A.
 \label{eq:fisher-basic}
\end{equation}
For a confidence level \(1-\alpha\), the joint confidence region for \(n\)
coefficients is the ellipsoid
\begin{equation}
 (\bm k-\widehat{\bm k})^T\mathbf F_k
 (\bm k-\widehat{\bm k})
 \leq \chi^2_{n,1-\alpha}.
 \label{eq:confidence-ellipsoid}
\end{equation}
Its \(n\)-dimensional volume is
\begin{equation}
 V_{n,1-\alpha}
 =\frac{\pi^{n/2}}{\Gamma(n/2+1)}
 \frac{[\chi^2_{n,1-\alpha}]^{n/2}}
 {\sqrt{\det\mathbf F_k}}.
 \label{eq:confidence-volume}
\end{equation}
For \(n=2\), Eq.~\eqref{eq:confidence-volume} becomes the area of the error
ellipse,
\begin{equation}
 S_{2,1-\alpha}
 =\pi\chi^2_{2,1-\alpha}\sqrt{\det\bm\Sigma_k}.
 \label{eq:error-ellipse-area}
\end{equation}
In the square, unit-covariance case,
\begin{equation}
 \begin{aligned}
 \det\bm\Sigma_k
 &=\frac{1}{N_{\rm seg}^{n}(\det\mathbf A)^2},\\
 S_{2,1-\alpha}
 &=\frac{\pi\chi^2_{2,1-\alpha}}
 {N_{\rm seg}|\det\mathbf A|}.
 \end{aligned}
 \label{eq:square-covariance-determinant}
\end{equation}
Thus, only in this square, unit-covariance special case does
\(|\det\mathbf A|\) directly control the joint error region.  For a general
rectangular response matrix or nontrivial covariance, identifiability is
instead determined by the weighted Gram matrix.  At least one coefficient
combination is unidentifiable when
\(\det(\mathbf A^T\mathbf C^{-1}\mathbf A)=0\), equivalently when
\(\det\mathbf F_k=0\).

The experimental model can also include nuisance amplitudes.  Stacking all
configurations into \(\bm y\), we write
\begin{equation}
 \bm y=\mathbf X\bm k+\mathbf Z\bm\eta+\bm n,
 \qquad \bm n\sim{\cal N}(\bm0,\mathbf C),
 \label{eq:linear-model}
\end{equation}
where \(\bm\eta\) contains Newtonian and instrumental nuisance parameters and
\(\mathbf Z\) contains their templates.  With a Gaussian prior
\(\bm\eta\sim{\cal N}(\bm0,\mathbf P)\), profiling over \(\bm\eta\) gives the
Schur-complement information matrix
\begin{equation}
\begin{aligned}
 \mathbf F_k={}&\mathbf X^T\mathbf C^{-1}\mathbf X
 \\
 &-\mathbf X^T\mathbf C^{-1}\mathbf Z
 \\
 &\quad\times
 \left(\mathbf Z^T\mathbf C^{-1}\mathbf Z
 +\mathbf P^{-1}\right)^{-1}
 \\
 &\quad\times
 \mathbf Z^T\mathbf C^{-1}\mathbf X.
\end{aligned}
\label{eq:profiled-fisher}
\end{equation}
Equation~\eqref{eq:confidence-volume} remains applicable after replacing the
ideal information matrix by Eq.~\eqref{eq:profiled-fisher}.

\subsection{Response determinants and geometry criteria}
\label{subsec:response-criteria}

The harmonic pairs in Sect.~\ref{sec:theory} lead naturally to a two-parameter response block.
For
\begin{equation}
 \mathbf A_2=\begin{pmatrix}a&b\\c&d\end{pmatrix},
 \label{eq:two-response-block}
\end{equation}
the estimator has covariance
\begin{equation}
 \bm\Sigma_2
 =\frac{1}{N_{\rm seg}(ad-bc)^2}
 \begin{pmatrix}
 b^2+d^2&-ab-cd\\
 -ab-cd&a^2+c^2
 \end{pmatrix},
 \label{eq:two-response-covariance}
\end{equation}
and the confidence-ellipse area is
\begin{equation}
 S_{2,1-\alpha}
 =\frac{\pi\chi^2_{2,1-\alpha}}
 {N_{\rm seg}|ad-bc|}.
 \label{eq:two-response-area}
\end{equation}
This motivates the squared two-vector determinant
\begin{equation}
 s(\bm u,\bm v)
 \equiv [\det(\bm u,\bm v)]^2
 =\bigl(u_1v_2-u_2v_1\bigr)^2,
 \label{eq:paired-determinant}
\end{equation}
which vanishes when two apparatus settings probe the same coefficient
direction.

For a four-parameter harmonic subspace, consider first the unit-response
rotation blocks
\begin{align}
 \mathbf A_{4,0}&=
 \begin{pmatrix}\mathbf Q_1&\mathbf Q_2\\
                 \mathbf Q_3&\mathbf Q_4\end{pmatrix},
 \nonumber\\
 \mathbf Q_i=\mathbf Q(\vartheta_i)&=
 \begin{pmatrix}
 \cos\vartheta_i&-\sin\vartheta_i\\
 \sin\vartheta_i& \cos\vartheta_i
 \end{pmatrix}.
 \label{eq:unit-four-response-block}
\end{align}
Using
\(\mathbf Q^T(\vartheta)=\mathbf Q^{-1}(\vartheta)\), its Gram matrix is
\begin{equation}
 \begin{aligned}
 \mathbf A_{4,0}^T\mathbf A_{4,0}
 &=\begin{pmatrix}2\mathbf I_2&\mathbf M\\
                   \mathbf M^T&2\mathbf I_2\end{pmatrix},\\
 \mathbf M&=\mathbf Q(\vartheta_2-\vartheta_1)\\
 &\quad+\mathbf Q(\vartheta_4-\vartheta_3).
 \end{aligned}
 \label{eq:unit-four-gram-matrix}
\end{equation}
It follows that
\begin{equation}
 \det\mathbf A_{4,0}
 =4\sin^2\!\left(\frac{\Delta\vartheta}{2}\right),
 \quad
 \Delta\vartheta=\vartheta_1-\vartheta_2-
 \vartheta_3+\vartheta_4.
 \label{eq:unit-four-response-determinant}
\end{equation}
Thus the response becomes singular whenever
\(\Delta\vartheta=0\pmod{2\pi}\).

The amplitude-weighted blocks that occur in the torque response are equivalent
up to fixed row and column sign conventions and can be parameterized as
\begin{equation}
 \begin{aligned}
 \mathbf A_4&=
 \begin{pmatrix}\mathbf R_1&\mathbf R_2\\
                  \mathbf R_3&\mathbf R_4\end{pmatrix},\\
 \mathbf R_i&=
 \begin{pmatrix}
 r_i\sin\vartheta_i&r_i\cos\vartheta_i\\
 r_i\cos\vartheta_i&-r_i\sin\vartheta_i
 \end{pmatrix}.
 \end{aligned}
 \label{eq:four-response-block}
\end{equation}
Equivalently, if
\(\mathbf R_i=\left(\begin{smallmatrix}a_i&b_i\\b_i&-a_i\end{smallmatrix}\right)\),
then
\begin{equation}
 r_i=\sqrt{a_i^2+b_i^2},\qquad
 \sin\vartheta_i=\frac{a_i}{r_i},\qquad
 \cos\vartheta_i=\frac{b_i}{r_i}.
 \label{eq:block-polar-parameters}
\end{equation}
The determinant controlling the confidence hypervolume is
\begin{equation}
\begin{split}
 \det\mathbf A_4={}&r_1^2r_4^2+r_2^2r_3^2
 -2r_1r_2r_3r_4\cos\Delta\vartheta,\\
 \Delta\vartheta={}&\vartheta_1-\vartheta_2-
 \vartheta_3+\vartheta_4.
\end{split}
 \label{eq:four-response-determinant}
\end{equation}
For \(r_i=1\), Eq.~\eqref{eq:four-response-determinant} reduces to
Eq.~\eqref{eq:unit-four-response-determinant}; hence complementary angular
phases are as important as the individual response amplitudes.  In Cartesian
form the same determinant is
\begin{align}
 \det\mathbf A_4={}&
 2a_1a_4b_2b_3+2a_2a_3b_1b_4
 \nonumber\\
 &+(a_2^2+b_2^2)(a_3^2+b_3^2)
 \nonumber\\
 &+(a_1^2+b_1^2)(a_4^2+b_4^2)
 \nonumber\\
 &-2a_3a_4b_1b_2-2a_2a_4b_1b_3
 \nonumber\\
 &-2a_1a_3b_2b_4-2a_1a_2b_3b_4
 \nonumber\\
 &-2a_1a_2a_3a_4-2b_1b_2b_3b_4.
 \label{eq:four-response-determinant-cartesian}
\end{align}

The two- and four-parameter determinants above reproduce the harmonic-block
criteria used for \(d=6\) in Ref.~\cite{Jin2026Optimization}.  For the
complete coefficient vector, we instead rank a geometry by the dimensionless,
noise-weighted D-optimal score
\begin{equation}
 {\cal D}_d
 =\left[
 \frac{\det\mathbf F_k}
      {\det\mathbf F_{k,\mathrm{ref}}}
 \right]^{1/(2n_d)},
 \label{eq:doptimal}
\end{equation}
where \(n_d\) is the number of estimated coefficients and the reference matrix
fixes the normalization.  Section~\ref{sec:results} introduces complementary
unweighted response products for displaying which harmonic block drives a
change of geometry.  These products are diagnostic quantities rather than
coefficient limits.  Physical limits follow only from the covariance in
Eq.~\eqref{eq:mle-covariance} or, with nuisance parameters included, from
Eq.~\eqref{eq:profiled-fisher}.

\section{Geometry optimization and coefficient sensitivity}
\label{sec:results}

\subsection{\texorpdfstring{\(d=6\) response}{d=6 response}}
\label{subsec:d6results}

\subsubsection{Dimensionless response measure}

We adopt the five-stripe \(d=6\) baseline established in
Ref.~\cite{Jin2026Optimization} and compare only its centerline registrations.
The 14 transfer functions in Eq.~\eqref{eq:gamma-volume-definition} have units
of torque per \(d=6\) coefficient.  For an apparatus azimuth
\(\alpha\in\{\theta,\phi\}\), we remove this unit by defining
\begin{equation}
 \begin{aligned}
 \widehat\Gamma_i^{(6)}(\alpha)
 &=\frac{\Gamma_i^{(6)}(\alpha)}{\Gamma_{\rm ref}^{(6)}},\\
 \Gamma_{\rm ref}^{(6)}
 &=10^{-9}\ {\rm N\,m\,m^{-2}}.
 \end{aligned}
 \label{eq:d6-normalized-gamma}
\end{equation}
Here \(\theta\) and \(\phi\) label two independently selected apparatus-azimuth
settings. An actual two-orientation measurement uses distinct values, whereas
the maps below scan the full pair \((\theta,\phi)\), including the diagonal as
 a null check. The first two-coefficient block provides the simplest measure of
 whether the selected settings probe independent directions.
\begin{align}
 s_0^{(6)}(\theta,\phi)
 &\equiv
 \left[
 \det
 \begin{pmatrix}
  \widehat\Gamma_1^{(6)}(\theta)&
  \widehat\Gamma_2^{(6)}(\theta)\\
  \widehat\Gamma_1^{(6)}(\phi)&
  \widehat\Gamma_2^{(6)}(\phi)
 \end{pmatrix}
 \right]^2
 \nonumber\\
 &=\left[
 \widehat\Gamma_1^{(6)}(\theta)
 \widehat\Gamma_2^{(6)}(\phi)
 -\widehat\Gamma_2^{(6)}(\theta)
 \widehat\Gamma_1^{(6)}(\phi)
 \right]^2.
 \label{eq:d6-s0}
\end{align}
Thus \(s_0^{1/2}=0\) when the two response vectors are parallel even if their
individual amplitudes are large.

The remaining factors follow the ordered \(d=6\) basis of Sec.~\ref{sec:theory}.
The quantities \(s_2^{1/2}\) and \(s_1^{1/2}\) are the absolute four-response
determinants in Eq.~\eqref{eq:four-response-determinant-cartesian}, formed from
\(\widehat\Gamma_{3\text{--}6}^{(6)}\) and
\(\widehat\Gamma_{9\text{--}12}^{(6)}\), respectively.
The two remaining two-component channels are
\begin{align}
 s_4+s_{44}
 &=\sum_{i=7}^{8}\left[
 (\widehat\Gamma_i^{(6)}(\theta))^2+
 (\widehat\Gamma_i^{(6)}(\phi))^2\right],
 \nonumber\\
 s_3+s_{33}
 &=\sum_{i=13}^{14}\left[
 (\widehat\Gamma_i^{(6)}(\theta))^2+
 (\widehat\Gamma_i^{(6)}(\phi))^2\right].
 \label{eq:d6-pair-responses}
\end{align}
Each of these highest-frequency coefficient pairs can be resolved at a single
apparatus azimuth.
We combine the five blocks into the dimensionless total transfer coefficient
\begin{equation}
 {\cal R}_6(\theta,\phi)=
 \left[s_0^{1/2}s_2^{1/2}(s_4+s_{44})
 s_1^{1/2}(s_3+s_{33})\right]^{1/7}.
 \label{eq:s6metric}
\end{equation}
The exponent \(1/7\) is part of the definition of this aggregate transfer
function. The construction contains the seven elementary contributions
\(s_0,s_1,s_2,s_3,s_{33},s_4,s_{44}\).  This convention
rewards both response amplitude and linear independence, and the aggregate
vanishes if any required harmonic block becomes singular.  It is an
unweighted geometry diagnostic and should not be identified with the
noise-weighted score \({\cal D}_6\) in Eq.~\eqref{eq:doptimal}.

\subsubsection{Centerline registration}

\begin{figure*}[t]
 \centering
 \includegraphics[width=0.80\textwidth]{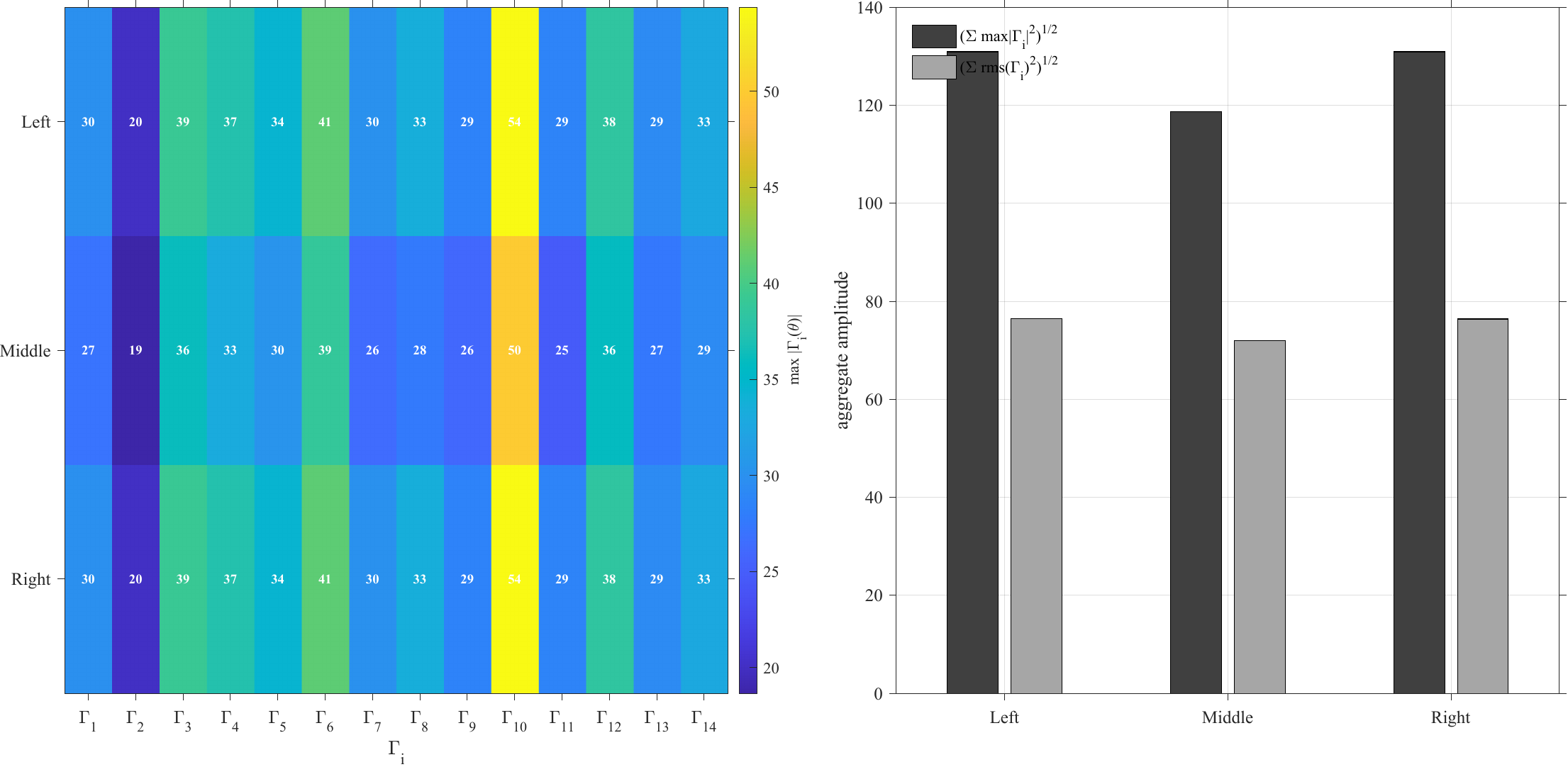}
 \par\vspace{0.6ex}
 \includegraphics[width=0.88\textwidth]{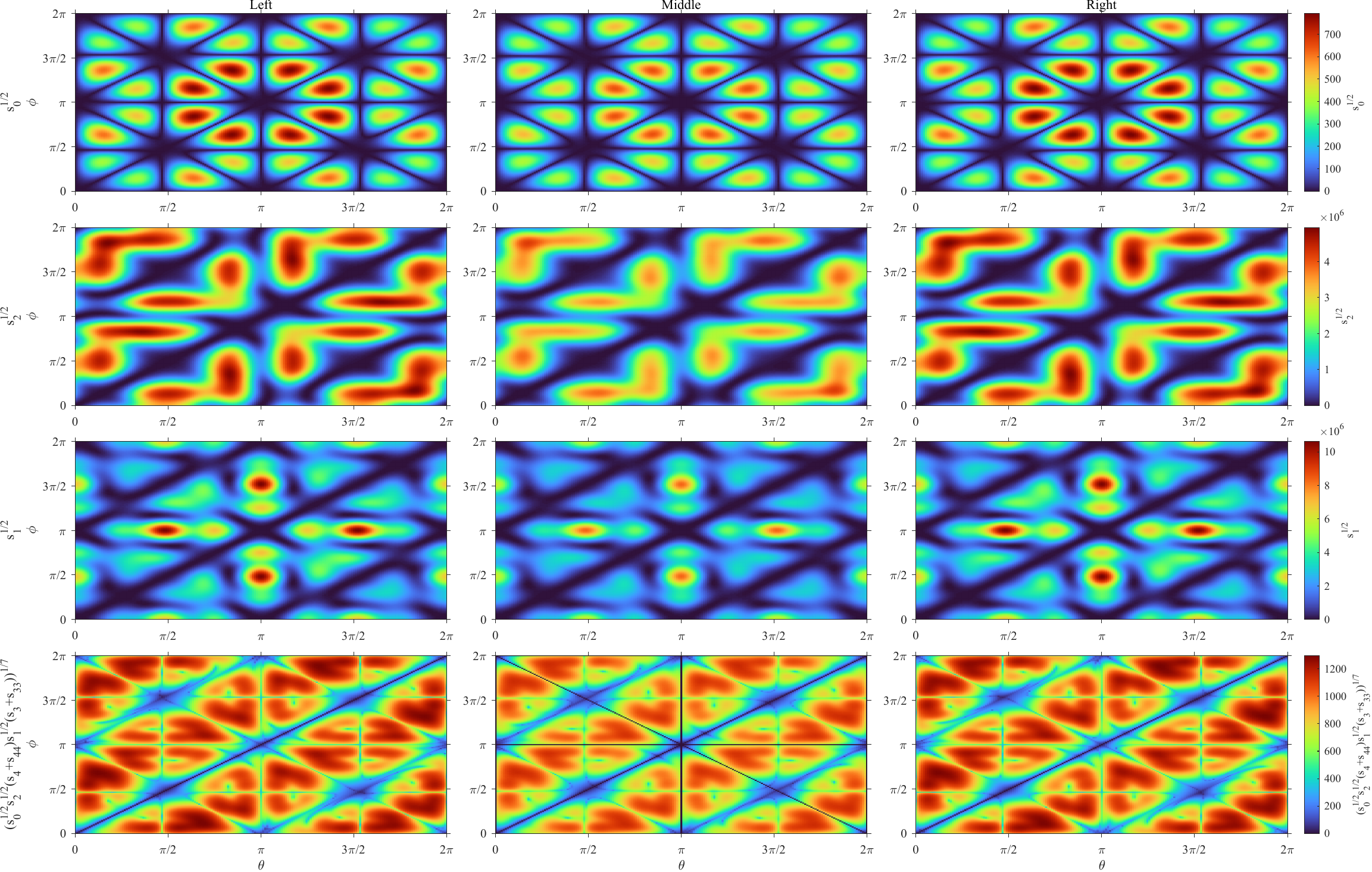}
 \caption{Centerline-registration comparison for the \(d=6\), five-stripe
 vertical family.  The upper plot compares the 14 coefficient amplitudes and
 their aggregate measures for the left, middle, and right registrations.  The
 lower plot shows the corresponding maps of \(s_0^{1/2}\), \(s_2^{1/2}\),
 \(s_1^{1/2}\), and \({\cal R}_6\), with a common color scale across the
 three columns in each row.}
 \label{fig:d6-midline-vertical}
\end{figure*}

\begin{figure*}[t]
 \centering
 \includegraphics[width=0.80\textwidth]{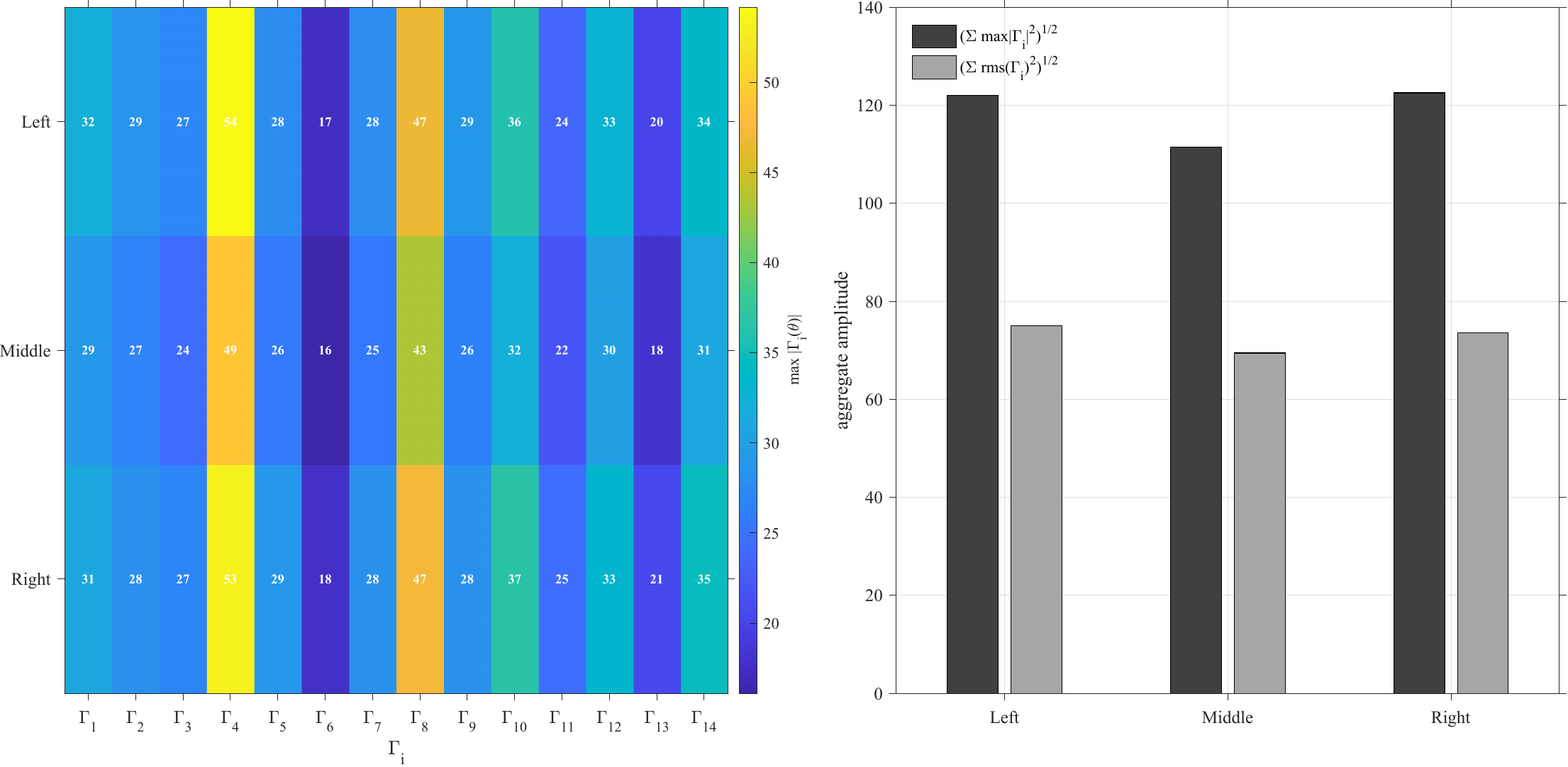}
 \par\vspace{0.6ex}
 \includegraphics[width=0.88\textwidth]{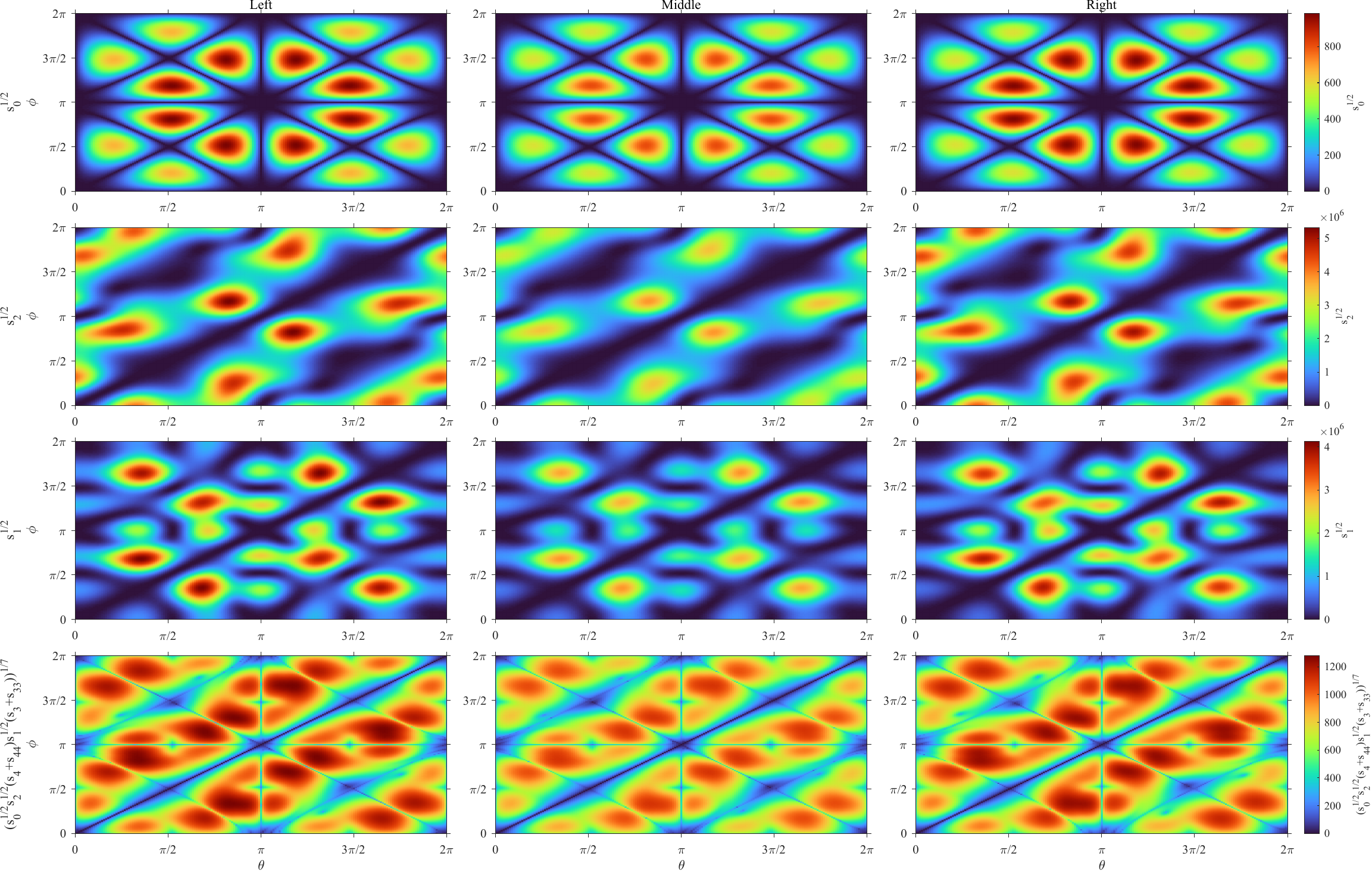}
 \caption{Centerline-registration comparison for the \(d=6\), five-stripe
 horizontal family, in the same format as Fig.~\ref{fig:d6-midline-vertical}.
 The columns are ordered left, middle, and right.}
 \label{fig:d6-midline-horizontal}
\end{figure*}

With \(N_s=5\) fixed, we move the connector and stripe registration from
the middle to the left or right side.  The resulting coefficient amplitudes
and two-angle response maps are shown for the vertical-stripe family in
Fig.~\ref{fig:d6-midline-vertical} and for the horizontal-stripe family in
Fig.~\ref{fig:d6-midline-horizontal}.  The columns are ordered left, middle,
and right in both figures, matching the physical centerline registration.

Either off-center registration improves the displayed total response.  For
vertical stripes, \(\max{\cal R}_6\) is \(1.297\times10^3\),
\(1.151\times10^3\), and \(1.297\times10^3\) for the left, middle, and right
registrations, respectively.  This gives a \(12.6\%\) gain and an almost exact
left--right reflection.  For horizontal stripes the corresponding maxima are
\(1.279\times10^3\), \(1.108\times10^3\), and \(1.239\times10^3\), and the
best value is \(15.4\%\) above the middle result.  The small left--right
difference in the horizontal family is consistent with the finite angular
sampling used to locate the maxima.  Centerline placement therefore
provides a useful secondary optimization without changing the outer envelope
or stripe count. The choice between the two reflected off-center layouts should also account
for fabrication access, electrostatic shielding, and the systematic
derivatives in Eq.~\eqref{eq:systematic-linear}.
The individual \(\Gamma_i^{(6)}(\alpha)\) curves for both stripe orientations
are collected in Appendix~\ref{app:gamma-curves}.

\subsection{\texorpdfstring{\(d=8\) response}{d=8 response}}
\label{subsec:d8results}

\subsubsection{Dimensionless response measure}

For \(d=8\), the corresponding dimensionless transfer functions at
\(\alpha\in\{\theta,\phi\}\) are
\begin{equation}
 \begin{aligned}
 \widehat\Gamma_i^{(8)}(\alpha)
 &=\frac{\Gamma_i^{(8)}(\alpha)}{\Gamma_{\rm ref}^{(8)}},\\
 \Gamma_{\rm ref}^{(8)}
 &=10^{-4}\ {\rm N\,m\,m^{-4}}.
 \end{aligned}
 \label{eq:d8-normalized-gamma}
\end{equation}
The same fixed \(\Gamma_{\rm ref}^{(8)}\) is used for every geometry.  This is
 a transfer-coefficient scale rather than a torque sensitivity. Multiplication
by a representative coefficient of \(10^{-12}\,\mathrm{m}^{4}\) gives a torque
scale of \(10^{-16}\,\mathrm{N\,m}\).  We retain the SI-valued transfer
functions and apply Eq.~\eqref{eq:d8-normalized-gamma} uniformly.  A separate
maximum or rms normalization for each configuration would erase precisely the
amplitude differences that the geometry comparison is intended to preserve.
The first block is again
\begin{equation}
 s_0^{(8)}(\theta,\phi)=
 \left[
 \widehat\Gamma_1^{(8)}(\theta)
 \widehat\Gamma_2^{(8)}(\phi)
 -\widehat\Gamma_2^{(8)}(\theta)
 \widehat\Gamma_1^{(8)}(\phi)
 \right]^2.
 \label{eq:d8-s0}
\end{equation}
The four-response determinants \(s_2^{1/2}\), \(s_4^{1/2}\),
\(s_1^{1/2}\), and \(s_3^{1/2}\) are formed from the consecutive groups
\(\widehat\Gamma_{3\text{--}6}^{(8)}\),
\(\widehat\Gamma_{7\text{--}10}^{(8)}\),
\(\widehat\Gamma_{13\text{--}16}^{(8)}\), and
\(\widehat\Gamma_{17\text{--}20}^{(8)}\), respectively.  The highest
two-component channels are
\begin{align}
 s_6+s_{66}
 &=\sum_{i=11}^{12}\left[
 (\widehat\Gamma_i^{(8)}(\theta))^2+
 (\widehat\Gamma_i^{(8)}(\phi))^2\right],
 \nonumber\\
 s_5+s_{55}
 &=\sum_{i=21}^{22}\left[
 (\widehat\Gamma_i^{(8)}(\theta))^2+
 (\widehat\Gamma_i^{(8)}(\phi))^2\right].
 \label{eq:d8-pair-responses}
\end{align}
The total dimensionless transfer coefficient is therefore
\begin{equation}
 \begin{split}
 {\cal R}_8(\theta,\phi)=\bigl[&
 s_0^{1/2}s_2^{1/2}s_4^{1/2}(s_6+s_{66})\\
 &\times s_1^{1/2}s_3^{1/2}(s_5+s_{55})\bigr]^{1/9}.
 \end{split}
 \label{eq:s8metric}
\end{equation}
The exponent \(1/9\) analogously reflects the nine elementary contributions
\(s_0,s_1,s_2,s_3,s_4,s_5,s_{55},s_6,s_{66}\) entering this definition.  As
for \({\cal R}_6\), \({\cal R}_8\) is an unweighted aggregate transfer
function that diagnoses response strength and linear independence before a
noise model is imposed; it is not the Fisher-information score
\({\cal D}_8\).

\subsubsection{Stripe-number comparison}

\begin{figure*}[t]
 \centering
 \includegraphics[width=0.80\textwidth]{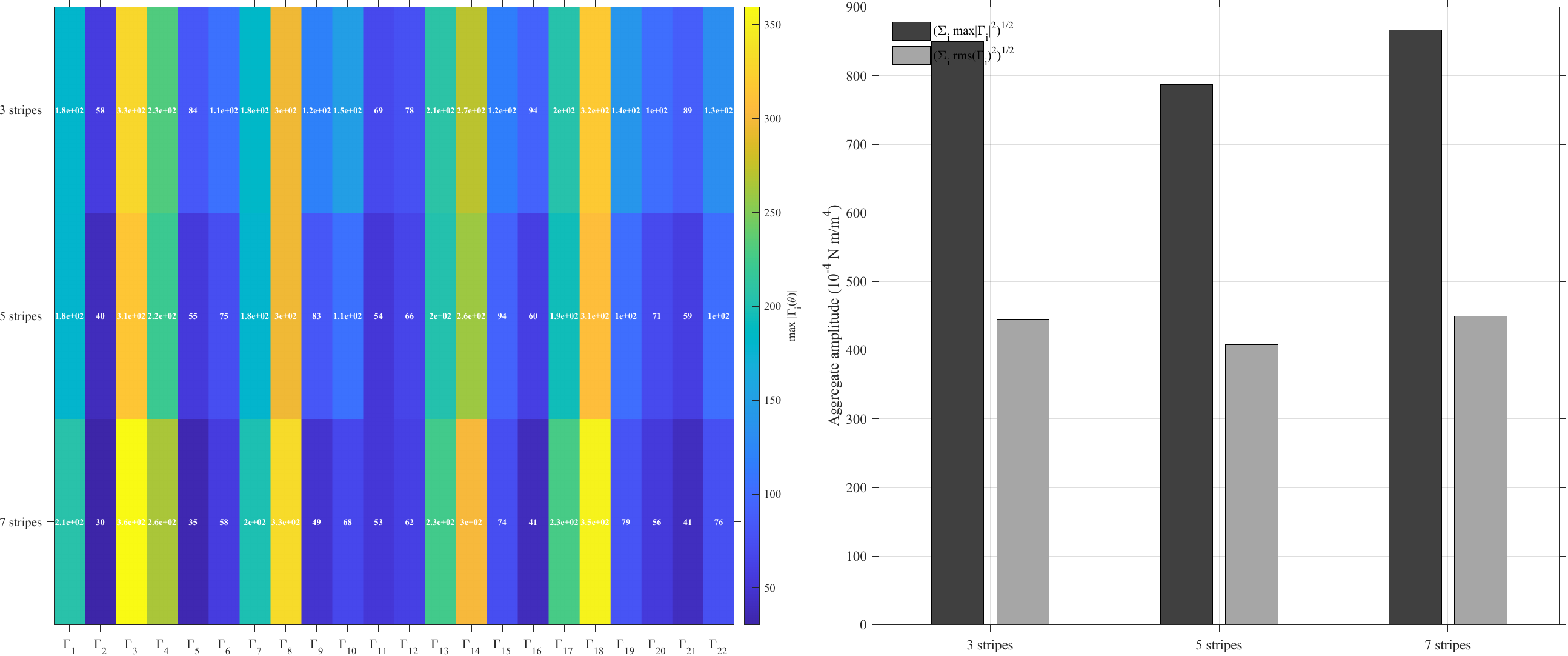}
 \par\vspace{0.6ex}
 \includegraphics[width=0.88\textwidth]{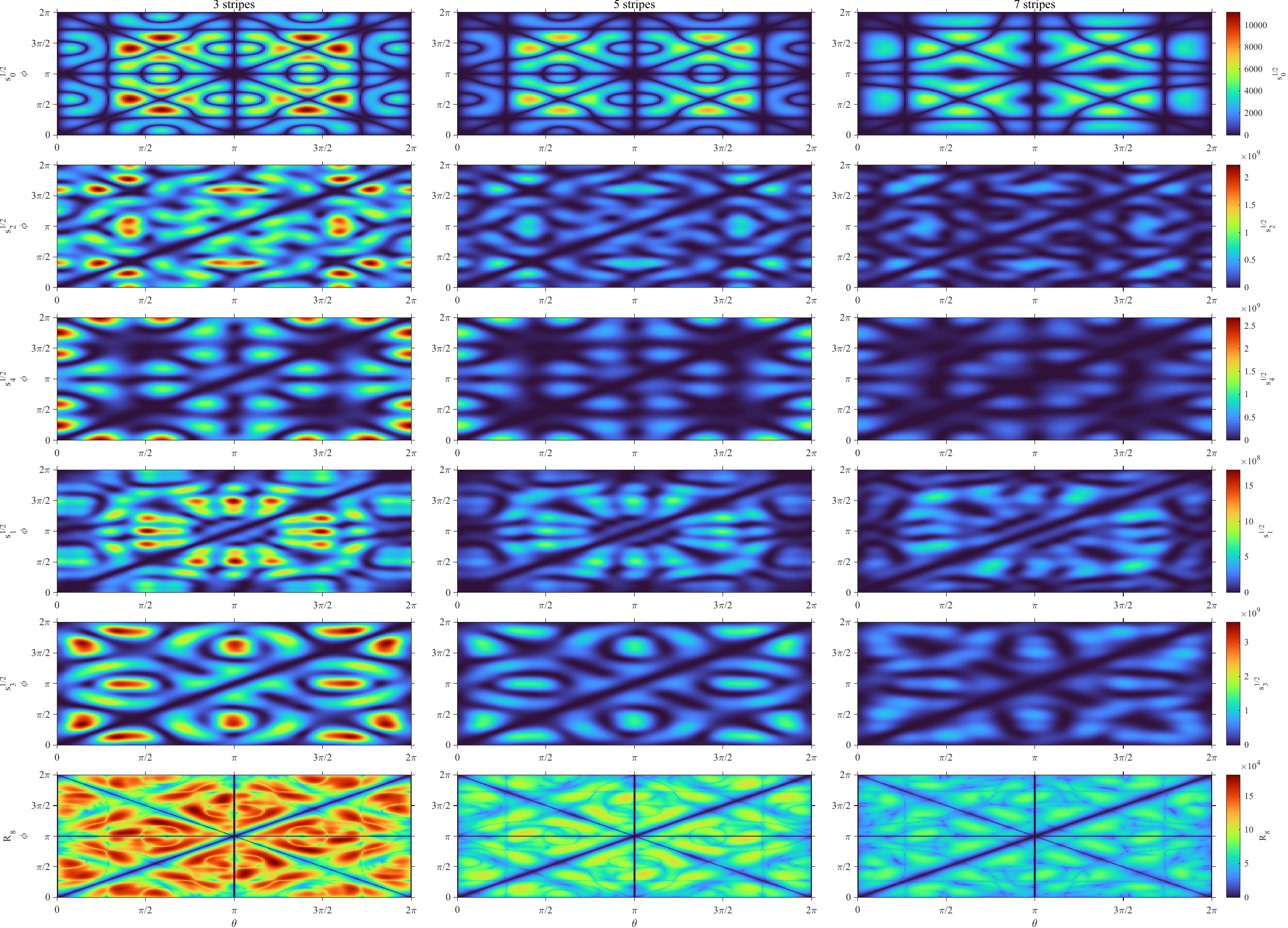}
 \caption{Stripe-number comparison for the \(d=8\) vertical-stripe family.
 The upper coefficient summary gives
 \(\max_\alpha|\widehat\Gamma_i^{(8)}(\alpha)|\) for all 22 coefficients and
 the aggregate maximum and rms norms.  The lower response montage gives
 \(s_0^{1/2}\), \(s_2^{1/2}\), \(s_4^{1/2}\), \(s_1^{1/2}\),
 \(s_3^{1/2}\), and \({\cal R}_8\), with columns ordered
 \(N_s=3,5,7\).  Each metric row uses a common color scale across the three
 stripe counts.}
 \label{fig:d8-stripes-vertical}
\end{figure*}

\begin{figure*}[t]
 \centering
 \includegraphics[width=0.80\textwidth]{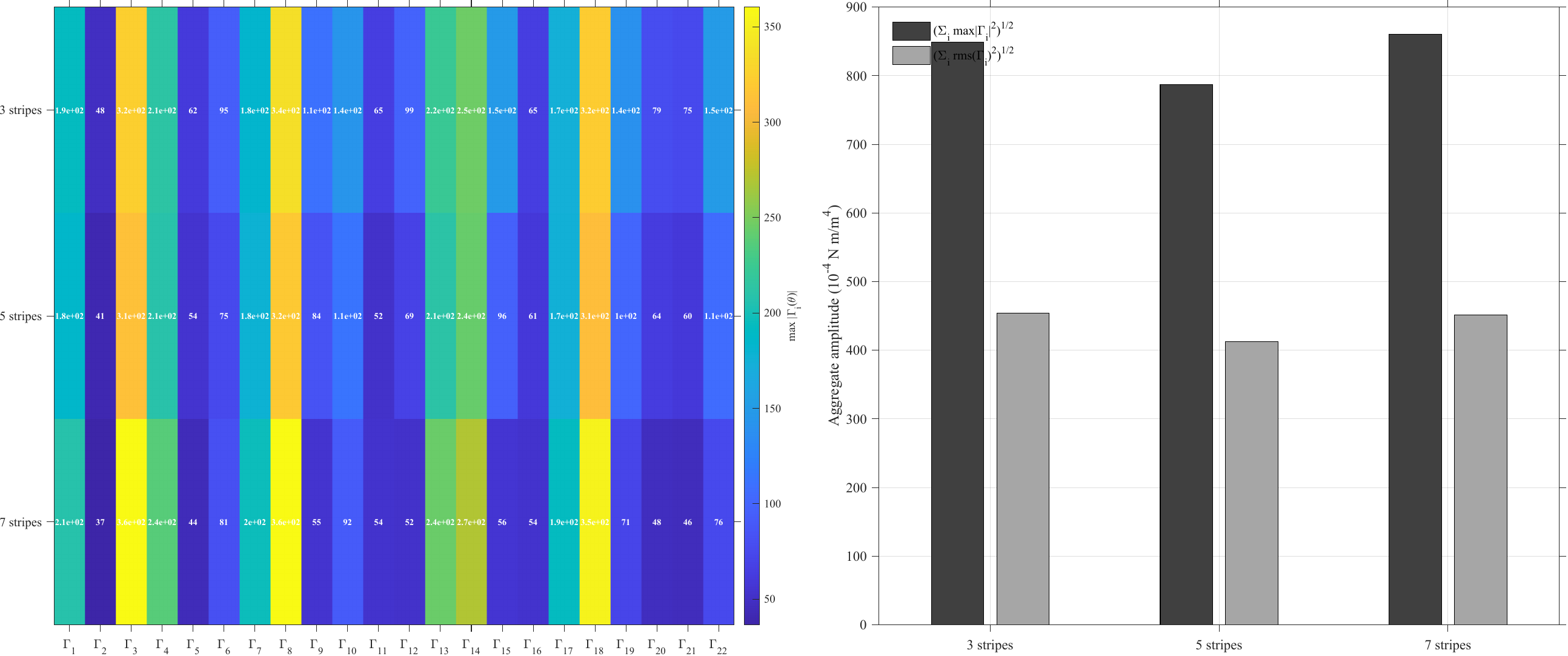}
 \par\vspace{0.6ex}
 \includegraphics[width=0.88\textwidth]{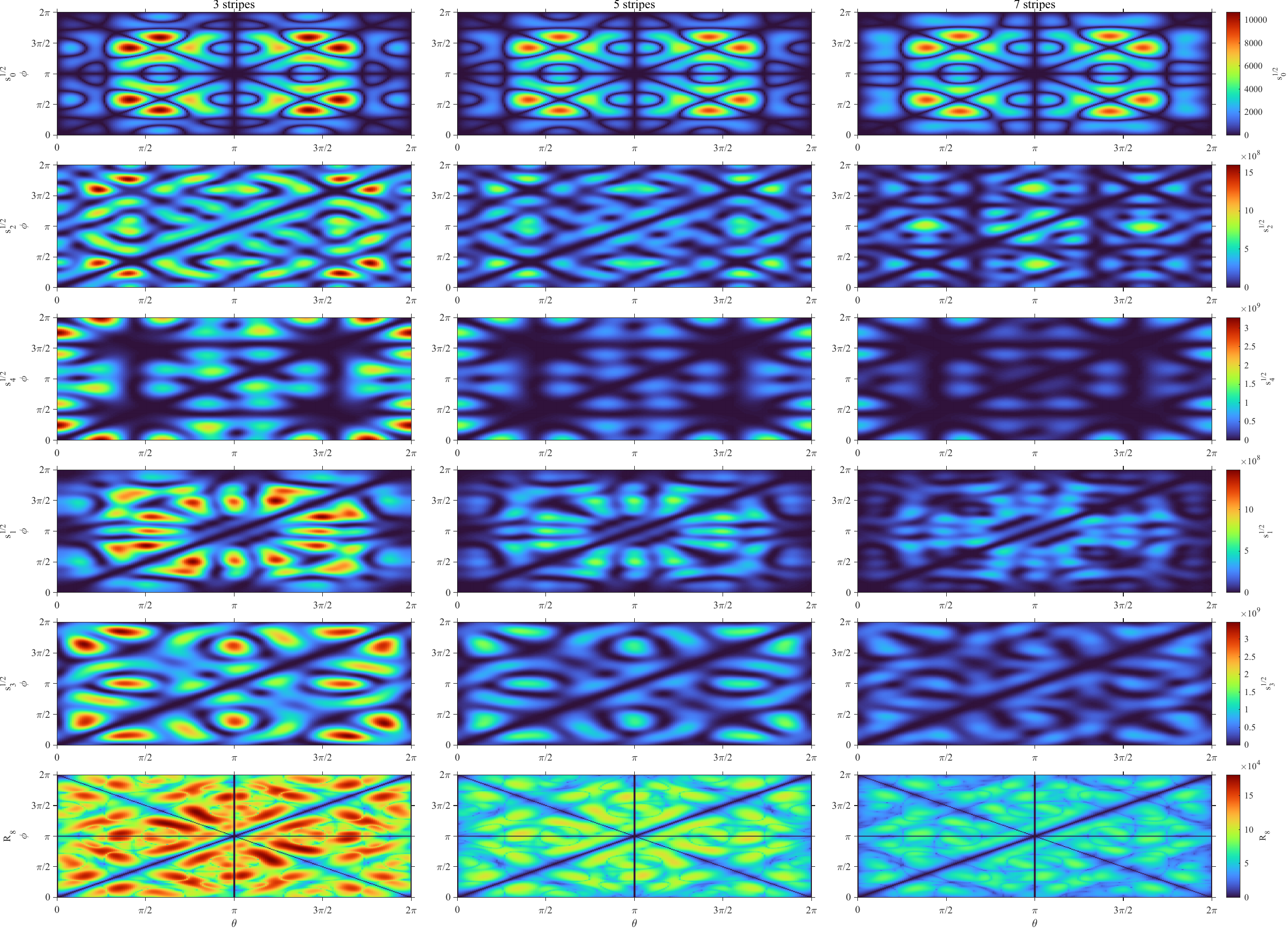}
 \caption{Stripe-number comparison for the \(d=8\) horizontal-stripe family,
 in the same format as Fig.~\ref{fig:d8-stripes-vertical}.  The upper plot
 compares all 22 normalized transfer-coefficient amplitudes and their
 aggregate norms; the lower plot shows the five displayed response blocks and
 \({\cal R}_8\) for \(N_s=3,5,7\).}
 \label{fig:d8-stripes-horizontal}
\end{figure*}

Earlier work evaluated the complete \(d=8\) coefficient response for a fixed
multistripe baseline \cite{Jin2024d8}.  Here the stripe count is varied under
one common normalization so that the three candidate geometries can be ranked
without rescaling away their absolute response differences.  The \(d=8\)
kernel is more strongly concentrated at the nearest density
boundaries than its \(d=6\) counterpart and contains both \(j=4\) and \(j=6\)
angular moments.  Consequently, stripe orientation and stripe count affect
the harmonic blocks more unevenly.  Figures~\ref{fig:d8-stripes-vertical} and
\ref{fig:d8-stripes-horizontal} show a compact summary of all 22 normalized
transfer coefficients together with the two-angle response maps.  The coefficient
summary reports \(\max_\alpha|\widehat\Gamma_i^{(8)}(\alpha)|\) and the two
aggregate norms constructed from the coefficient maxima and rms values.  For
clarity, the density montage displays
\(s_0^{1/2}\), \(s_2^{1/2}\), \(s_4^{1/2}\), \(s_1^{1/2}\),
\(s_3^{1/2}\), and \({\cal R}_8\). Within each figure, every metric row uses
one color scale for all three stripe counts.

For the vertical-stripe family, the maxima of \({\cal R}_8\) are
\(1.811\times10^5\), \(1.118\times10^5\), and \(8.186\times10^4\) for
\(N_s=3,5,7\), respectively.  The horizontal-stripe family gives the closely
corresponding sequence \(1.806\times10^5\), \(1.124\times10^5\), and
\(8.156\times10^4\).  Thus, in both orientations the five- and seven-stripe
responses are about 0.62 and 0.45 times the three-stripe response.  The common
normalization removes the earlier apparent enhancement of selected \(j=6\)
channels and yields the same stripe-count ordering in the two orientations.

The coefficient maxima alone do not give this ordering.  For example, the
aggregate maximum norms for \(N_s=3,5,7\) are \(850.1\), \(787.4\), and
\(867.0\) in the vertical family and \(848.6\), \(786.9\), and \(860.1\) in
the horizontal family.  Seven stripes can therefore produce a slightly larger
Euclidean coefficient amplitude while giving a substantially smaller
\({\cal R}_8\).  The additional response is concentrated in a less balanced
set of channels and does not compensate for the loss of angular independence
in the determinant product.  Increasing the number of density boundaries is
therefore not, by itself, a monotonic route to improved simultaneous
sensitivity.

\subsubsection{Centerline registration}

\begin{figure*}[t]
 \centering
 \includegraphics[width=0.80\textwidth]{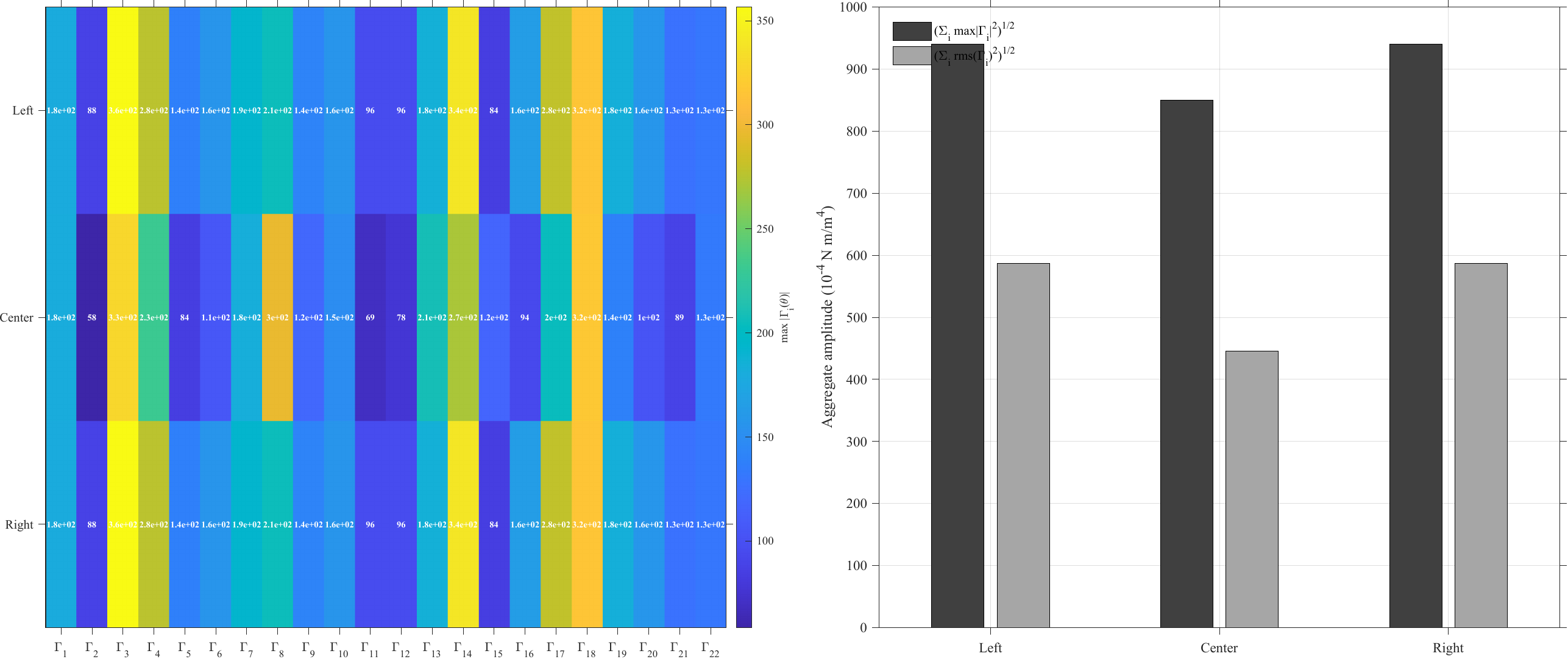}
 \par\vspace{0.6ex}
 \includegraphics[width=0.88\textwidth]{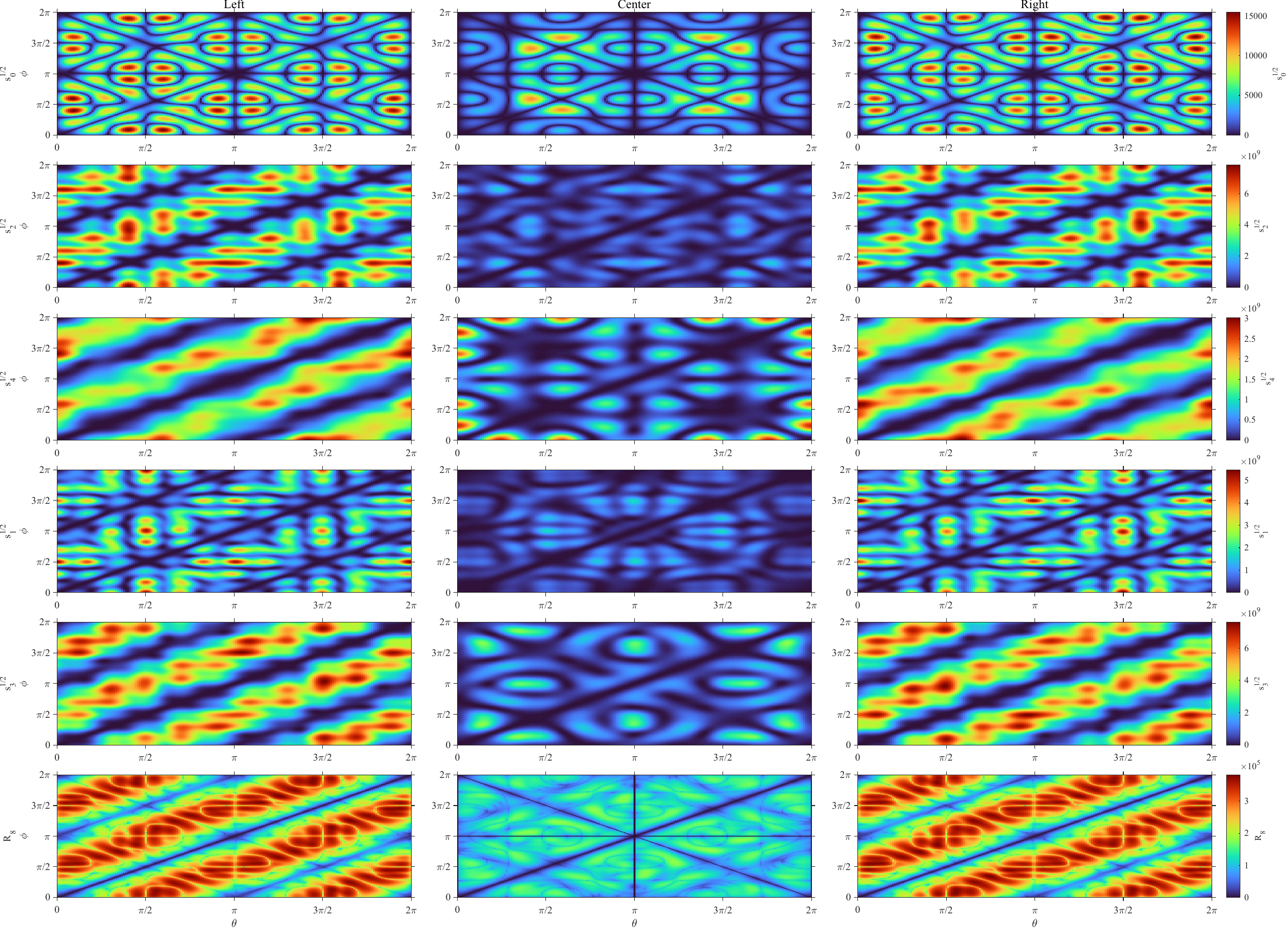}
 \caption{Centerline-registration comparison for the \(d=8\), three-stripe
 vertical family.  The upper coefficient summary gives all 22 normalized
 coefficient amplitudes and the aggregate maximum and rms norms.  The lower
 response montage gives \(s_0^{1/2}\), \(s_2^{1/2}\), \(s_4^{1/2}\),
 \(s_1^{1/2}\), \(s_3^{1/2}\), and \({\cal R}_8\), with columns ordered left,
 center, and right.  Each row uses one color scale for the three
 registrations.  The left and right registrations give equal aggregate
 amplitudes and equal maxima of \({\cal R}_8\), as required by the reflected
 geometry in this orientation.}
 \label{fig:d8-midline-vertical}
\end{figure*}

\begin{figure*}[t]
 \centering
 \includegraphics[width=0.80\textwidth]{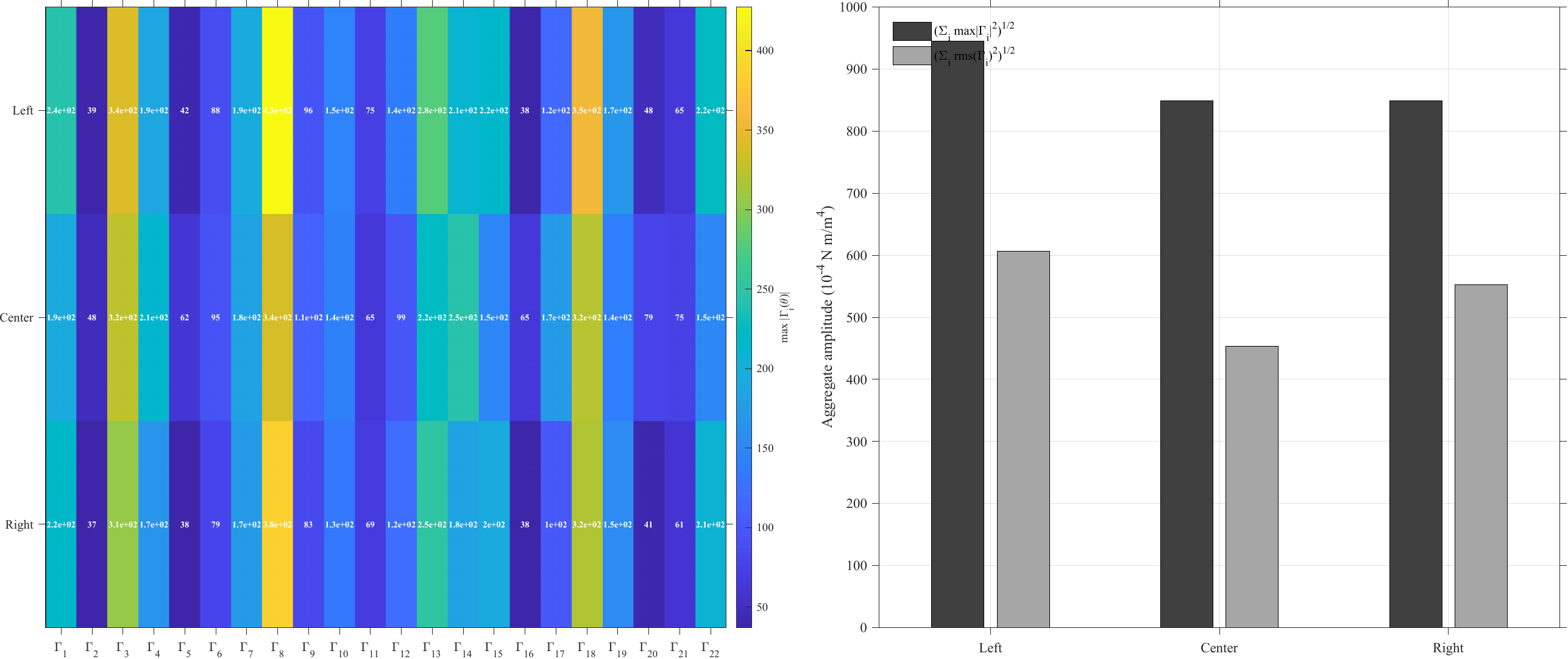}
 \par\vspace{0.6ex}
 \includegraphics[width=0.88\textwidth]{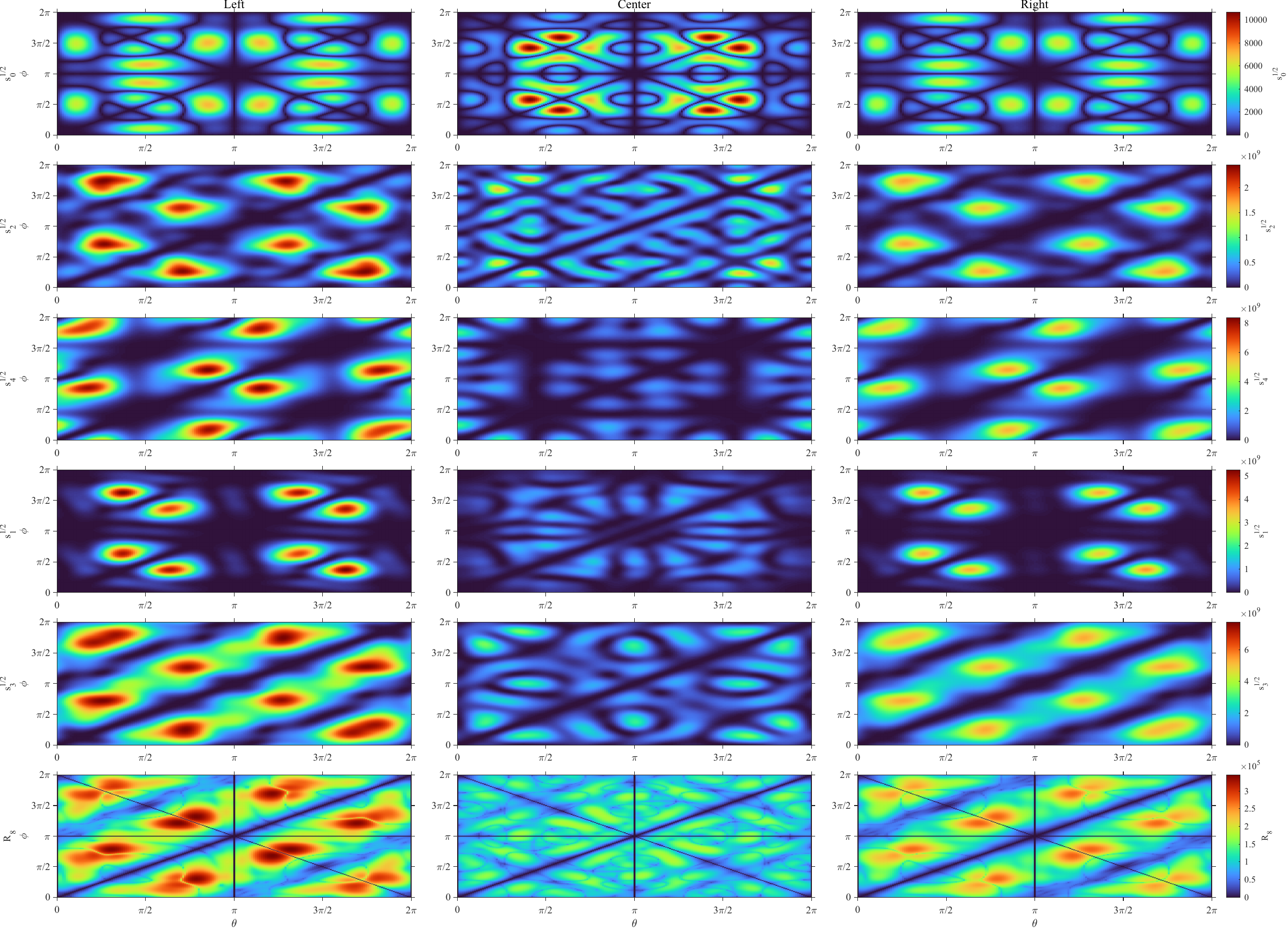}
 \caption{Centerline-registration comparison for the \(d=8\), three-stripe
 horizontal family, in the same format as
 Fig.~\ref{fig:d8-midline-vertical}.  The two off-center registrations produce
 different response maxima for this orientation.}
 \label{fig:d8-midline-horizontal}
\end{figure*}

Having identified \(N_s=3\) as the preferred stripe count, we next move the
connector away from the center while holding all other geometric parameters
fixed.  This changes both the amplitude distribution and the angular phase of
the \(d=8\) response.
Figures~\ref{fig:d8-midline-vertical} and
\ref{fig:d8-midline-horizontal} display the coefficient amplitudes and all
six two-angle response measures shown for the left, center, and right
 registrations.  The centered columns reproduce, within numerical precision,
 the three-stripe baselines in Figs.~\ref{fig:d8-stripes-vertical} and
 \ref{fig:d8-stripes-horizontal}.  Color ranges are shared within a figure but
not between the stripe-number and centerline figures; cross-figure comparisons below
therefore use the numerical maxima rather than color saturation.

For the vertical-stripe family, the left, center, and right registrations give
\(\max {\cal R}_8=3.818\times10^5\), \(1.811\times10^5\), and
\(3.818\times10^5\), respectively.  The equal left and right values, together
with their equal aggregate coefficient norms, provide the expected reflection
check; either off-center registration increases the unweighted response by a
factor of about 2.11.  For the horizontal-stripe family, the corresponding
values are \(3.453\times10^5\), \(1.806\times10^5\), and
\(2.714\times10^5\).  The left displacement is consequently about 1.91 times
stronger than the centered design, while the right displacement gives a factor
of 1.50.

Among the \(d=8\) configurations evaluated here, the largest value of
\({\cal R}_8\) is therefore obtained with three vertical stripes and an
off-center registration, with the three-stripe horizontal-left configuration a
close alternative.  This ranking concerns the unweighted geometric response.
The noise-weighted information determinant can still select between the two
orientations once measured harmonic covariances and machining priors are
included.  The complete \(\Gamma_i^{(8)}(\alpha)\) curves for all four
comparisons are collected in Appendix~\ref{app:gamma-curves}.

The contrast between the two operator dimensions is useful for apparatus
design. The five-stripe \(d=6\) baseline of Ref.~\cite{Jin2026Optimization}
differs from the three-stripe \(d=8\) geometry selected here. In both sectors an
off-center registration gives the strongest unweighted response among the
geometries studied.  A
single apparatus intended to constrain both sectors must therefore balance
these different geometric optima.  A joint design maximizes
\(w_6\ln{\cal D}_6+w_8\ln{\cal D}_8\), with weights set by the scientific
priorities and measured noise.
\subsection{Coefficient sensitivity and operating azimuths}
\label{subsec:sensitivity}

The response measures above rank geometries without assuming an experimental
noise level.  A coefficient-level forecast requires the orientation
degeneracies to be removed and the transfer matrix to be combined with an
explicit covariance for the fitted sidereal amplitudes.  A single apparatus
azimuth supplies only 9 amplitudes for 14 real \(d=6\) coefficient components
and 13 amplitudes for 22 real \(d=8\) components.  At least two independently
chosen azimuths are therefore required for a simultaneous determination in
either sector.  Two-azimuth operation and sector-specific sensitivity estimates
were considered in the preceding \(d=6\) and \(d=8\) design studies
\cite{Li2025CQG,Jin2026Optimization,Jin2024d8}.  The analysis below extends that treatment
by applying one covariance model to every centerline registration and by
reoptimizing the two azimuths for each geometry.

For a common treatment, let \(p_6=9\), \(p_8=13\), \(n_6=14\), and
\(n_8=22\).  The amplitude vectors at two azimuths are stacked as
\begin{equation}
 \begin{aligned}
 \bm y_d&=\mathbf H_d(\theta_d,\phi_d)\bm k^{(d)}+\bm n_d,\\
 \mathbf H_d(\theta_d,\phi_d)&=
 \begin{pmatrix}
  \mathbf G^{(d)}(\theta_d)\\
  \mathbf G^{(d)}(\phi_d)
 \end{pmatrix},
 \qquad d\in\{6,8\},
 \end{aligned}
 \label{eq:two-angle-sensitivity-model}
\end{equation}
where \(\mathbf H_d\) has dimensions \(2p_d\times n_d\).  Its row and column
orderings follow Eqs.~\eqref{eq:amplitude-vector-general},
\eqref{eq:d8-amplitude-vector}, \eqref{eq:d6-coeff-basis}, and
\eqref{eq:d8-coeff-basis}.  The SI-valued transfer coefficients are used in
\(\mathbf H_d\).  The normalization factors in
Eqs.~\eqref{eq:d6-normalized-gamma} and \eqref{eq:d8-normalized-gamma} are used
only for the dimensionless geometry measures and are not retained in the
covariance inversion.

The constant term and the sidereal-harmonic amplitudes are distinct
observables with different error mechanisms.  The amplitude \(C_0\) is
independent of sidereal time, so a static Newtonian torque generated by
machining, metrology, or alignment errors contributes directly to this
channel.  The striped-balance error budget of Ref.~\cite{Shao2016Enhanced}
gives a total uncertainty of approximately
\(3.8\times10^{-16}\,{\rm N\,m}\) for \(C_0\), dominated by these systematic
effects, while the thermal-noise contribution is about
\(0.4\times10^{-16}\,{\rm N\,m}\).

By contrast, \(C_m\) and \(S_m\) are the cosine and sine quadratures at
\(m\omega_\oplus\).  They are obtained from a harmonic fit over the
\(2\pi/\omega_\oplus\simeq23.93~{\rm h}\) sidereal cycle and, in practice, over
several such cycles.  Static offsets do not project onto these harmonics.  For
the reference integration time of about 10 days,
 Ref.~\cite{Shao2016Enhanced} estimates an uncertainty below
 \(0.45\times10^{-16}\,{\rm N\,m}\) for each quadrature, dominated by thermal
 noise.  We use rounded values as effective one-standard-deviation design
 uncertainties and represent the expected null outcome as
\begin{equation}
 \begin{aligned}
 \widehat C_0^{(d)}&=(0\pm4.00)\times10^{-16}\ {\rm N\,m},\\
 \widehat C_m^{(d)}&=(0\pm0.50)\times10^{-16}\ {\rm N\,m},\\
  \widehat S_m^{(d)}&=(0\pm0.50)\times10^{-16}\ {\rm N\,m}.
 \end{aligned}
 \label{eq:benchmark-amplitude-noise}
\end{equation}
These relations apply for \(m=1,\ldots,d-2\).  The zeros are assumed
central values and not measured results.  The
corresponding standard deviations are
\(\sigma_0=4.00\times10^{-16}\,{\rm N\,m}\) and
\(\sigma_h=0.50\times10^{-16}\,{\rm N\,m}\).  At one azimuth, the amplitude
covariance is
\begin{equation}
 \mathbf C_{\rm amp}^{(d)}=
 \operatorname{diag}\!\left(
 \sigma_0^2,
 \underbrace{\sigma_h^2,\ldots,\sigma_h^2}_{p_d-1}
 \right).
 \label{eq:benchmark-amplitude-covariance}
\end{equation}
The two-azimuth covariance is
\begin{equation}
 \mathbf C_a^{(d)}=\operatorname{diag}\!\left(
 \mathbf C_{\rm amp}^{(d)},\mathbf C_{\rm amp}^{(d)}\right).
 \label{eq:two-azimuth-amplitude-covariance}
\end{equation}
In the notation used here, \(s_0\) is the two-azimuth response determinant associated with
\(C_0\).  Increasing \(s_0\) can therefore compensate for the larger systematic
floor in the constant channel, but it does not define a universal optimum for
all coefficients.
The same error budget is assigned at the two azimuths, and correlations
between fitted amplitudes are neglected at this design stage.  The Fisher
matrix and marginalized coefficient covariance are
\begin{equation}
 \mathbf F_d=\mathbf H_d^T
 \left(\mathbf C_a^{(d)}\right)^{-1}\mathbf H_d,
 \qquad
 \bm\Sigma_d=\mathbf F_d^{-1}.
 \label{eq:projected-coefficient-covariance}
\end{equation}

Equation~\eqref{eq:projected-coefficient-covariance} describes a simultaneous
fit of all 14 or 22 real components.  The SME coefficient components are
independent parameters, but their estimators can be correlated because several
components enter the same measured harmonic.  An angle optimum is therefore
defined only after an estimation objective has been specified.  For a joint
fit, we use the \(D\)-optimal criterion
\begin{equation}
 (\theta_{D,d},\phi_{D,d})=
 \underset{\theta_d,\phi_d\in[0,2\pi)}{\operatorname{arg\,max}}
 \log\det\mathbf F_d.
 \label{eq:d-optimal-azimuths}
\end{equation}
This choice minimizes the joint covariance volume.  It need not maximize any
individual \(\widehat\Gamma_i^{(d)}\), minimize every diagonal element of
\(\bm\Sigma_d\), or maximize the unweighted response \({\cal R}_d\).  If one
component \(k_a^{(d)}\) is prioritized, a targeted choice would instead obey
\begin{equation}
 (\theta_a,\phi_a)=
 \underset{\theta,\phi\in[0,2\pi)}{\operatorname{arg\,min}}
 \left[\mathbf F_d^{-1}(\theta,\phi)\right]_{aa},
 \label{eq:coefficient-targeted-azimuths}
\end{equation}
 and generally changes with \(a\).  The Fisher pair is consequently a
 noise-model-dependent compromise for simultaneous estimation rather than a
 coefficient-independent optimum.

We first apply the joint-fit criterion to the centered baseline geometries used
for the sensitivity forecast, namely five horizontal stripes for \(d=6\) and
three horizontal stripes for \(d=8\).  Representative Fisher \(D\)-optimal
pairs are
\begin{equation}
 \begin{aligned}
 (\theta_{D,6}^{\rm H},\phi_{D,6}^{\rm H})
   &=(2.6161098,\,1.0595183)\ {\rm rad},\\
 (\theta_{D,8}^{\rm H},\phi_{D,8}^{\rm H})
   &=(1.9265034,\,3.8009980)\ {\rm rad}.
 \end{aligned}
 \label{eq:d-optimal-angle-pairs-horizontal}
\end{equation}
The superscript \({\rm H}\) denotes horizontal stripes.  The corresponding
stacked response matrices have ranks 14 and 22, respectively.  Their complete
transfer coefficients at these azimuths are listed in
Table~\ref{tab:horizontal-angle-gamma}.

\begin{table*}[t]
\caption{Transfer coefficients for the centered horizontal-stripe baselines at
the Fisher angle pairs in Eq.~\eqref{eq:d-optimal-angle-pairs-horizontal}.
The \(d=6\) entries are in units of
\(10^{-9}\,{\rm N\,m/m^{2}}\), and the \(d=8\) entries are in units of
\(10^{-4}\,{\rm N\,m/m^{4}}\).}
\label{tab:horizontal-angle-gamma}

\centering
\small

\begin{tabular}{crr@{\hspace{2.4em}}crr}
\toprule
\multicolumn{3}{c}{\(d=6\), five horizontal stripes} &
\multicolumn{3}{c}{\(d=8\), three horizontal stripes}\\
\(i\) & \(\Gamma_i^{(6)}(\theta_{D,6}^{\rm H})\) &
\(\Gamma_i^{(6)}(\phi_{D,6}^{\rm H})\) &
\(i\) & \(\Gamma_i^{(8)}(\theta_{D,8}^{\rm H})\) &
\(\Gamma_i^{(8)}(\phi_{D,8}^{\rm H})\)\\
\midrule
 1 & -16.63 & -22.82 &  1 & -182.45 &  -35.35\\
 2 & -12.23 &   9.12 &  2 &  -15.86 &  -34.08\\
 3 &   8.66 &  23.47 &  3 &   55.41 & -225.19\\
 4 & -38.48 &  21.07 &  4 &  -11.42 &  146.78\\
 5 & -22.91 & -25.60 &  5 &  -53.28 &  -27.69\\
 6 & -10.68 &  -1.78 &  6 &   67.03 &   -3.80\\
 7 &  18.96 &   9.06 &  7 & -127.12 &   90.17\\
 8 & -19.81 &  -8.17 &  8 &  -10.51 &  181.76\\
 9 &  10.88 &  24.76 &  9 &  -83.60 &   18.53\\
10 &  25.97 & -15.88 & 10 &  -11.89 & -113.62\\
11 &   9.46 &  14.04 & 11 &  -11.87 &  -65.07\\
12 &  -6.73 &  24.98 & 12 &   -1.00 &  -42.82\\
13 &  13.36 &  -4.48 & 13 & -223.38 &   68.45\\
14 &  22.93 & -26.73 & 14 &   89.91 &  194.51\\
   &         &         & 15 & -126.65 &   34.87\\
   &         &         & 16 &   26.35 &  -23.55\\
   &         &         & 17 &  -70.96 &    7.43\\
   &         &         & 18 & -142.49 & -104.24\\
   &         &         & 19 &   74.19 & -101.32\\
   &         &         & 20 &   12.21 &  -59.45\\
   &         &         & 21 &  -71.03 &  -19.13\\
   &         &         & 22 &   55.19 &   84.96\\
\bottomrule
\end{tabular}

\end{table*}

The projected horizontal-stripe uncertainties obtained from
Eq.~\eqref{eq:projected-coefficient-covariance} are given in
Table~\ref{tab:horizontal-angle-sensitivity}.  A grouped \({\rm Re/Im}\) entry
applies separately to two independent real components whose uncertainties are
equal under the symmetric quadrature-noise model.

\begin{table*}[t]
\caption{Projected marginalized coefficient uncertainties for the centered
horizontal-stripe baselines under Eq.~\eqref{eq:benchmark-amplitude-noise}.
The \(d=6\) entries are in \(10^{-9}\,{\rm m}^{2}\), and the \(d=8\) entries
are in \(10^{-14}\,{\rm m}^{4}\).}
\label{tab:horizontal-angle-sensitivity}

\centering
\small

\begin{tabular}{lcc@{\hspace{2.2em}}lcc}
\toprule
\multicolumn{3}{c}{\(d=6\), five horizontal stripes} &
\multicolumn{3}{c}{\(d=8\), three horizontal stripes}\\
Coefficient & \(\delta k_D\) & \(2\delta k_D\) &
Coefficient & \(\delta k_D\) & \(2\delta k_D\)\\
\midrule
\(k_{2,0}^{(6)}\)                   & 14.2 & 28.3 &
\(k_{4,0}^{(8)}\)                   &  2.7 &  5.3\\
\(k_{4,0}^{(6)}\)                   & 26.2 & 52.5 &
\(k_{6,0}^{(8)}\)                   & 13.1 & 26.3\\
\({\rm Re/Im}\,k_{2,2}^{(6)}\)      &  1.1 &  2.1 &
\({\rm Re/Im}\,k_{4,2}^{(8)}\)      &  0.2 &  0.4\\
\({\rm Re/Im}\,k_{4,2}^{(6)}\)      &  1.5 &  3.0 &
\({\rm Re/Im}\,k_{6,2}^{(8)}\)      &  0.6 &  1.2\\
\({\rm Re/Im}\,k_{4,4}^{(6)}\)      &  1.7 &  3.3 &
\({\rm Re/Im}\,k_{4,4}^{(8)}\)      &  0.2 &  0.5\\
\({\rm Re/Im}\,k_{2,1}^{(6)}\)      &  1.4 &  2.7 &
\({\rm Re/Im}\,k_{6,4}^{(8)}\)      &  0.4 &  0.8\\
\({\rm Re/Im}\,k_{4,1}^{(6)}\)      &  1.8 &  3.6 &
\({\rm Re/Im}\,k_{6,6}^{(8)}\)      &  0.6 &  1.3\\
\({\rm Re/Im}\,k_{4,3}^{(6)}\)      &  1.3 &  2.6 &
\({\rm Re/Im}\,k_{4,1}^{(8)}\)      &  0.2 &  0.4\\
                                      &        &         &
\({\rm Re/Im}\,k_{6,1}^{(8)}\)      &  0.5 &  1.0\\
                                      &        &         &
\({\rm Re/Im}\,k_{4,3}^{(8)}\)      &  0.3 &  0.5\\
                                      &        &         &
\({\rm Re/Im}\,k_{6,3}^{(8)}\)      &  0.4 &  0.7\\
                                      &        &         &
\({\rm Re/Im}\,k_{6,5}^{(8)}\)      &  0.4 &  0.8\\
\bottomrule
\end{tabular}

\end{table*}

The horizontal Fisher pairs give marginalized \(1\sigma\) ranges of
\((1.06\times10^{-9}\text{--}2.62\times10^{-8})\,{\rm m}^{2}\) for \(d=6\)
and \((1.93\times10^{-15}\text{--}1.31\times10^{-13})\,{\rm m}^{4}\) for
\(d=8\).  The broad spread reflects the weaker constant channel as well as
the different transfer strengths of the harmonic blocks.

The vertical-stripe experiment currently under development at HUST uses the
centered five-stripe \(d=6\) geometry and the apparatus azimuths \(\pi/6\) and
\(3\pi/5\).  It strengthens the constant-channel determinant in response to
the larger \(C_0\) error.  To separate this experimental objective from a joint-fit
criterion, we compare it below with the Fisher pair for the same vertical
geometry.  We also perform the analogous comparison for the centered
three-stripe \(d=8\) vertical geometry, for which the \(s_0\)-oriented angles
must be determined independently.

The \(s_0\)-oriented vertical angle pairs are
\begin{equation}
 \begin{aligned}
 (\theta_{S,6}^{\rm V},\phi_{S,6}^{\rm V})&\simeq
  \left(\frac{\pi}{6},\frac{3\pi}{5}\right),\\
 (\theta_{S,8}^{\rm V},\phi_{S,8}^{\rm V})
   &=(1.3080799,\,1.8582241)\ {\rm rad}.
 \end{aligned}
 \label{eq:s0-oriented-angle-pairs}
\end{equation}
Here the subscript \(S\) denotes a choice that prioritizes the constant
response block \(s_0\).  Such a choice is experimentally motivated because
the assumed uncertainty of \(C_0\) is eight times that of each sidereal
quadrature.  The first line follows the high-response operating pair used for
the five-stripe \(d=6\) design.  The second line is obtained independently by
maximizing Eq.~\eqref{eq:d8-s0} for the three-stripe vertical \(d=8\)
geometry.

The \(d=8\) response has four symmetry-related \(s_0\) maxima in
\([0,2\pi)\).  Besides the representative pair in
Eq.~\eqref{eq:s0-oriented-angle-pairs}, the other pairs are approximately
\((1.30808,4.42496)\), \((1.85822,4.97511)\), and
\((4.42496,4.97511)\) rad.  They give the same maximum \(s_0\), but the
responses in the other harmonic blocks are not identical.  We use the
representative pair because it has the larger Fisher determinant within this
degenerate set.  The Fisher \(D\)-optimal pairs for the same vertical
baselines are
\begin{equation}
 \begin{aligned}
 (\theta_{D,6}^{\rm V},\phi_{D,6}^{\rm V})
   &=(0.3512740,\,2.1858087)\ {\rm rad},\\
 (\theta_{D,8}^{\rm V},\phi_{D,8}^{\rm V})
   &=(1.7941852,\,3.8540167)\ {\rm rad}.
 \end{aligned}
 \label{eq:d-optimal-angle-pairs-vertical}
\end{equation}

To compare an \(s_0\)-oriented pair with the Fisher pair, we define
\begin{equation}
 \eta_{D,d}=\left[
 \frac{\det\mathbf F_d(\theta_{S,d}^{\rm V},\phi_{S,d}^{\rm V})}
 {\det\mathbf F_d(\theta_{D,d}^{\rm V},\phi_{D,d}^{\rm V})}
 \right]^{1/n_d}.
 \label{eq:angle-pair-D-efficiency}
\end{equation}
For \(d=6\), the two values of \(s_0^{1/2}\) are 481.126 and 444.054
for the \(S\) and \(D\) pairs, respectively, and \(\eta_{D,6}=0.912\).
Thus the \(S\) pair increases \(s_0^{1/2}\) by 8.3\% while retaining 91.2\%
of the Fisher-pair \(D\)-efficiency.  For \(d=8\), the corresponding values
are 11196.611 and 6811.546, and \(\eta_{D,8}=0.769\).  The \(S\) pair therefore
increases \(s_0^{1/2}\) by 64.4\% while retaining 76.9\% of the Fisher-pair
 \(D\)-efficiency.  All four response matrices have full column rank, 14 for
 \(d=6\) and 22 for \(d=8\).

Equations~\eqref{eq:d-optimal-angle-pairs-horizontal} and
\eqref{eq:d-optimal-angle-pairs-vertical} also show that a Fisher pair found
for one stripe orientation cannot be transferred to the other.  Rotating the
stripe boundaries changes the finite-volume geometric moments and therefore
changes every angular response block.  The comparison between the \(S\) and
\(D\) pairs tests two operating objectives within the vertical geometry,
whereas the comparison between the superscripts \({\rm H}\) and \({\rm V}\)
tests two distinct stripe orientations.  Thus neither comparison supplies a
coefficient-independent angle pair that is optimal in all circumstances.

Table~\ref{tab:vertical-angle-gamma} lists the transfer coefficients used in
the comparisons.  The same units are used for all azimuths and geometries at a
given \(d\), preserving the response differences that determine the inferred
coefficient uncertainties.

\begin{table*}[t]
\caption{Transfer coefficients for the centered vertical-stripe baselines.
Each operator sector compares its own \(s_0\)-oriented pair in
Eq.~\eqref{eq:s0-oriented-angle-pairs} with the corresponding Fisher pair in
Eq.~\eqref{eq:d-optimal-angle-pairs-vertical}.  The \(d=6\) entries are in
units of \(10^{-9}\,{\rm N\,m/m^{2}}\), and the \(d=8\) entries are in units of
\(10^{-4}\,{\rm N\,m/m^{4}}\).}
\label{tab:vertical-angle-gamma}

\centering
\small

\begin{tabular}{crrrr}
\toprule
\multicolumn{5}{c}{\(d=6\), five vertical stripes}\\
\(i\) & \(\Gamma_i^{(6)}(\theta_{S,6}^{\rm V})\) &
\(\Gamma_i^{(6)}(\phi_{S,6}^{\rm V})\) &
\(\Gamma_i^{(6)}(\theta_{D,6}^{\rm V})\) &
\(\Gamma_i^{(6)}(\phi_{D,6}^{\rm V})\)\\
\midrule
 1 &  19.60 & -17.83 &  14.73 & -26.10\\
 2 &  13.76 &  12.04 &  14.63 &   4.23\\
 3 &  30.96 & -17.32 &  23.02 & -29.80\\
 4 &  18.65 & -24.01 &  26.30 & -12.45\\
 5 &  23.19 &  27.77 &  27.16 &  18.01\\
 6 & -19.06 &  20.12 &  -6.96 & -12.68\\
 7 &  25.84 &   7.04 &  24.79 &  -0.19\\
 8 &  -2.08 &  -1.26 &  11.66 & -13.10\\
 9 &  22.13 & -10.92 &  16.42 & -19.76\\
10 &  20.49 & -36.78 &  32.54 & -13.70\\
11 &   1.76 &  21.96 &   6.17 &  21.69\\
12 & -11.66 &   2.32 &   6.77 & -26.80\\
13 &  19.35 &  23.70 &  22.81 &  19.29\\
14 & -12.06 &   6.68 &   6.48 & -18.88\\
\midrule
\multicolumn{5}{c}{\(d=8\), three vertical stripes}\\
\(i\) & \(\Gamma_i^{(8)}(\theta_{S,8}^{\rm V})\) &
\(\Gamma_i^{(8)}(\phi_{S,8}^{\rm V})\) &
\(\Gamma_i^{(8)}(\theta_{D,8}^{\rm V})\) &
\(\Gamma_i^{(8)}(\phi_{D,8}^{\rm V})\)\\
\midrule
 1 &  -24.62 & -180.59 & -176.66 &  -25.03\\
 2 &   56.86 &  -37.71 &  -36.21 &  -43.69\\
 3 &  303.10 &  -38.22 &  -13.11 & -177.67\\
 4 &  -78.05 &  -62.46 &  -97.01 &  180.81\\
 5 &   72.40 &  -82.74 &  -82.54 &  -62.88\\
 6 &   -9.49 &   49.04 &   27.73 &  -21.97\\
 7 &   57.87 & -124.83 & -105.27 &   62.52\\
 8 &  -43.93 &  -45.02 &  -67.19 &  169.03\\
 9 &   48.52 &  -98.90 &  -95.28 &  -25.39\\
10 &   31.51 &  -11.57 &  -32.18 &  -90.85\\
11 &   68.19 &  -36.60 &  -32.72 &  -62.73\\
12 &   -6.20 &  -13.53 &  -28.74 &  -13.06\\
13 &  -36.93 & -204.31 & -193.88 &   51.59\\
14 & -158.53 &   -4.66 &  -58.47 &  247.94\\
15 &  -76.86 & -109.03 & -115.48 &   12.13\\
16 &  -22.85 &    7.42 &  -24.21 &  -11.80\\
17 &  176.43 & -105.09 &  -77.14 &   -2.58\\
18 &  -12.06 & -145.65 & -166.34 &    9.26\\
19 &  120.55 &   24.76 &   22.43 & -115.61\\
20 &   17.03 &    5.77 &  -20.65 &  -57.69\\
21 &   51.68 &  -88.90 &  -84.48 &  -41.15\\
22 &  -24.39 &   18.02 &  -12.23 &   66.48\\
\bottomrule
\end{tabular}

\end{table*}

The projected marginalized uncertainty of component \(k_i^{(d)}\) is
\(\delta k_i^{(d)}=[(\bm\Sigma_d)_{ii}]^{1/2}\).  The values obtained with
Eqs.~\eqref{eq:benchmark-amplitude-noise} and
\eqref{eq:benchmark-amplitude-covariance} are listed in
Table~\ref{tab:vertical-angle-sensitivity}.  These entries are uncertainties
and not coefficient measurements.  Under the assumed null outcome, they can
equivalently be written as
\(\widehat k_i^{(d)}=0\pm\delta k_i^{(d)}\) at \(1\sigma\), or
\(\widehat k_i^{(d)}=0\pm2\delta k_i^{(d)}\) at \(2\sigma\).  The latter is a
95.45\% Gaussian interval for one coefficient and not a simultaneous 95.45\%
region for the full 14- or 22-dimensional vector.

\begin{table*}[t]
\caption{Projected marginalized coefficient uncertainties for the centered
vertical-stripe baselines under Eq.~\eqref{eq:benchmark-amplitude-noise}.
Each sector compares its own \(s_0\)-oriented pair with the Fisher pair.  The
\(d=6\) entries are in \(10^{-9}\,{\rm m}^{2}\), and the \(d=8\) entries are
in \(10^{-14}\,{\rm m}^{4}\).  A grouped \({\rm Re/Im}\) entry applies
separately to the two independent real coefficient components.  Their equal
uncertainties do not identify the components or their values.}
\label{tab:vertical-angle-sensitivity}

\centering
\small

\begin{tabular}{lrrrr}
\toprule
\multicolumn{5}{c}{\(d=6\), five vertical stripes}\\
Coefficient & \(\delta k_S\) & \(2\delta k_S\) &
\(\delta k_D\) & \(2\delta k_D\)\\
\midrule
\(k_{2,0}^{(6)}\)                   & 15.2 & 30.4 & 13.7 & 27.4\\
\(k_{4,0}^{(6)}\)                   & 22.0 & 44.1 & 27.0 & 54.0\\
\({\rm Re/Im}\,k_{2,2}^{(6)}\)      &  1.2 &  2.4 &  1.1 &  2.1\\
\({\rm Re/Im}\,k_{4,2}^{(6)}\)      &  1.2 &  2.4 &  1.4 &  2.8\\
\({\rm Re/Im}\,k_{4,4}^{(6)}\)      &  1.9 &  3.7 &  1.6 &  3.3\\
\({\rm Re/Im}\,k_{2,1}^{(6)}\)      &  1.3 &  2.5 &  1.3 &  2.6\\
\({\rm Re/Im}\,k_{4,1}^{(6)}\)      &  2.5 &  4.9 &  1.6 &  3.2\\
\({\rm Re/Im}\,k_{4,3}^{(6)}\)      &  1.5 &  3.0 &  1.4 &  2.8\\
\midrule
\multicolumn{5}{c}{\(d=8\), three vertical stripes}\\
Coefficient & \(\delta k_S\) & \(2\delta k_S\) &
\(\delta k_D\) & \(2\delta k_D\)\\
\midrule
\(k_{4,0}^{(8)}\)                   &  2.4 &  4.9 &  3.3 &  6.7\\
\(k_{6,0}^{(8)}\)                   &  6.5 & 13.0 & 10.5 & 21.0\\
\({\rm Re/Im}\,k_{4,2}^{(8)}\)      &  0.2 &  0.4 &  0.2 &  0.4\\
\({\rm Re/Im}\,k_{6,2}^{(8)}\)      &  0.5 &  1.0 &  0.5 &  1.0\\
\({\rm Re/Im}\,k_{4,4}^{(8)}\)      &  0.6 &  1.2 &  0.2 &  0.5\\
\({\rm Re/Im}\,k_{6,4}^{(8)}\)      &  0.8 &  1.5 &  0.4 &  0.7\\
\({\rm Re/Im}\,k_{6,6}^{(8)}\)      &  0.6 &  1.3 &  0.6 &  1.3\\
\({\rm Re/Im}\,k_{4,1}^{(8)}\)      &  0.4 &  0.9 &  0.2 &  0.4\\
\({\rm Re/Im}\,k_{6,1}^{(8)}\)      &  0.8 &  1.6 &  0.5 &  1.0\\
\({\rm Re/Im}\,k_{4,3}^{(8)}\)      &  0.3 &  0.5 &  0.3 &  0.6\\
\({\rm Re/Im}\,k_{6,3}^{(8)}\)      &  0.5 &  1.0 &  0.4 &  0.8\\
\({\rm Re/Im}\,k_{6,5}^{(8)}\)      &  0.5 &  0.9 &  0.4 &  0.9\\
\bottomrule
\end{tabular}

\end{table*}

For \(d=6\), the \(s_0\)-oriented pair gives a marginalized
\(1\sigma\) range of
\((1.19\times10^{-9}\text{--}2.20\times10^{-8})\,{\rm m}^{2}\), whereas the
Fisher pair gives
\((1.07\times10^{-9}\text{--}2.70\times10^{-8})\,{\rm m}^{2}\).  The Fisher
pair reduces the uncertainties of \(k_{2,0}^{(6)}\),
\(k_{2,2}^{(6)}\), \(k_{4,4}^{(6)}\), \(k_{4,1}^{(6)}\), and
\(k_{4,3}^{(6)}\), with the largest reduction, 34\%, occurring for
\(k_{4,1}^{(6)}\).  The \(s_0\)-oriented pair instead gives smaller
uncertainties for \(k_{4,0}^{(6)}\), \(k_{4,2}^{(6)}\), and
\(k_{2,1}^{(6)}\).  Neither pair therefore dominates component by component.
This is the expected distinction between strengthening a selected response
block and minimizing a joint covariance volume.

For \(d=8\), the \(s_0\)-oriented and Fisher pairs give marginalized
\(1\sigma\) ranges of
\((1.94\times10^{-15}\text{--}6.51\times10^{-14})\,{\rm m}^{4}\) and
\((1.89\times10^{-15}\text{--}1.05\times10^{-13})\,{\rm m}^{4}\),
respectively.  The \(s_0\)-oriented pair reduces the uncertainties of
\(k_{4,0}^{(8)}\) and \(k_{6,0}^{(8)}\) by approximately 27\% and 38\%
relative to the Fisher pair.  It also gives slightly smaller uncertainties for
\(k_{6,6}^{(8)}\) and \(k_{4,3}^{(8)}\), while the Fisher pair is favored by
the remaining harmonic components.  The two angle strategies therefore show
the same component-dependent tradeoff as in the \(d=6\) sector.

\subsection{Centerline-dependent coefficient sensitivity}
\label{subsec:centerline-sensitivity}

The response maxima in Secs.~\ref{subsec:d6results} and
\ref{subsec:d8results} do not by themselves determine coefficient
sensitivity.  We therefore repeat the noise-weighted optimization for the
left, centered, and right registrations, using five stripes for \(d=6\) and
three stripes for \(d=8\).  Every row in Table~\ref{tab:midline-fisher}
uses its own Fisher-optimal pair rather than the angles optimized for the
centered geometry.  Under the equal quadrature errors in
Eq.~\eqref{eq:benchmark-amplitude-noise}, a common shift of both azimuths
rotates the sine--cosine basis without changing the information determinant.
The table therefore reports the optimized separation
\(\Delta\alpha_D=|\phi_D-\theta_D|\) for one representative pair.

\begin{figure*}[t]
\includegraphics[width=0.94\textwidth]{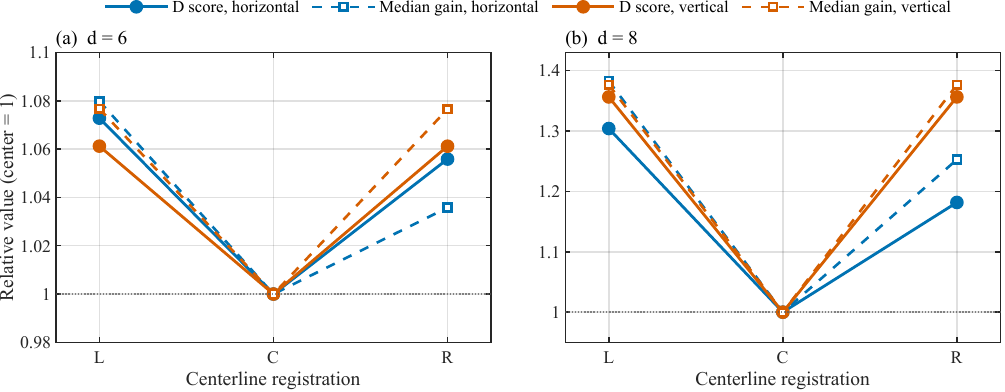}
\caption{Dependence of the covariance-weighted \(D\)-optimal score (solid
circles) and the coefficientwise median sensitivity gain (dashed open squares)
on centerline registration.  Each series is normalized to its centered
geometry and each point is reoptimized over the two apparatus azimuths.
Horizontal and vertical stripe axes are shown in blue and orange,
respectively.  Panel (a) uses five stripes at \(d=6\), and panel (b) uses three
stripes at \(d=8\).}
\label{fig:centerline-sensitivity-summary}
\end{figure*}

Figure~\ref{fig:centerline-sensitivity-summary} summarizes the determinant
score and the coefficientwise median gain, which answer different design
questions and need not change by the same amount.

\begin{table*}[t]
\caption{Noise-weighted sensitivity comparison for left (L), centered (C),
and right (R) registrations.  The score \({\cal D}_d\) is defined by
Eq.~\eqref{eq:doptimal} with the centered geometry as the reference separately
for each dimension and stripe orientation.  The coefficientwise median gain is
\(\widetilde g=\operatorname{median}_i[\delta k_{i,C}/\delta k_i]\), so values
above unity indicate an improvement over the centered design.  The final
column gives the range of marginalized \(1\sigma\) uncertainties in
\(10^{-9}\,{\rm m}^{2}\) for \(d=6\) and \(10^{-14}\,{\rm m}^{4}\) for
\(d=8\).}
\label{tab:midline-fisher}

\centering
\small

\begin{tabular}{cccccc}
\toprule
\(d\) & Stripe axis & Registration &
\(\Delta\alpha_D\) (rad) & \({\cal D}_d\) &
\(\widetilde g\quad[\min\delta k,\max\delta k]\)\\
\midrule
6 & horizontal & L & 1.541 & 1.073 & \(1.080\quad[0.98,23.83]\)\\
6 & horizontal & C & 1.557 & 1.000 & \(1.000\quad[1.06,26.23]\)\\
6 & horizontal & R & 1.349 & 1.056 & \(1.036\quad[0.97,26.70]\)\\
6 & vertical   & L & 1.770 & 1.061 & \(1.077\quad[0.99,24.07]\)\\
6 & vertical   & C & 1.835 & 1.000 & \(1.000\quad[1.07,27.00]\)\\
6 & vertical   & R & 1.770 & 1.061 & \(1.077\quad[0.99,24.07]\)\\
\midrule
8 & horizontal & L & 1.672 & 1.304 & \(1.383\quad[0.13,18.25]\)\\
8 & horizontal & C & 1.874 & 1.000 & \(1.000\quad[0.19,13.14]\)\\
8 & horizontal & R & 1.667 & 1.182 & \(1.253\quad[0.14,22.12]\)\\
8 & vertical   & L & 1.247 & 1.356 & \(1.377\quad[0.12,8.37]\)\\
8 & vertical   & C & 2.060 & 1.000 & \(1.000\quad[0.19,10.48]\)\\
8 & vertical   & R & 1.247 & 1.356 & \(1.377\quad[0.12,8.37]\)\\
\bottomrule
\end{tabular}

\end{table*}

The covariance-weighted comparison confirms the centerline trend while
quantifying it more conservatively than the unweighted response maps.  At
\(d=6\), the off-center registrations increase \({\cal D}_6\) by
\(5.6\%\)--\(7.3\%\), with coefficientwise median sensitivity gains between
\(3.6\%\) and \(8.0\%\).  At \(d=8\), the left horizontal registration raises
\({\cal D}_8\) by \(30.4\%\), and either off-center vertical registration
raises it by \(35.6\%\) with a median coefficient sensitivity gain of
\(37.7\%\).  The two vertical results are equal within the integration
precision, as required by reflection symmetry.

The equality of the left and right vertical results permits one of them to be
used as a representative off-center design.  For the left-offset, three-stripe
vertical geometry at \(d=8\), a representative Fisher pair is
\begin{equation}
 (\theta_{D,8}^{{\rm V},L},\phi_{D,8}^{{\rm V},L})
 =(3.9434473,\,5.1900041)\ \mathrm{rad}.
 \label{eq:d8-vertical-left-angle-pair}
\end{equation}
The common-azimuth degeneracy noted above leaves the Fisher determinant
unchanged, so Eq.~\eqref{eq:d8-vertical-left-angle-pair} fixes the phase
convention for the transfer coefficients in
Table~\ref{tab:d8-vertical-left-gamma}.  The corresponding coefficient
uncertainties are given in Table~\ref{tab:d8-vertical-left-sensitivity}.

\begin{table*}[t]
\caption{Transfer coefficients for the representative left-offset,
three-stripe vertical \(d=8\) geometry at the Fisher pair in
Eq.~\eqref{eq:d8-vertical-left-angle-pair}.  Entries are in units of
\(10^{-4}\,{\rm N\,m/m^{4}}\).}
\label{tab:d8-vertical-left-gamma}

\centering
\small

\begin{tabular}{crr@{\hspace{2.8em}}crr}
\toprule
\(i\) & \(\Gamma_i^{(8)}(\theta_{D,8}^{{\rm V},L})\) &
\(\Gamma_i^{(8)}(\phi_{D,8}^{{\rm V},L})\) &
\(i\) & \(\Gamma_i^{(8)}(\theta_{D,8}^{{\rm V},L})\) &
\(\Gamma_i^{(8)}(\phi_{D,8}^{{\rm V},L})\)\\
\midrule
 1 &    6.17 & -116.94 & 12 &   42.40 &   64.62\\
 2 &  -45.67 &  -41.71 & 13 &   26.39 & -124.29\\
 3 &   -2.05 & -332.28 & 14 &  317.17 &   95.03\\
 4 &  272.25 &   87.53 & 15 &  -49.46 &   36.82\\
 5 & -117.85 &  -34.90 & 16 &   56.41 &  119.14\\
 6 &  -36.65 &   75.40 & 17 &   31.82 & -262.27\\
 7 &   35.69 & -159.70 & 18 &  254.16 &  111.75\\
 8 &  193.19 &   54.95 & 19 & -143.68 &    1.27\\
 9 & -121.63 &  -27.14 & 20 &   -4.67 &  129.18\\
10 &  -15.06 &  121.27 & 21 &  -96.76 &  -51.26\\
11 &  -55.56 &  -62.88 & 22 &   60.23 &   98.85\\
\bottomrule
\end{tabular}

\end{table*}

\begin{table*}[t]
\caption{Projected marginalized coefficient uncertainties for the
left-offset, three-stripe vertical \(d=8\) geometry under
Eq.~\eqref{eq:benchmark-amplitude-noise}.  Entries are in
\(10^{-14}\,{\rm m}^{4}\).  A grouped \({\rm Re/Im}\) entry applies
separately to two independent real components.}
\label{tab:d8-vertical-left-sensitivity}

\centering
\small
\renewcommand{\arraystretch}{1.12}

\begin{tabular}{@{}lrr@{\hspace{4em}}lrr@{}}
\toprule
Coefficient & \(\delta k_D\) & \(2\delta k_D\) &
Coefficient & \(\delta k_D\) & \(2\delta k_D\)\\
\midrule
\(k_{4,0}^{(8)}\)              & 4.4 &  8.8 & \({\rm Re/Im}\,k_{6,6}^{(8)}\) & 0.4 & 0.9\\
\(k_{6,0}^{(8)}\)              & 8.4 & 16.7 & \({\rm Re/Im}\,k_{4,1}^{(8)}\) & 0.2 & 0.3\\
\({\rm Re/Im}\,k_{4,2}^{(8)}\) & 0.1 &  0.2 & \({\rm Re/Im}\,k_{6,1}^{(8)}\) & 0.4 & 0.8\\
\({\rm Re/Im}\,k_{6,2}^{(8)}\) & 0.3 &  0.7 & \({\rm Re/Im}\,k_{4,3}^{(8)}\) & 0.1 & 0.3\\
\({\rm Re/Im}\,k_{4,4}^{(8)}\) & 0.2 &  0.4 & \({\rm Re/Im}\,k_{6,3}^{(8)}\) & 0.3 & 0.5\\
\({\rm Re/Im}\,k_{6,4}^{(8)}\) & 0.3 &  0.6 & \({\rm Re/Im}\,k_{6,5}^{(8)}\) & 0.3 & 0.6\\
\bottomrule
\end{tabular}

\end{table*}

For this off-center geometry, the marginalized \(1\sigma\) uncertainties span
\(1.16\times10^{-15}\) to \(8.37\times10^{-14}\,{\rm m}^{4}\), with a
median of \(2.89\times10^{-15}\,{\rm m}^{4}\).  Relative to the centered
vertical geometry, \({\cal D}_8\) increases by 35.6\% and the coefficientwise
median sensitivity improves by 37.7\%.  The improvement is not uniform.  For
example, the uncertainty of \(k_{4,0}^{(8)}\) is larger than in the centered
design, whereas most sidereal components are better constrained.  This
component-dependent behavior is why the registration comparison is based on
the complete covariance rather than a single transfer coefficient.

The Fisher determinant measures the joint covariance volume and does not imply
componentwise dominance.  In the \(d=8\) horizontal family, for example, the
off-center layouts improve the determinant and the median coefficient
sensitivity even though the largest individual uncertainty can increase.  The
factor of 2.11 in the unweighted vertical \({\cal R}_8\) response likewise
becomes a factor of 1.356 in \({\cal D}_8\) after the unequal channel errors
and coefficient correlations are included.  Centerline registration is thus
a genuine sensitivity variable, but its experimental value must be assessed
with the full covariance rather than inferred from a single response maximum.

In both sectors, the largest uncertainties occur for coefficients entering
\(C_0\), whose assumed error is eight times that of a sidereal quadrature.  The
equality of the real- and imaginary-part uncertainties follows from the common,
uncorrelated error assigned to each \(C_m,S_m\) pair and the paired response
structure.  It does not imply equal coefficient values.  Unequal sine and
cosine errors or correlated systematics will generally split the two
uncertainties.

The constant and harmonic uncertainties must be rescaled separately when the
observing time or systematic control changes.  If the harmonic fit remains
thermal-noise limited, its uncertainty approximately follows
\begin{equation}
 \sigma_h(T)\simeq\sigma_{h,{\rm ref}}
 \sqrt{\frac{T_{\rm ref}}{T}},
 \qquad T_{\rm ref}\simeq10\ {\rm days},
 \label{eq:sensitivity-rescaling}
\end{equation}
provided that the noise is stationary over the fitted sidereal cycles.  The
constant channel instead contains a systematic floor and is more appropriately
written as
\begin{equation}
 \begin{aligned}
 \sigma_0^2(T)&=\sigma_{0,{\rm sys}}^2+
 \sigma_{0,{\rm stat}}^2(T),\\
 \sigma_{0,{\rm stat}}(T)&\propto T^{-1/2},
 \end{aligned}
 \label{eq:constant-channel-rescaling}
\end{equation}
with \(\sigma_{0,{\rm sys}}\) set mainly by metrology and alignment.  The
dominant part of \(\sigma_0\) does not average down with the number of sidereal
cycles.  Whenever either uncertainty changes,
Eq.~\eqref{eq:projected-coefficient-covariance} must be reevaluated with the
updated amplitude covariance rather than rescaling every coefficient by one
common factor.

The fixed-angle transfer coefficients in
Tables~\ref{tab:horizontal-angle-gamma} and
\ref{tab:vertical-angle-gamma} were checked against the directly evaluated
angular responses.  The horizontal weighted forecasts were also reconstructed
from the exported transfer coefficients, and the blockwise and full Fisher
 formulations give the same covariance to numerical precision.

An analysis of data acquired for an earlier gravitational inverse-square-law
test constrained individual \(d=6\) combinations at approximately the
\(10^{-8}\,{\rm m}^{2}\) level \cite{Shao2015ISL}.  Subsequent combined
short-range analyses reached approximately \(10^{-9}\,{\rm m}^{2}\) for
\(d=6\) and \(10^{-12}\,{\rm m}^{4}\) for \(d=8\)
\cite{Shao2016Combined,Shao2019InverseSixth}.  For the most favorable
coefficient components, the present projections improve on the constraints
extracted from the earlier inverse-square-law data by about two orders of
magnitude.  The gain is coefficient dependent because the transfer strengths
and the constant-channel contribution differ among components.
Tables~\ref{tab:horizontal-angle-sensitivity} and
\ref{tab:vertical-angle-sensitivity} should therefore be interpreted as
transparent baseline forecasts and not as reported experimental bounds.  A
final projection should replace
Eqs.~\eqref{eq:benchmark-amplitude-covariance} and
\eqref{eq:two-azimuth-amplitude-covariance} with the measured harmonic
covariance and use the profiled Fisher matrix in
Eq.~\eqref{eq:profiled-fisher} for the selected geometry, including machining
tolerances and Newtonian-background nuisance parameters.

\section{Conclusion}
\label{sec:conclusion}

We have extended striped torsion-balance optimization to include centerline
registration and dimension-dependent geometric responses. Finite-volume
transfer matrices and coefficient covariances allow the response strength and
angular independence to be assessed together, with distinct uncertainties for
the constant and sidereal channels. The \(d=8\) comparison selects three
stripes rather than five or seven. Moving the pattern off center increases the
noise-weighted \(D\)-optimal score by up to \(35.6\%\) at \(d=8\) and
\(7.3\%\) at \(d=6\), after the two operating azimuths are reoptimized.

The two-azimuth response matrices permit simultaneous estimation of all 14
\(d=6\) or 22 \(d=8\) coefficient components. Under the stated amplitude
uncertainties, the best marginalized \(1\sigma\) sensitivities reach
\(10^{-9}\,\mathrm{m}^{2}\) and \(10^{-15}\,\mathrm{m}^{4}\), respectively.
The registration and azimuths that minimize the joint covariance need not
favor every component, especially those entering the less precise constant
channel. This distinction provides a practical basis for selecting the mass
pattern and operating angles according to the targeted Lorentz-violating
coefficients.

\backmatter

\bmhead{Acknowledgments}

This work was supported by the National Natural Science Foundation of China (Grant No. 12405060).

\bmhead{Statements and Declarations}
\mbox{}\par

\noindent\textbf{Competing interests.} The authors declare no competing interests.

\noindent\textbf{Data availability.}

No new experimental data were acquired in this study.  The numerical data and
Mathematica notebooks supporting the figures and sensitivity forecasts are not
publicly available because they form part of an ongoing experimental-design
study.  They are available from the corresponding authors upon reasonable
request.

\appendix
\setcounter{equation}{0}
\renewcommand{\theequation}{\thesection\arabic{equation}}

\section{Cartesian form of the \texorpdfstring{\(j=6\)}{j=6} spherical harmonics}
\label{app:j6-harmonics}

For completeness, this appendix gives the Cartesian expressions used to
obtain the \(j=6\) terms in Eq.~\eqref{eq:d8-gamma-basis}. With
\(r^2=x^2+y^2+z^2\) and the same Condon--Shortley convention as in
Sec.~\ref{subsec:finite-response}, define the following \(j=6\) harmonic
polynomials.
\begin{align}
 P_0(x,y,z)={}&5x^6+5y^6-90y^4z^2
 \nonumber\\
 &+120y^2z^4-16z^6
 \nonumber\\
 &+15x^4(y^2-6z^2)
 \nonumber\\
 &+15x^2(y^4-12y^2z^2+8z^4),
 \label{eq:j6-P0}\\
 P_1(x,y,z)={}&5x^4+5y^4-20y^2z^2
 \nonumber\\
 &+8z^4+10x^2(y^2-2z^2),
 \label{eq:j6-P1}\\
 P_2(x,y,z)={}&x^4+y^4-16y^2z^2
 \nonumber\\
 &+16z^4+2x^2(y^2-8z^2).
 \label{eq:j6-P2}
\end{align}
The negative-\(m\) components are
\begin{align}
Y_{6,-6}&=\frac{1}{64}\sqrt{\frac{3003}{\pi}}
\frac{(x-iy)^6}{r^6},
\nonumber\\
Y_{6,-5}&=\frac{3}{32}\sqrt{\frac{1001}{\pi}}
\frac{(x-iy)^5z}{r^6}.
\label{eq:j6-negative-harmonics}
\end{align}
\begin{align}
Y_{6,-4}&=-\frac{3}{32}\sqrt{\frac{91}{2\pi}}
\frac{(x-iy)^4(x^2+y^2-10z^2)}{r^6},
\nonumber\\
Y_{6,-3}&=-\frac{1}{32}\sqrt{\frac{1365}{\pi}}
\frac{(x-iy)^3z(3x^2+3y^2-8z^2)}{r^6}.
\end{align}
\begin{align}
Y_{6,-2}&=\frac{1}{64}\sqrt{\frac{1365}{\pi}}
\frac{(x-iy)^2P_2(x,y,z)}{r^6},
\nonumber\\
Y_{6,-1}&=\frac{1}{16}\sqrt{\frac{273}{2\pi}}
\frac{(x-iy)zP_1(x,y,z)}{r^6}.
\end{align}
 The zonal harmonic can be written in either polynomial or compact radial form.
\begin{equation}
\begin{split}
Y_{6,0}={}&-\frac{1}{32}\sqrt{\frac{13}{\pi}}
\frac{P_0(x,y,z)}{r^6}\\
={}&-\frac{1}{32}\sqrt{\frac{13}{\pi}}
\left(5-105\frac{z^2}{r^2}
+315\frac{z^4}{r^4}\right.\\
&\left.\quad-231\frac{z^6}{r^6}\right).
\end{split}
\label{eq:j6-zonal-harmonic}
\end{equation}
The positive-\(m\) components are
\begin{align}
Y_{6,1}&=-\frac{1}{16}\sqrt{\frac{273}{2\pi}}
\frac{(x+iy)zP_1(x,y,z)}{r^6},
\nonumber\\
Y_{6,2}&=\frac{1}{64}\sqrt{\frac{1365}{\pi}}
\frac{(x+iy)^2P_2(x,y,z)}{r^6}.
\end{align}
\begin{align}
Y_{6,3}&=\frac{1}{32}\sqrt{\frac{1365}{\pi}}
\frac{(x+iy)^3z(3x^2+3y^2-8z^2)}{r^6},
\nonumber\\
Y_{6,4}&=-\frac{3}{32}\sqrt{\frac{91}{2\pi}}
\frac{(x+iy)^4(x^2+y^2-10z^2)}{r^6}.
\end{align}
\begin{align}
Y_{6,5}&=-\frac{3}{32}\sqrt{\frac{1001}{\pi}}
\frac{(x+iy)^5z}{r^6},
\nonumber\\
Y_{6,6}&=\frac{1}{64}\sqrt{\frac{3003}{\pi}}
\frac{(x+iy)^6}{r^6}.
\label{eq:j6-positive-harmonics}
\end{align}
These expressions satisfy
\(Y_{6,-m}=(-1)^mY_{6,m}^{*}\). Substitution of
Eqs.~\eqref{eq:j6-negative-harmonics}--\eqref{eq:j6-positive-harmonics} into
the combinations in Eq.~\eqref{eq:d8-gamma-basis} yields the Cartesian
\(\gamma_i^{(8)}\) functions displayed in
Eq.~\eqref{eq:d8-gamma-functions}.

\section{Individual transfer coefficients}
\label{app:gamma-curves}

\begin{figure*}[p]
 \centering
 \textbf{(a) Vertical stripes}\par
 \includegraphics[width=0.90\textwidth]{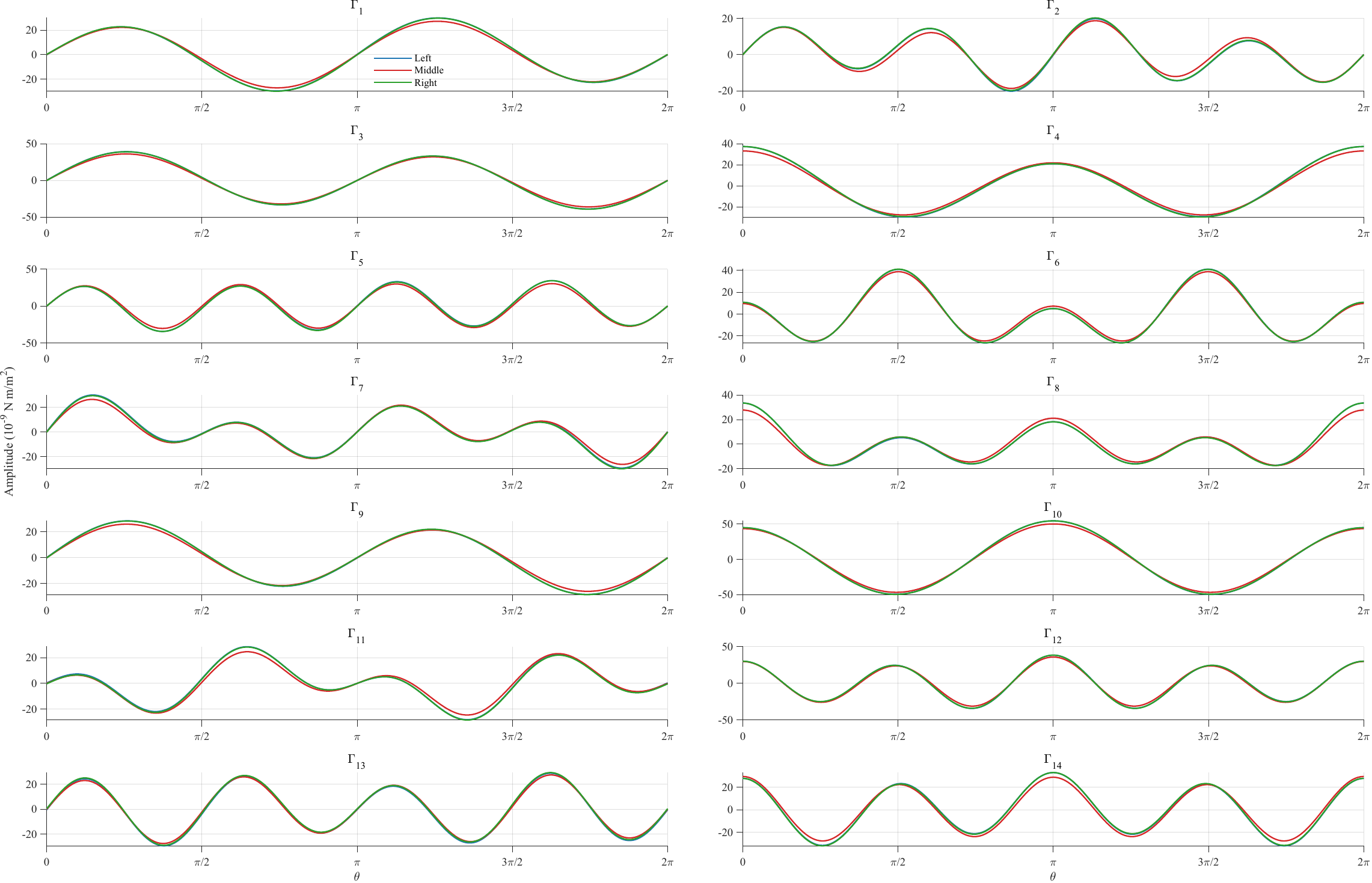}
 \vspace{0.2em}

 \textbf{(b) Horizontal stripes}\par
 \includegraphics[width=0.90\textwidth]{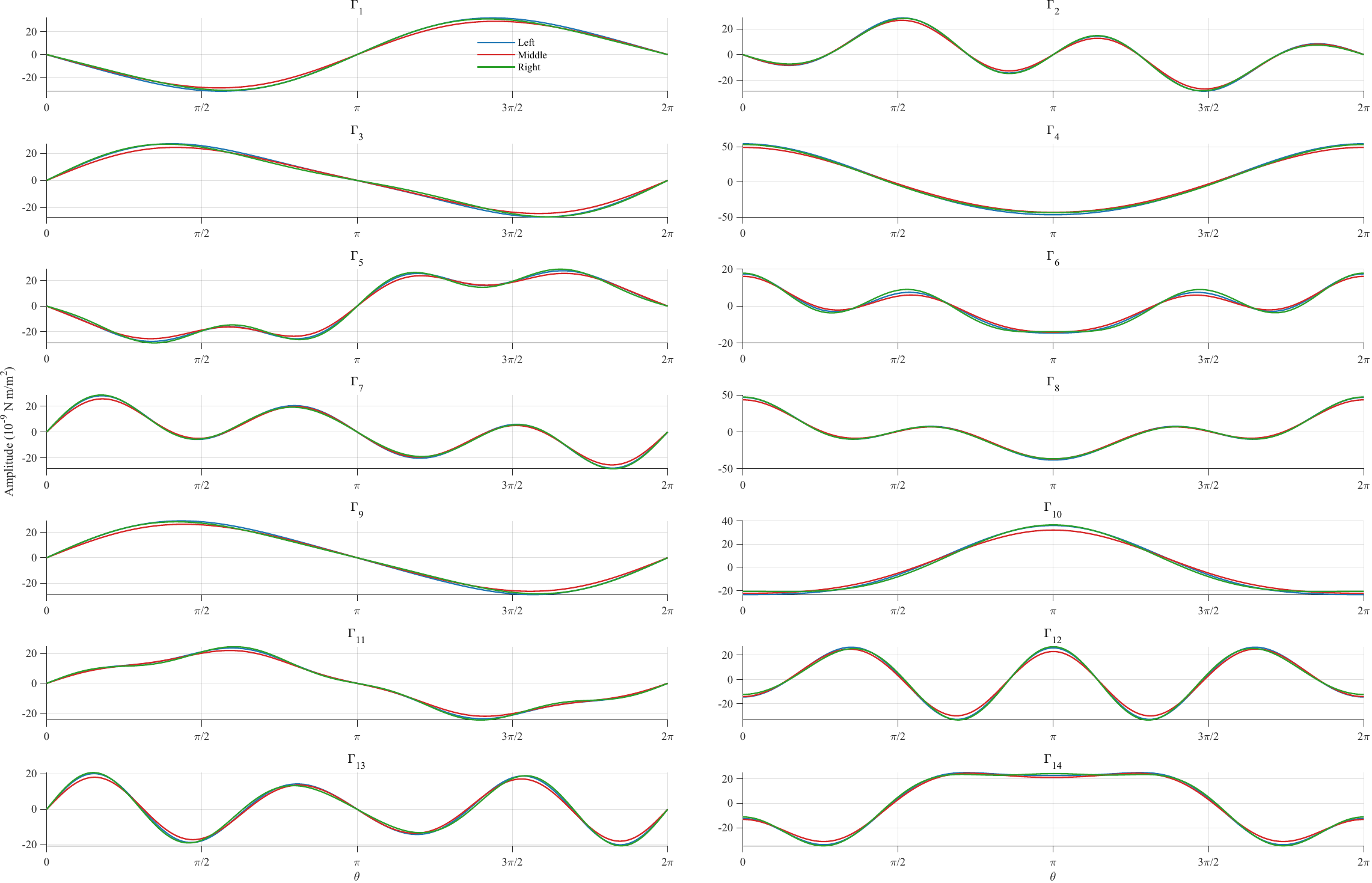}
 \caption{Individual \(d=6\) transfer coefficients at \(N_s=5\), comparing
 left, middle, and right centerline registrations.  Panels (a) and (b) show
 the vertical- and horizontal-stripe families, respectively.}
 \label{fig:app-d6-gamma-midline}
\end{figure*}

\begin{figure*}[p]
 \centering
 \textbf{(a) Vertical stripes}\par
 \includegraphics[width=0.82\textwidth]{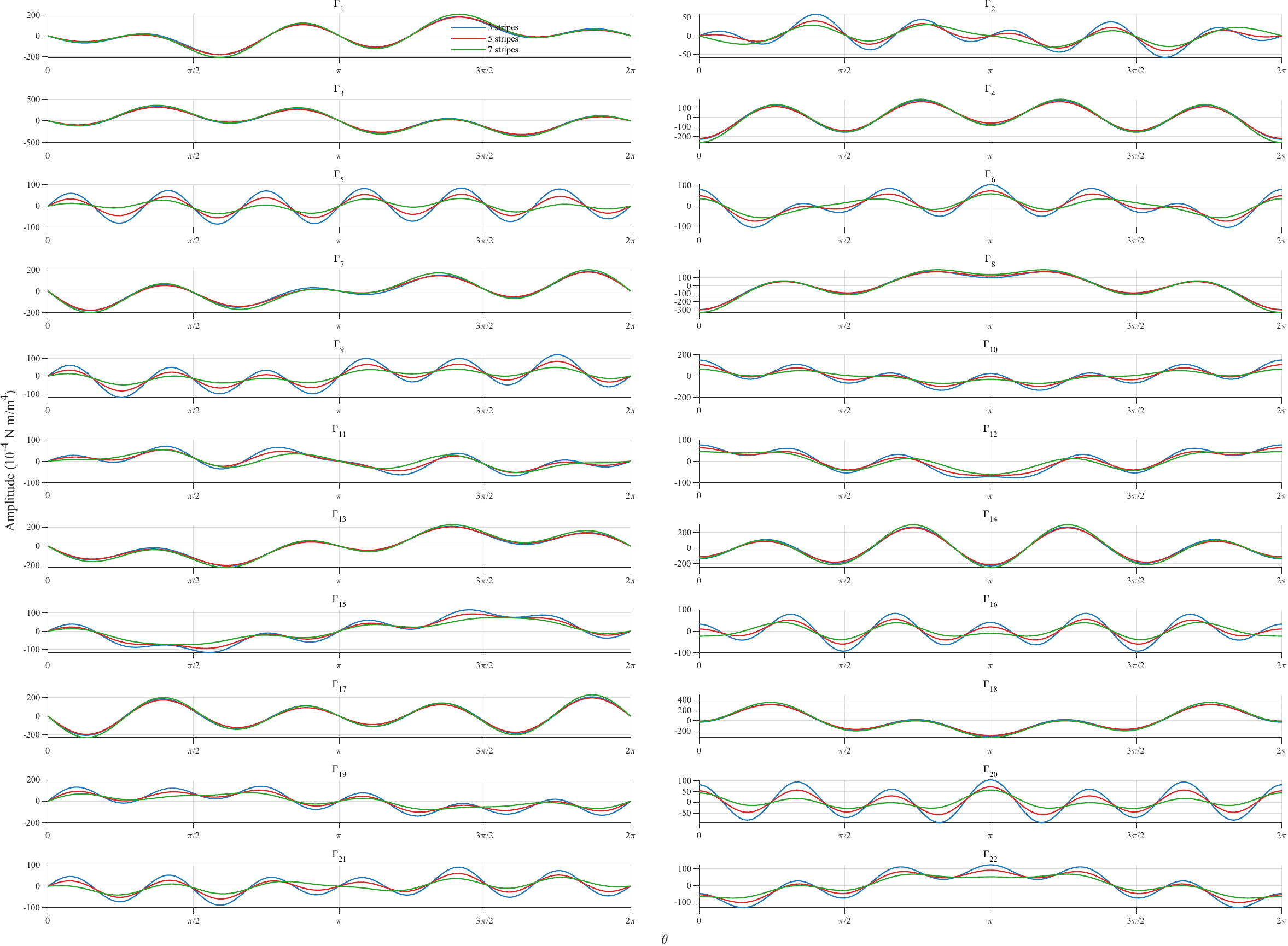}
 \vspace{0.2em}

 \textbf{(b) Horizontal stripes}\par
 \includegraphics[width=0.82\textwidth]{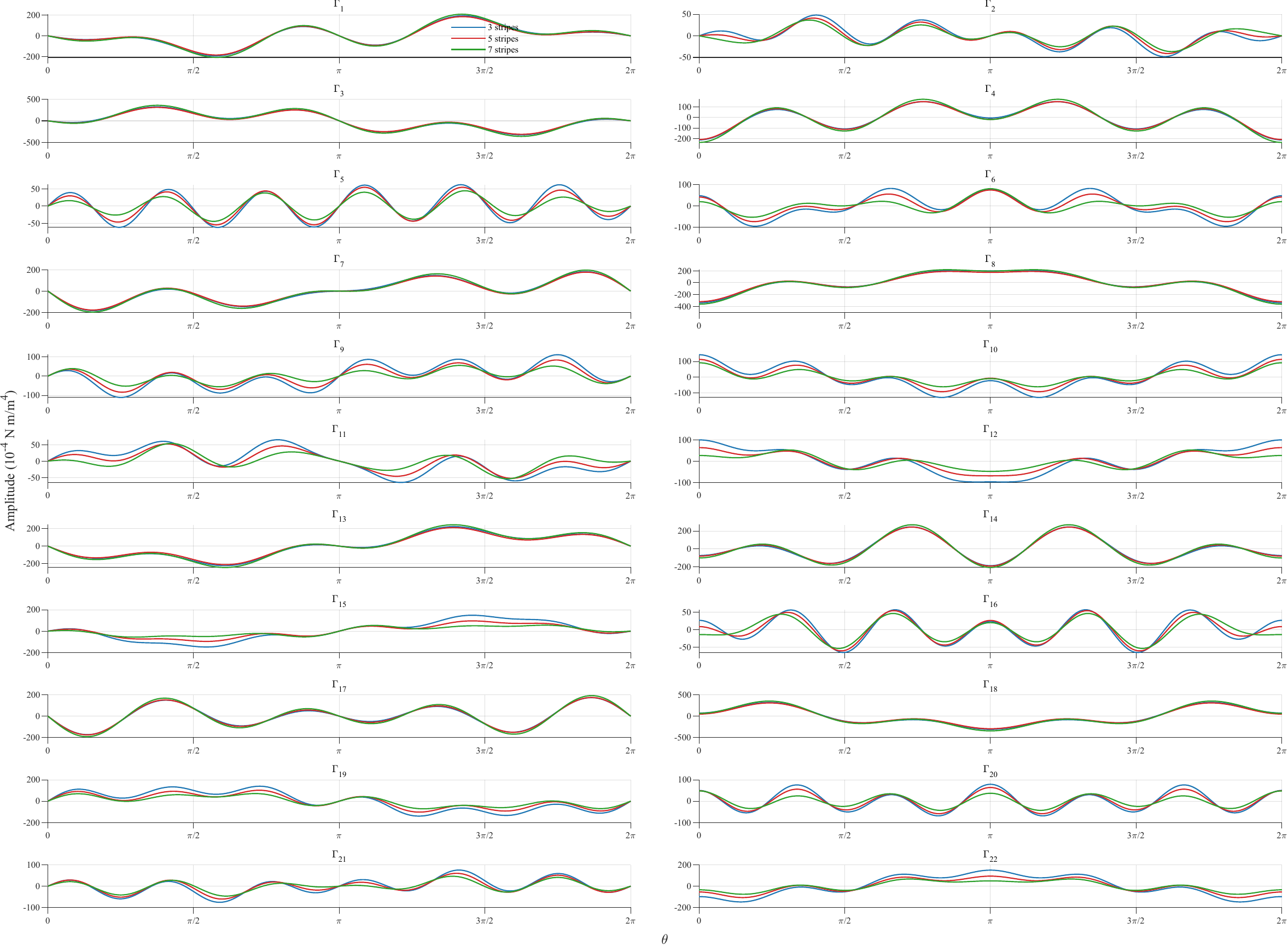}
 \caption{Individual \(d=8\) transfer coefficients for \(N_s=3,5,7\),
 using the common normalization of Eq.~\eqref{eq:d8-normalized-gamma}.
 Panels (a) and (b) show the vertical- and horizontal-stripe families,
 respectively.}
 \label{fig:app-d8-gamma-stripes}
\end{figure*}

\begin{figure*}[p]
 \centering
 \textbf{(a) Vertical stripes}\par
 \includegraphics[width=0.82\textwidth]{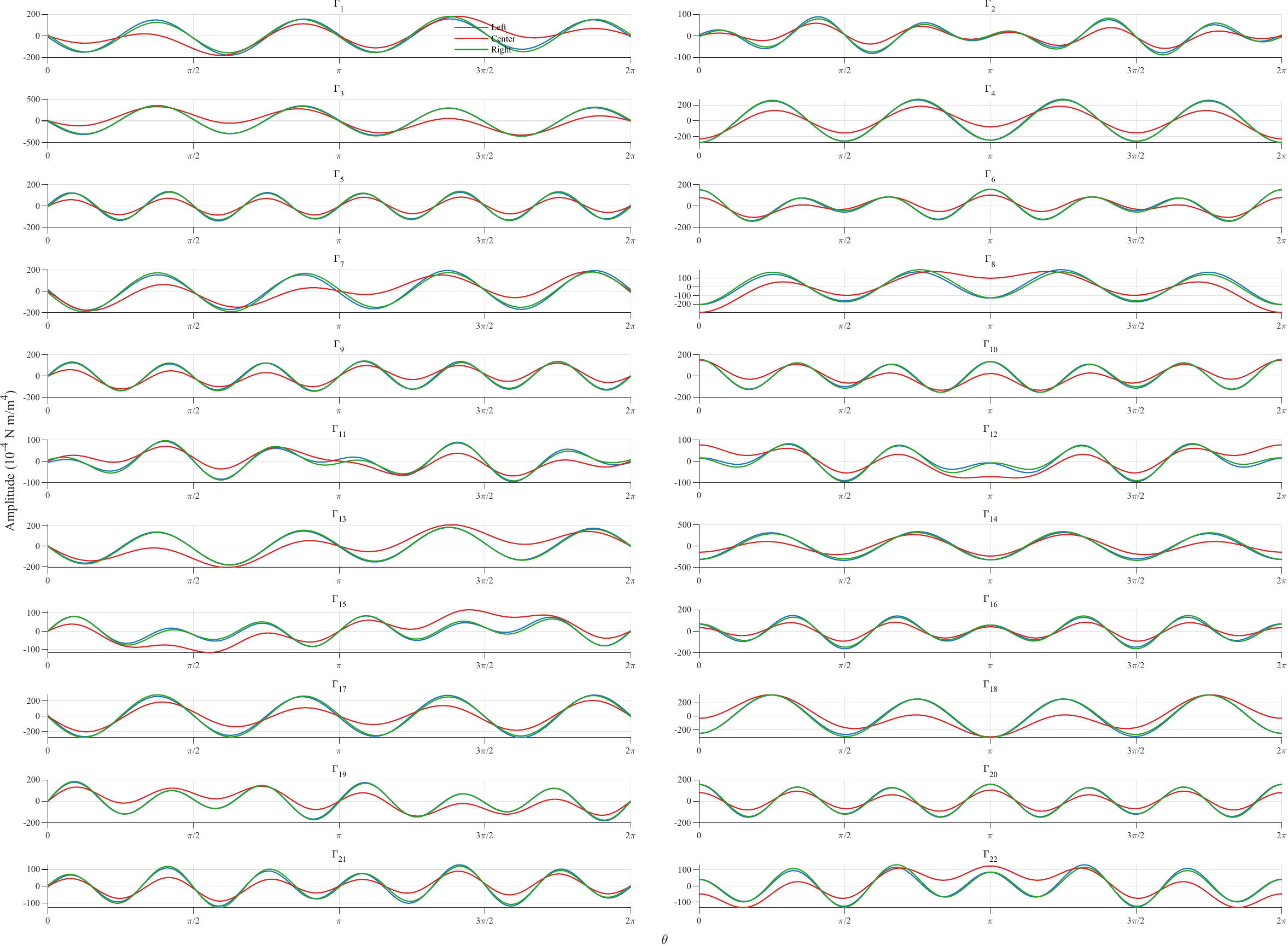}
 \vspace{0.2em}

 \textbf{(b) Horizontal stripes}\par
 \includegraphics[width=0.82\textwidth]{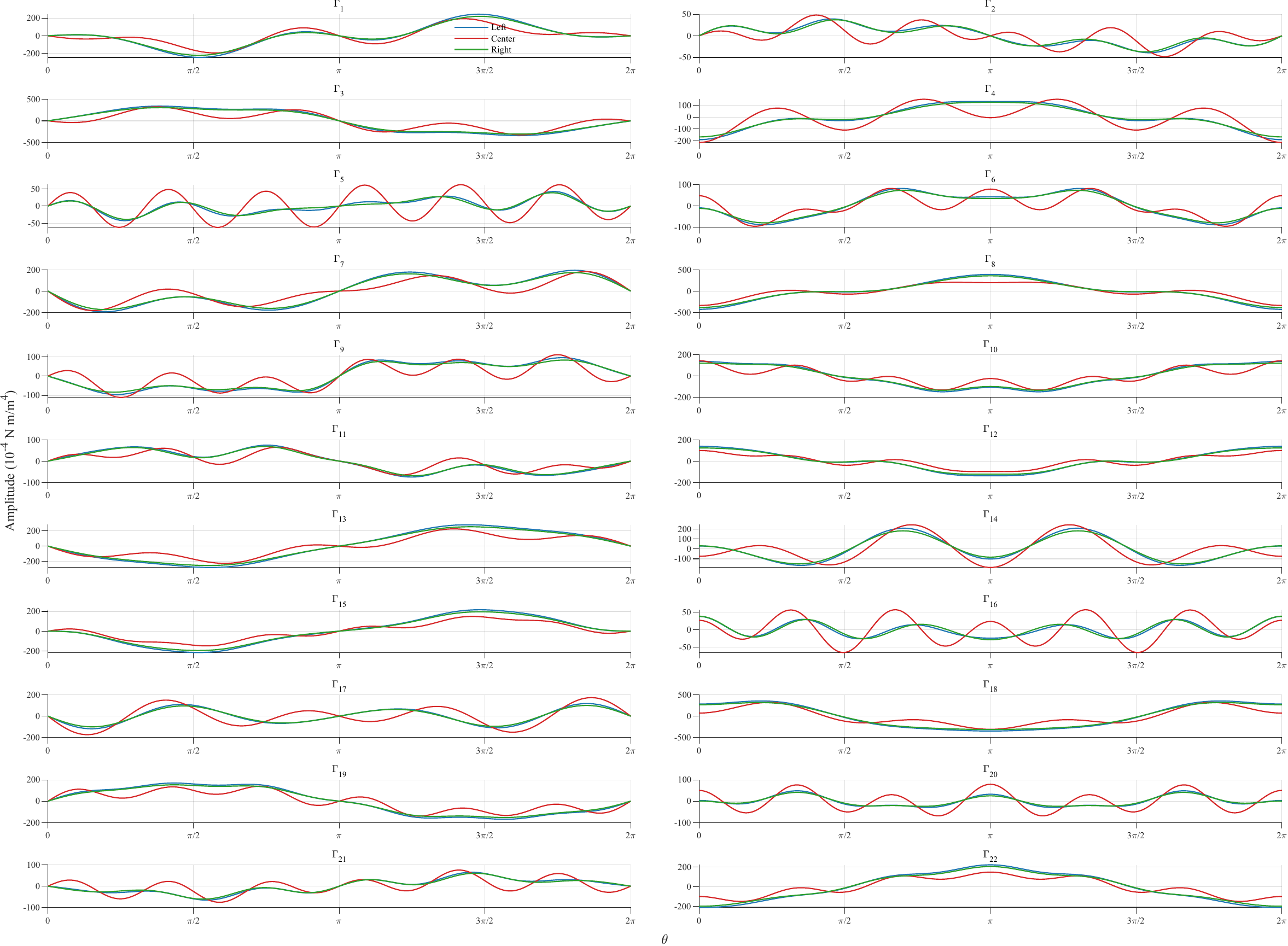}
 \caption{Individual \(d=8\) transfer coefficients at \(N_s=3\), comparing
 left, center, and right centerline registrations.  Panels (a) and (b) show
 the vertical- and horizontal-stripe families, respectively.}
 \label{fig:app-d8-gamma-midline}
\end{figure*}

The curves span \(0\leq\alpha\leq2\pi\), with a common vertical scale for
the geometries compared in each panel. Figure~\ref{fig:app-d6-gamma-midline}
shows the \(d=6\) centerline comparison. Figures~\ref{fig:app-d8-gamma-stripes} and
\ref{fig:app-d8-gamma-midline} give the corresponding \(d=8\) results in
units of \(\Gamma_{\rm ref}^{(8)}\) from
Eq.~\eqref{eq:d8-normalized-gamma}.

\clearpage

\clearpage

\clearpage
\bibliography{Lorentz_torsion_references}

@article{Will2014,
  author  = {Will, Clifford M.},
  title   = {The Confrontation between General Relativity and Experiment},
  journal = {Living Rev. Relativ.},
  volume  = {17},
  pages   = {4},
  year    = {2014},
  doi     = {10.12942/lrr-2014-4}
}

@article{KosteleckySamuel1989,
  author  = {Kosteleck{\'y}, V. Alan and Samuel, Stuart},
  title   = {Spontaneous Breaking of Lorentz Symmetry in String Theory},
  journal = {Phys. Rev. D},
  volume  = {39},
  pages   = {683--685},
  year    = {1989},
  doi     = {10.1103/PhysRevD.39.683}
}

@article{Kostelecky2004,
  author  = {Kosteleck{\'y}, V. Alan},
  title   = {Gravity, Lorentz Violation, and the Standard Model},
  journal = {Phys. Rev. D},
  volume  = {69},
  pages   = {105009},
  year    = {2004},
  doi     = {10.1103/PhysRevD.69.105009}
}

@article{KosteleckyRussell2011,
  author        = {Kosteleck{\'y}, V. Alan and Russell, Neil},
  title         = {Data Tables for Lorentz and {CPT} Violation},
  journal       = {Rev. Mod. Phys.},
  volume        = {83},
  pages         = {11--31},
  year          = {2011},
  doi           = {10.1103/RevModPhys.83.11},
  eprint        = {0801.0287},
  archiveprefix = {arXiv},
  note          = {2026 edition}
}

@article{Tasson2016,
  author  = {Tasson, Jay D.},
  title   = {The Standard-Model Extension and Gravitational Tests},
  journal = {Symmetry},
  volume  = {8},
  pages   = {111},
  year    = {2016},
  doi     = {10.3390/sym8110111}
}

@article{ColladayKostelecky1997,
  author  = {Colladay, Don and Kosteleck{\'y}, V. Alan},
  title   = {{CPT} Violation and the Standard Model},
  journal = {Phys. Rev. D},
  volume  = {55},
  pages   = {6760--6774},
  year    = {1997},
  doi     = {10.1103/PhysRevD.55.6760}
}

@article{ColladayKostelecky1998,
  author  = {Colladay, Don and Kosteleck{\'y}, V. Alan},
  title   = {Lorentz-Violating Extension of the Standard Model},
  journal = {Phys. Rev. D},
  volume  = {58},
  pages   = {116002},
  year    = {1998},
  doi     = {10.1103/PhysRevD.58.116002}
}

@article{BaileyKostelecky2006,
  author  = {Bailey, Quentin G. and Kosteleck{\'y}, V. Alan},
  title   = {Signals for Lorentz Violation in Post-Newtonian Gravity},
  journal = {Phys. Rev. D},
  volume  = {74},
  pages   = {045001},
  year    = {2006},
  doi     = {10.1103/PhysRevD.74.045001}
}

@article{KosteleckyMewes2017,
  author  = {Kosteleck{\'y}, V. Alan and Mewes, Matthew},
  title   = {Testing Local Lorentz Invariance with Short-Range Gravity},
  journal = {Phys. Lett. B},
  volume  = {766},
  pages   = {137--143},
  year    = {2017},
  doi     = {10.1016/j.physletb.2016.12.062}
}

@article{LedesmaMewes2020,
  author  = {{Gonzalez Ledesma}, Francisco and Mewes, Matthew},
  title   = {Spherical-Harmonic Tensors},
  journal = {Phys. Rev. Research},
  volume  = {2},
  pages   = {043061},
  year    = {2020},
  doi     = {10.1103/PhysRevResearch.2.043061}
}

@article{Shao2019InverseSixth,
  author  = {Shao, Cheng-Gang and Chen, Ya-Fen and Tan, Yu-Jie
             and Yang, Shan-Qing and Luo, Jun and Tobar, Michael Edmund
             and Long, J. C. and Weisman, E. and Kosteleck{\'y}, V. Alan},
  title   = {Combined Search for a Lorentz-Violating Force in Short-Range
             Gravity Varying as the Inverse Sixth Power of Distance},
  journal = {Phys. Rev. Lett.},
  volume  = {122},
  pages   = {011102},
  year    = {2019},
  doi     = {10.1103/PhysRevLett.122.011102}
}

@article{LongKostelecky2015,
  author  = {Long, J. C. and Kosteleck{\'y}, V. Alan},
  title   = {Search for Lorentz Violation in Short-Range Gravity},
  journal = {Phys. Rev. D},
  volume  = {91},
  pages   = {092003},
  year    = {2015},
  doi     = {10.1103/PhysRevD.91.092003}
}

@article{Shao2015ISL,
  author  = {Shao, Cheng-Gang and Tan, Yu-Jie and Tan, Wen-Hai
             and Yang, Shan-Qing and Luo, Jun and Tobar, Michael Edmund},
  title   = {Search for Lorentz Invariance Violation through Tests of the
             Gravitational Inverse Square Law at Short Ranges},
  journal = {Phys. Rev. D},
  volume  = {91},
  pages   = {102007},
  year    = {2015},
  doi     = {10.1103/PhysRevD.91.102007}
}

@article{Shao2016Combined,
  author  = {Shao, Cheng-Gang and Tan, Yu-Jie and Tan, Wen-Hai
             and Yang, Shan-Qing and Luo, Jun and Tobar, Michael Edmund
             and Bailey, Quentin G. and Long, J. C. and Weisman, E.
             and Xu, Rui and Kosteleck{\'y}, V. Alan},
  title   = {Combined Search for Lorentz Violation in Short-Range Gravity},
  journal = {Phys. Rev. Lett.},
  volume  = {117},
  pages   = {071102},
  year    = {2016},
  doi     = {10.1103/PhysRevLett.117.071102}
}

@article{Shao2016Enhanced,
  author  = {Shao, Cheng-Gang and Chen, Ya-Fen and Tan, Yu-Jie
             and Luo, Jun and Yang, Shan-Qing and Tobar, Michael Edmund},
  title   = {Enhanced Sensitivity to Lorentz Invariance Violations in
             Short-Range Gravity Experiments},
  journal = {Phys. Rev. D},
  volume  = {94},
  pages   = {104061},
  year    = {2016},
  doi     = {10.1103/PhysRevD.94.104061}
}

@article{ChenTanShao2017,
  author  = {Chen, Ya-Fen and Tan, Yu-Jie and Shao, Cheng-Gang},
  title   = {Experimental Design for Testing Local Lorentz Invariance
             Violations in Gravity},
  journal = {Symmetry},
  volume  = {9},
  pages   = {219},
  year    = {2017},
  doi     = {10.3390/sym9100219}
}

@article{AdelbergerCollinsHoyle2006,
  author  = {Adelberger, Eric G. and Collins, N. A. and Hoyle, C. D.},
  title   = {Analytic Expressions for Gravitational Inner Multipole Moments
             of Elementary Solids and for the Force between Two Rectangular
             Solids},
  journal = {Class. Quantum Grav.},
  volume  = {23},
  pages   = {125--142},
  year    = {2006},
  doi     = {10.1088/0264-9381/23/1/007},
  note    = {Erratum, Class. Quantum Grav. 23, 5463(E) (2006)}
}

@article{Jin2024d8,
  author  = {Jin, Tao and Li, Jia-Rui and Tan, Yu-Jie and Wang, Pan-Pan
             and Qin, Cheng-Gang and Shao, Cheng-Gang},
  title   = {Constraint of {$d=8$} Lorentz Invariance Violation with New
             Experimental Design},
  journal = {Symmetry},
  volume  = {16},
  pages   = {1432},
  year    = {2024},
  doi     = {10.3390/sym16111432}
}

@article{Jin2026Optimization,
  author  = {Jin, Tao and Wang, Pan-Pan and Huang, Weisheng and Luo, Rui
             and Tan, Yu-Jie and Shao, Cheng-Gang},
  title   = {Parameter Optimization for Torsion-Balance Experiments Testing
             {$d=6$} Lorentz-Violating Effects in the Pure-Gravity Sector},
  journal = {Symmetry},
  volume  = {18},
  pages   = {559},
  year    = {2026},
  doi     = {10.3390/sym18040559}
}

@article{Li2025CQG,
  author  = {Li, Jia-Rui and Tan, Yu-Jie and Jin, Tao and Huang, Wei-Sheng
             and Huang, Hao and Qin, Cheng-Gang and Shao, Cheng-Gang},
  title   = {Experimental Design for Testing Local Lorentz Invariance
             Violation in Gravity},
  journal = {Class. Quantum Grav.},
  volume  = {42},
  number  = {2},
  pages   = {025012},
  year    = {2025},
  doi     = {10.1088/1361-6382/ad9c0c}
}

@article{DongWangShao2024,
  author  = {Dong, Yiming and Wang, Ziming and Shao, Lijing},
  title   = {New Limits on the Local Lorentz Invariance Violation of Gravity
             in the Standard Model Extension with Pulsars},
  journal = {Phys. Rev. D},
  volume  = {109},
  pages   = {084024},
  year    = {2024},
  doi     = {10.1103/PhysRevD.109.084024}
}

\end{document}